%% file: hd.tex
\documentclass[twoside,12pt]{article}

\usepackage{epsfig}

\usepackage{color}
\usepackage{here}
\usepackage{multirow}
\usepackage{bm}

\def\Journal#1#2#3#4{{#1} {#2} (#4) #3 }

\def\NPA{{\em Nucl. Phys.} A}

\def\PLB{{\em Phys. Lett.} B}

\def\PLB{{\em Phys. Lett.} B}
\def\PRL{\em Phys. Rev. Lett.}

\def\PREP{\em Phys. Rep.}

\def\PRD{{\em Phys. Rev.} D}
\def\PRC{{\em Phys. Rev.} C}

\newcommand{\be}{\begin{equation}}
\newcommand{\ee}{\end{equation}}
\newcommand{\bea}{\begin{eqnarray}}
\newcommand{\eea}{\end{eqnarray}}

\begin{document}

\title{ \vspace{1cm} Hadron Physics at J-PARC }
\author{H.\ Ohnishi,$^{1}$ F.\ Sakuma,$^2$ T.\ Takahashi,$^3$
\\
$^1$Research Center for Electron Photon Science (ELPH),\\
 Tohoku University, Sendai 982-0826, Japan\\
$^2$RIKEN Cluster for Pioneering Research (CPR),\\
 RIKEN, Wako 351-0198, Japan\\
$^3$Institute of Particle and Nuclear Study (IPNS),\\
 High Energy Accelerator Research Organization (KEK),\\
 Tsukuba 305-0801, Japan\\
}
\maketitle

\begin{abstract}

The aim of the hadron physics research programs conducted at J-PARC is to explore the
structure of hadronic matter using the world's highest-intensity meson
beams.
Since the first beam was extracted at the hadron experimental facility
(HEF) in February 2009, a wide variety of physics experiments have been
proposed and performed to address open questions regarding quantum
chromodynamics (QCD) at low energy.
The high-intensity $K^-$ and high-momentum beams available at J-PARC
open a new era in hadron and nuclear physics, in which strange and
charm quarks play an important role.
We review the programs focused on addressing the hadron structure
as strongly interacting composite particles, the origin of hadron mass,
and interactions between hadrons under broken flavor SU(3) symmetry.

\end{abstract}

\tableofcontents

\section{Introduction \label{sec:intro}}
Understanding the formation and interaction of a hadron is one of the fundamental goals of hadron physics. 
Nowadays, quantum chromodynamics (QCD) has succeeded in describing the interactions between quarks and gluons. 
However low energy phenomena, such as the formation of a hadron, are not clearly explained because perturbation theory does not work in the low energy regime.

The scientific programs at J-PARC aim to understand the formation and
interaction of hadrons based on QCD. 
Especially, experimental programs underway and planned are intending to answer the following questions and issues:
\begin{itemize}
 \item What are the effective degrees of freedom inside hadrons?
 \item How is the mass of the hadron acquired from the QCD vacuum?
 \item Can a meson exist in nuclear matter by holding the mesonic degrees of freedom?
  \item Reveal the hyperon-nucleon and hyperon-hyperon interaction to achieve a better understanding of nuclear forces. 
\end{itemize}

\begin{figure}[htb]
\centerline{\includegraphics[width=12cm,angle=-90]{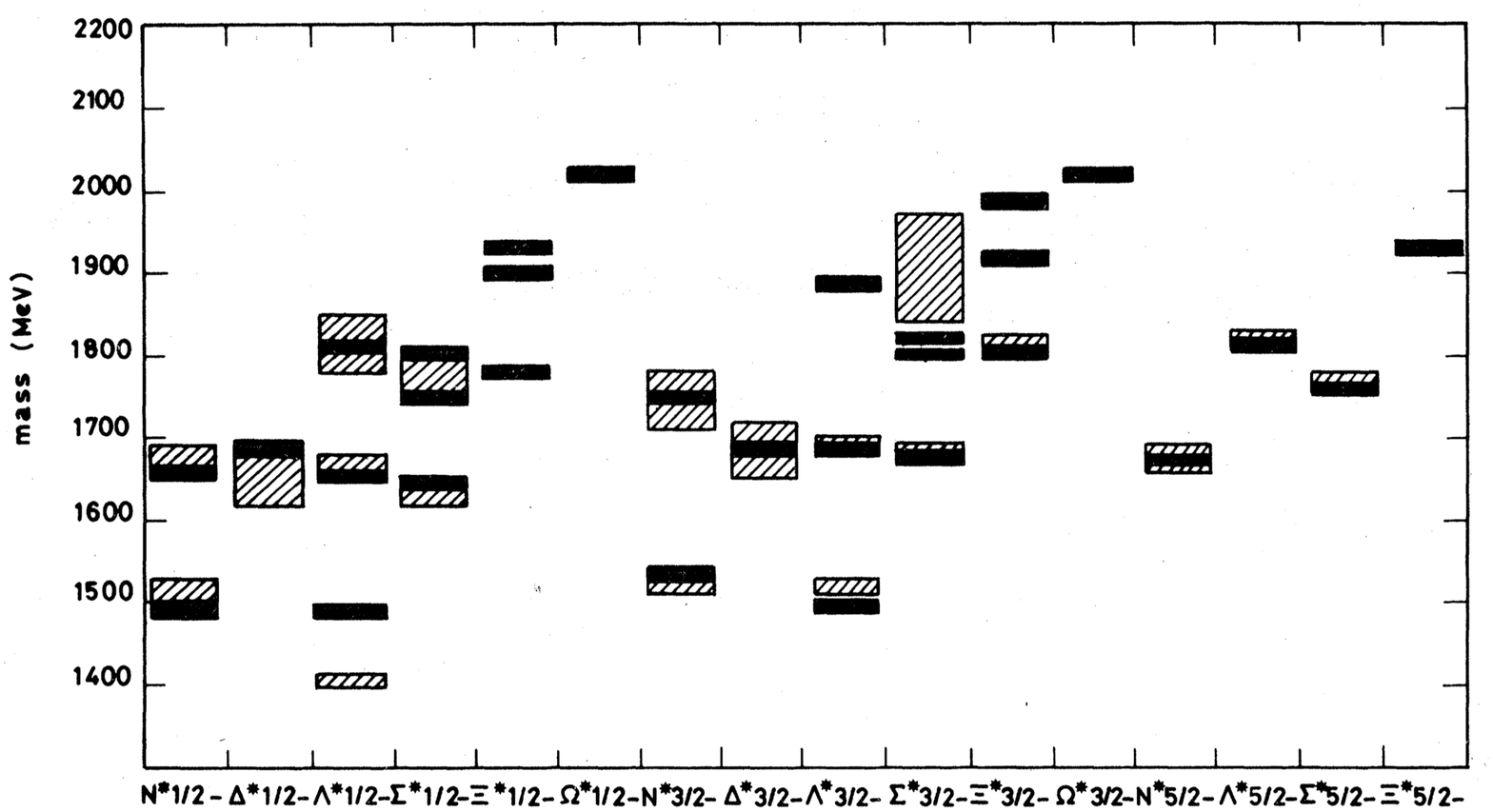}}
\vspace{-1.0cm}
\caption{Mass spectra for negative-parity baryons. The hatched area shows experimental results and black bars represent the theoretical predictions.
The figure is taken from Ref.~\cite{Isgur:1978xj}.}
\label{fig:Isgur-karl}
\end{figure}

QCD is supposed to generate constituent quarks dynamically as low-energy effective degrees of freedom.
The constituent quark model assumes that the low-lying hadrons are composed of a few constituent quarks that carry color, spin, and flavor quantum numbers. The model successfully gives the number of states, their quantum numbers, and the flavor multiplet structure of their spectrum. 
The meson and baryon spectra suggest that the constituent quark has a dynamical mass of order 300 MeV/$c^2$, and that they are confined in a hadron.
The model works very well for the ground state hadrons, while its prediction for the excited states is not satisfactory.
Figure.~\ref{fig:Isgur-karl}~\cite{Isgur:1978xj} shows a prediction of the negative-parity baryon spectrum in a constituent quark model. We notice that most of the low-lying states are fairly well reproduced, but there is an exception -- the $\Lambda$($1/2^-$) state, where the prediction is off by 100 MeV from the observed state. It is also noted that many predicted states are not yet observed.    

Another deficiency of the constituent quark model is that the dynamics of confinement is not clear for multi-quark systems. QCD restricts hadrons to be color singlet but does not forbid multi-quark states, such as $qq\bar{q}\bar{q}$, or $qqqq\bar{q}$, referred to the tetra-quark and pentaquark baryons, respectively. 
The reason why the number of exotic hadrons is so small compared to ordinary 
hadrons with $qqq$ or $q\bar{q}$ configuration is not yet known. 

The study of light hadrons in QCD sum rules revealed that their masses are generated by quark and gluon condensates in the QCD vacuum.  The quark condensates are the direct consequence of spontaneous chiral symmetry breaking predicted by Y.~Nambu and G.~Jona-Lasinio~\cite{Nambu:1961tp,Nambu:1961fr}.
It is fascinating to confirm the relations between the hadron mass and the quark condensates from experiments.
QCD also predicts that the chiral symmetry might be restored at high temperature or high baryon density.
The embedded meson in a nucleus is a powerful  tool to access the information on the QCD vacuum. 
One of the methods is to observe the change of the meson properties in nuclear matter, {\it i.e.}, modification of the spectral function.
Vector mesons are used as a ``probe'' for this purpose.
The other is trying to form a meson-nuclear bound state 
and draw out the information on binding energy and decay width of the state which will be connected 
with the QCD vacuum structure in the nucleus. 
Hereafter we are focusing on the research on anti-kaon ($\bar{K}$) interactions with nucleons and nuclei, because the world's highest-intensity kaon beam is available at J-PARC.
The interaction between a $\bar{K}$ and a nucleon is known to be strongly attractive, and there is a hyperon resonance below the sum of the $\bar{K}$ and nucleon masses, {\it i.e.}, $\Lambda(1405)$.
Thus $\Lambda(1405)$ is expected to be a quasi-bound state of a $\bar{K}$ and a nucleon. 
If so, it is reasonable to extend the idea that the $\bar{K}$ embedded into a
nucleus may form a kaonic nuclear quasi-bound state. 
It should be noted that the density of the kaonic nuclear quasi-bound state could reach higher values than normal nuclear density, 
which will be connected to the physics inside neutron stars.

In the world of nuclei, a nucleon, {\it i.e}., a proton or a neutron, is used as an effective 
degree of freedom to describe them, such as their structures and properties.
One of the outstanding problems is the origin of short-range repulsion and spin-orbit force
in nucleon-nucleon ($NN$) interactions. 
Because in short-range parts, where nucleons and the exchanged mesons overlap each other, 
it is reasonable to consider quark and gluon degrees of freedom. 
The repulsive force in the short-range is essential to explain nuclear saturation properties due to the exquisite balance
between the middle and long-range attractions. 
The spin-orbit force in nucleon-nucleus potentials and $NN$ interactions are also essential to explain the periodical stability of nuclei -- nuclear magic numbers. 
These important features which are originated in the short-range
part must be understood based on QCD.
One of the most effective methods to investigate and understand
the short-range phenomena in nuclear forces is by introducing new degrees of
freedom to the system, 
{\it i.e.}, strange ($s$) quarks, which extends the $NN$ interaction 
under the isospin symmetry of SU(2) to the interactions 
between octet baryons ($B_8$$B_8$ interactions) 
under flavor SU(3) (SU(3)$_{\rm f}$) symmetry.
Although the $s$-quark is heavier than $u$ and $d$ quarks, 
the mass difference is still sufficiently small to treat the system 
under SU(3)$_{\rm f}$
considering the QCD energy scale $\Lambda_{QCD}$ of $\sim$ 200 MeV. 
These newly introduced hyperon-nucleon ($YN$) and $YY$ interactions should
be determined by experiments in close cooperation with
theoretical studies.
Due to the difficulty of direct scattering experiments,
spectroscopic studies on hypernuclei which contain one or more hyperons are essentially important to investigate $YN$ and $YY$ interactions.

The nature of QCD, {\it i.e.}, strong coupling phenomena in the low energy region, is thus a challenging problem to solve analytically. An approach to overcome this difficulty is lattice QCD simulation. 
From the effort of theorists together with the progress of supercomputers, 
some part of the hadron mass spectra and interaction between hadrons 
can be extracted from the QCD Lagrangian, 
and the results are consistent with the experimental data.
However, difficulties to reproduce or predict the properties of hadrons still remain, especially for the properties and interactions of excited hadrons. Moreover, it has been extremely difficult to solve QCD at finite density by lattice QCD simulation. Therefore, experimental efforts could represent essential inputs to understanding the nature of QCD.

In this article, we provide a brief introduction to the facility in Sec.~\ref{sec:hall}, including an overview of the J-PARC
accelerator.
In Sec.~\ref{sec:exp}, we review the hadron physics experiments at J-PARC, 
focusing on {\it ``QCD Vacuum and Hadron Structure''}, {\it
``Antikaon-Nucleon Interaction''}, and {\it``Baryon-Baryon Interaction''}.
The future prospects of the HEF extension are also presented in
Sec.~\ref{sec:exension}, which are expected to enhance the opportunities of hadron physics 
at J-PARC.
Finally, a summary is given in Sec.~\ref{sec:summary}.

\section{J-PARC Hadron Experimental Facility \label{sec:hall}}
\subsection{Overview of J-PARC}
\subsection{Overview of J-PARC}

\begin{figure}[htbp]
\centerline{\includegraphics[width=0.6\textwidth]{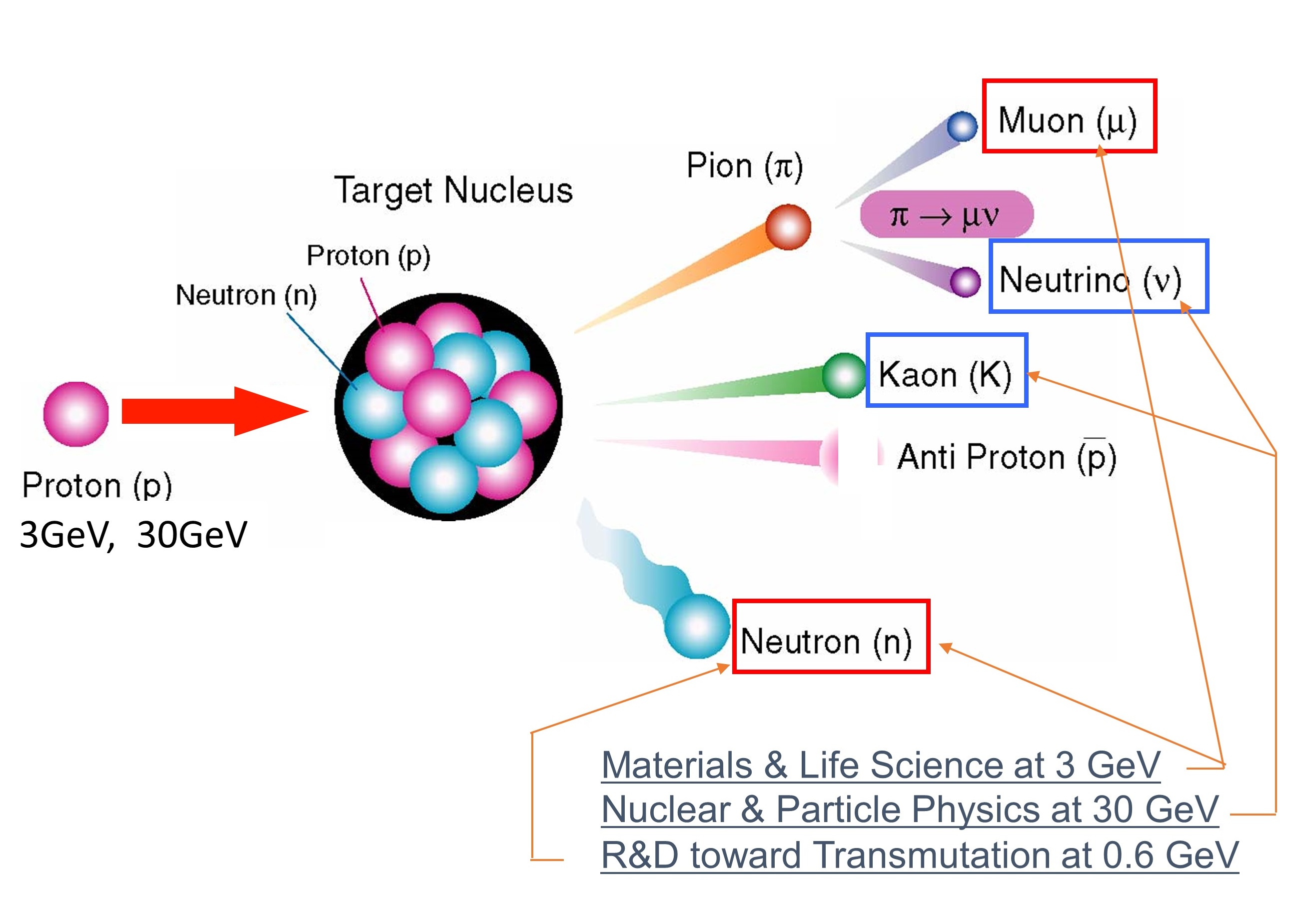}}
\caption{Multi purpose accelerator facility, J-PARC. A variety of particles
produced by a high-intensity proton beam are provided as beams
for a variety of experimental research on basic science to industrial applications.
.}
\label{fig:multi-use}
\end{figure}

\begin{figure}[htbp]
\centerline{\includegraphics[width=0.9\textwidth]{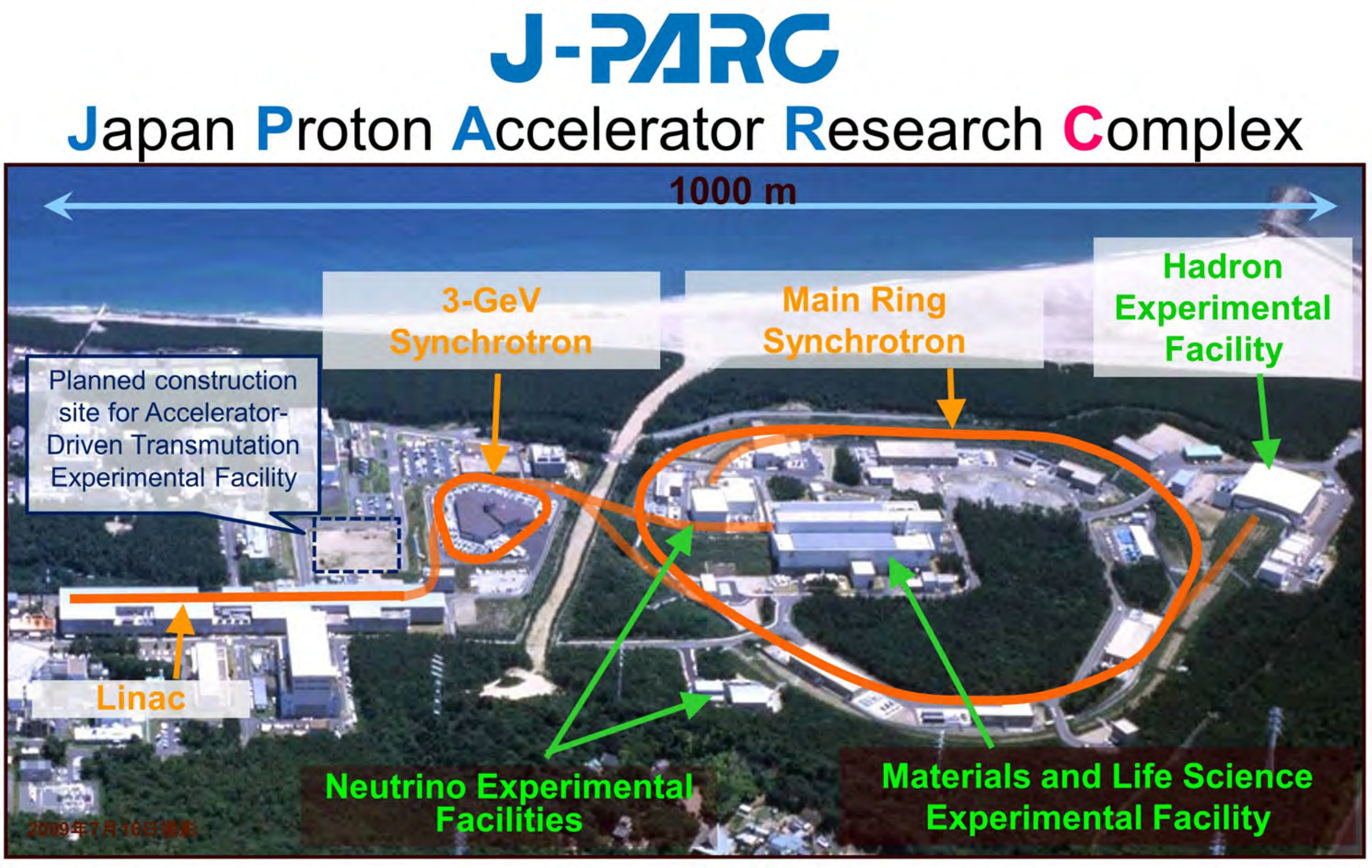}}
\caption{Aerial photograph of J-PARC.
J-PARC consists of three high-intensity proton accelerators: linac, 
3 GeV synchrotron and main ring synchrotron,
and three experimental facilities, the Material and Life Science Facility, 
the Hadron Experimental Facility, and the Neutrino Experimental Facility. 
The construction of an accelerator-driven transmutation facility is planned.}
\label{fig:J-PARC}
\end{figure}

The Japan Proton Accelerator Research Complex (J-PARC) is a
multi-purpose accelerator facility located in Tokai village,
Japan~\cite{Nagamiya:2006en,Nagamiya:2012tma}.
The aims of J-PARC are to promote a variety of scientific research
programs ranging from 
the basic science of particle, nuclear, atomic and condensed
matter physics and life science to applications for industry use and
future nuclear transmutation, 
using various types of high-intensity secondary beams of neutrinos,
muons, pions, kaons, protons, neutrons, and their antiparticles, as shown
in Fig.~\ref{fig:multi-use}.
The J-PARC accelerator consists of three high-intensity proton
accelerators, as shown in Fig.~\ref{fig:J-PARC}: a 400 MeV linac, a 3 GeV
rapid cycling synchrotron (RCS), and a main ring synchrotron
(MR).
Both the RCS and MR have a three-fold symmetry, and circumferences
of 348.3 m and 1567.5 m, respectively.

The linac is a beam injector to the RCS, which consists of a negative
hydrogen ion source, a 3 MeV radio frequency quadrupole (RFQ), a 50 MeV drift tube linac (DTL), a
191 MeV separated-type DTL (SDTL), and a 400 MeV annular-ring coupled
structure (ACS)~\cite{Ikegami:2012uma}.
At the linac, negative hydrogen ions instead of protons are accelerated
up to 400 MeV because the charge-exchange injection scheme is
adopted with a stripper foil at the RCS injection.
The RCS, which is the world's highest class of a high-power pulsed
proton driver, accelerates the injected protons up to 3 GeV at a
repetition rate of 25 Hz~\cite{Hotchi:2012vma}.
Most of the RCS beam pulses are delivered to the Material and Life
Science Facility (MLF), while only a portion of the pulses are injected
to the MR.
Muon and neutron beams are available at the MLF, where beams are
generated by colliding 3 GeV protons with carbon and mercury targets,
respectively.

The protons injected from the RCS to the MR are accelerated up to 30
GeV, and delivered to the Neutrino Experimental Facility (NEF) and the Hadron
Experimental Facility (HEF) with different extraction
modes~\cite{Koseki:2012wma}.
In the fast extraction mode (FX), all beam bunches are extracted within
a one-turn time period to the NEF.
Neutrino and anti-neutrino beams are used for the Tokai to Kamioka
long-baseline neutrino oscillation experiment,
T2K~\cite{Abe:2011ks}.
On the other hand, the beam is extracted to the HEF over several seconds
in the slow extraction mode (SX).
The uniform structure and low ripple noise of the slow extraction beam 
enables a wide variety of experiments to be performed,
which generally requires the coincidence of several detectors to
obtain the low-frequency occurrence events of interest among the large
number of background interactions.

User operation was started in December 2008 at the MLF.
In the MR, the beam was successfully accelerated to 30 GeV on
December 23, 2008. 
The first beam was extracted to the HEF on February 23, 2009.
Since then, the beam power has gradually increased by 536 kW at
the RCS, $\sim$ 500 kW at the MR-FX, and 51 kW at the MR-SX, as of the end of 2019.
Now, further improvement of the beam power is proceeding and planned for
each facility: a 1.2 MW equivalent test was successfully conducted with
the RCS in 2018 for future $\sim$ 1 MW operation at the MLF, and an upgrade of
the MR main-magnet power supplies planned for 2021 will realize $\sim$ 1 MW
operation of the MR-FX and over 100 kW operation of the MR-SX by
operation with a higher repetition rate than that at the present.

\subsection{Hadron Experimental Facility}

The HEF focuses on particle and nuclear physics 
using the primary 30 GeV proton beam and secondary beams of
pions, kaons, antiprotons, and muons.
The 30 GeV proton beam slowly extracted from the MR is transported
through a beam-switching yard (SY) to a secondary-particle production
target (T1) located in the hadron experimental hall (HD-hall) 
with a width of 60 m and 56 m in length~\cite{Agari:2012ana}.

In the SY, which is $\sim$ 200 m long along the primary beamline,
the primary beam is shifted up by 2.9~m to avoid beam halo originated from beam loss at the extraction devices. 
In the middle of the beam transport line, a beam branching device 
(a Lambertson magnet at present) is installed, so that 
a small fraction of the beam is transported to the new beamline under construction.
The new beamline is branched to the high momentum (high-p) and
COMET beamlines in the HD-hall.

\begin{figure}[htbp]
\centerline{\includegraphics[width=0.6\textwidth]{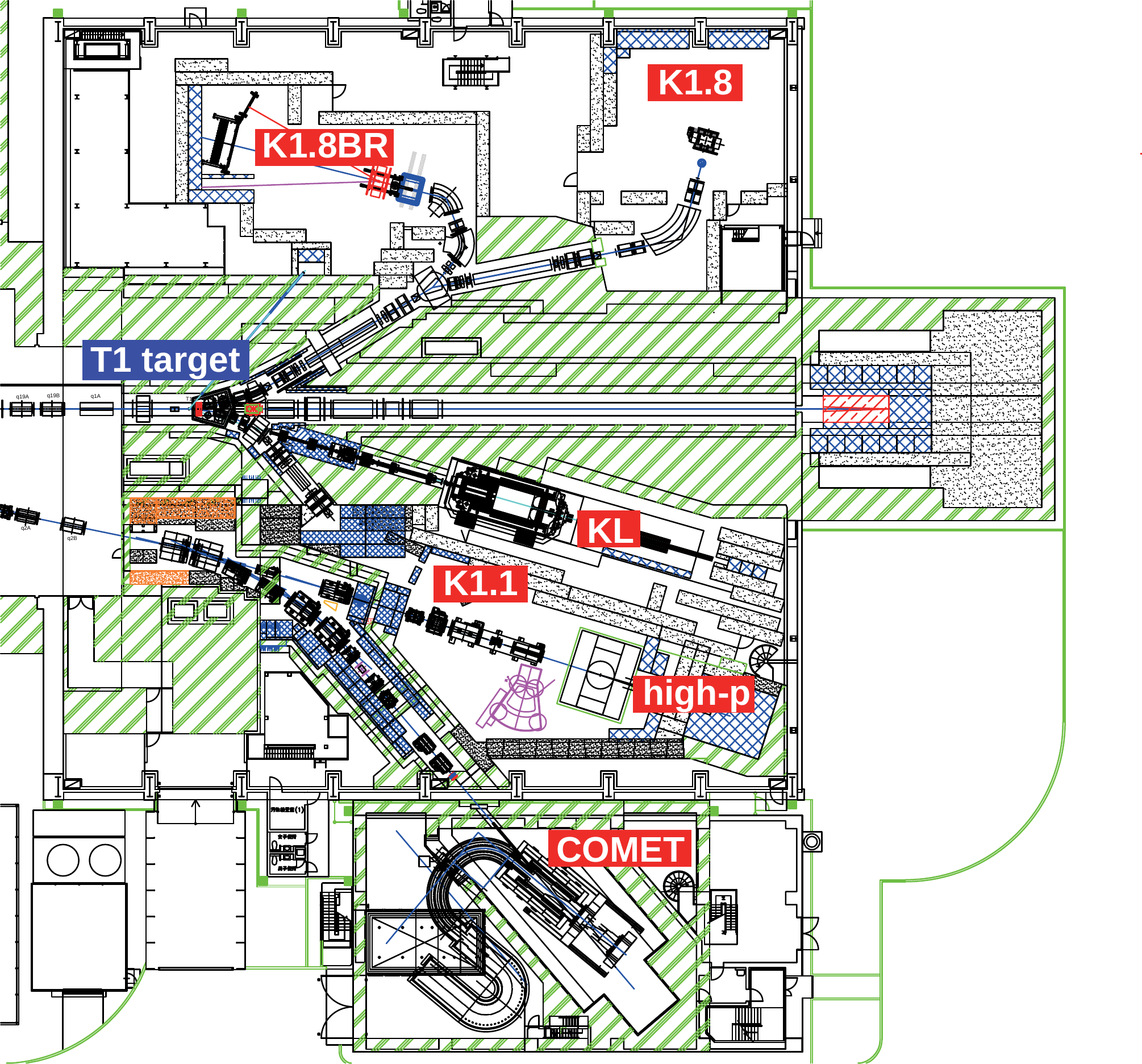}}
\caption{Layout of the hadron experimental (HD) hall and the south experimental hall.
Three secondary beamlines, K1.8, K1.8BR, and KL, are in operation.
The high-p beamline will be ready for operation in February 2020.
}
\label{fig:HDhall}
\end{figure}

Secondary beams produced at the T1 target are extracted from the primary
beamline to three charged secondary beamlines (K1.8, K1.8BR, and
K1.1)~\cite{Agari:2012kid} and one neutral secondary beamline (KL).
Figure.~\ref{fig:HDhall} shows a layout of the HD-hall and the south hall.

The K1.8 beamline is mainly designed for systematic studies of the
double strangeness systems via the $(K^-, K^+)$ reactions, such as a
spectroscopic study of $\Xi$ hypernuclei~\cite{Takahashi:2012cka}.
A maximum central momentum of $\sim$ 2 GeV/$c$ is available at K1.8
because the cross section of the $p(K^-, K^+)\Xi$ reaction is known to
be a maximum at 1.8 GeV/$c$.
The K1.8 beamline is composed of 4 sections: the front-end section, the
first mass separation section, the second mass separation section, and
the beam analyzer section.
The total length of the beamline is 45.8 m.
In the front-end section, secondary particles from the primary beam are
extracted with 6$^{\circ}$, where the kaon production cross section is
expected to be a maximum according to the Sanford-Wang
parametrization~\cite{Sanford:1967zza}. 
Two 6 m long electrostatic separators (ES) 
are employed to separate kaons from
pions in the first and second mass separation sections.
The beam trajectory and momentum are determined in the beam analyzer
section, called the K1.8 beam spectrometer, 
where point-to-point optics are realized between the entrance
and exit of the section with a QQDQQ magnet configuration.
A set of tracking detectors and beam timing counters
installed at the entrance and exit of the
QQDQQ provide precise beam information. 

The K1.8BR beamline is a branch line of the K1.8 beamline designed as a short beamline 
to deliver a low-momentum mass-separated kaon beam of 0.7--1.1
GeV/$c$, of which the total length is 31.3 m.
The cross sections of the quasi-elastic reactions of $K^- N \to \bar K N$
are maximal around 1 GeV/$c$; therefore, the K1.8BR beamline is suitable for
experimental studies of the $\bar K N$ interactions via the $(K^-, N)$
reactions with light nuclear targets~\cite{Agari:2012gj}.
The beamline shares the upstream components of K1.8, {\it i.e.} the
front-end section and the first mass separation section of K1.8.
The beam is bent to the opposite side of K1.8 at the D magnet after the
first mass separation section, and transported to the experimental
target through a momentum analyzer section composed of QDQD magnets.

The KL beamline is a neutral secondary beamline used to search for the 
rare CP-violating kaon decay, $K^0_L \to \pi^0 \nu \bar{\nu}$
(the KOTO experiment)~\cite{Yamanaka:2012yma}.
The beamline consists of a 1$^{\rm st}$ collimator, a sweeping magnet,
and a 2$^{\rm nd}$ collimator.
The $K^0_L$ beam is extracted with 16$^{\circ}$ and transported to a
detector system mainly composed of a CsI electromagnetic calorimeter and
a charged-particle veto counter.

The K1.1 beamline is constructed on the opposite side of the K1.8
beamline.
A maximum momentum of 1.1 GeV/$c$ is available, where mainly precise
measurements of $\Lambda$ hypernuclei are conducted.
The beamline consists of a front-end section, a mass-separation section
with two stages of cross field type ES with 2 m of length, 
and a beam analyzer section with a DQQ
magnet configuration.
The total length of the beamline is 27.1 m.
Experiments at K1.1 will be performed after the completion of the first
stage experiments at the high-p beamline, by rebuilding
the layout of the magnets and the experimental area at the south side of
the HD-hall.
This is due to interference of space between the high-p and the
K1.1 beamlines.

The construction of the high-p beamline was completed in 2020 January.
It will provide a primary 30 GeV proton beam with
an intensity of up to $\sim$ $10^{10}$ per pulse.
The beam commissioning will start from March 2020. 
The beamline was originally designed to investigate the in-medium
spectral change of vector mesons produced via the $p + A$ reactions.
The use of secondary high-momentum but mass-unseparated beams of 
pions, kaons, and antiprotons up to 20 GeV/$c$ is also planned in the future
by placing a thin production target at the branching point.
Charmed baryon spectroscopy will be performed via 
the $(\pi^-, D^{*-})$ reactions at $\sim$ 20 GeV/$c$ 
to explore the effective degrees of freedom to describe hadrons.
The COMET beamline at the south experimental hall
under construction
is dedicated to deliver 8 GeV protons 
in the bunched slow extraction mode for the experiment to
search for $\mu^-$ to $e^-$ conversion, 
COMET~\cite{Adamov:2018vin}.

As of April 2019, a beam power of 51 kW was achieved with a 2.0 s
beam spill length in a 5.2 repetition cycle, which corresponds to
5.4$\times$10$^{13}$ protons per pulse.
Up to the spring of 2019, the beam power was limited by the production
target system composed of a gold target and an indirect water-cooling
system~\cite{Takahashi:2015}.
To increase the beam power for the HEF, a new production target
system, allowable to 95 kW, was installed in November 2019.
A new target system up to $>$150 kW is also under
development, which employs a directly-cooled rotating-target system.

\section{Experiments at the J-PARC HEF \label{sec:exp}}
In this section, we describe details of individual experiments conducted
at the HEF.
Table~\ref{table:exp} summarizes the current status of the experiments
classified as ``completed'', ``ongoing'', ``forthcoming'', and
``planned''.

\begin{table}[htbp]
 \caption{Summary of the status of the experiments at the HEF described
 in this article.}
 \begin{center}
\begin{tabular}{lp{10cm}llp{3cm}}
\hline
 \multicolumn{2}{c}{\multirow{2}{*}{Experiment}}	& \multirow{2}{*}{Beamline}	& Beam
	 & \multirow{2}{*}{Status}	\\
 & & & particle & \\
\hline
 E03 & Measurement of X-rays from $\Xi$-atom & K1.8 & $K^-$ & forthcoming \\
 E05 & Spectroscopic study of $\Xi$-hypernucleus, $^{12}_{\Xi}$Be, via the $^{12}$C$(K^-,K^+)$ reaction & K1.8 & $K^-$ & completed \\
 E07 & Systematic study of double strangeness system with an emulsion-counter hybrid method & K1.8 & $K^-$ & completed \\
 E10 & Production of neutron-rich $\Lambda$-hypernuclei with the double charge-exchange reactions & K1.8 & $\pi^-$ & completed \\
 E13 & Gamma-ray spectroscopy of light hypernuclei & K1.8 & $K^-$ & completed \\
 E15 & A search for deeply-bound kaonic nuclear states by in-flight $^3$He$(K^-, n)$ reaction & K1.8BR & $K^-$ & completed \\
 E16 & Electron pair spectrometer at the J-PARC 50-GeV PS to explore the chiral symmetry in QCD & high-p & $p$ & ongoing \\
 E19 & High-resolution search for $\Theta^+$ pentaquark in $\pi^- p \to K^- X$ reactions & K1.8 & $\pi^-$ & completed \\
 E26 & Direct measurements of $\omega$ mass modification in A$(\pi^-,n)\omega$ reaction and $\omega \to \pi^0 \gamma$ decays & K1.8 & $\pi^-$ & planned \\
 E27 & Search for a nuclear $\bar K$ bound state $K^-pp$ in the d$(\pi^+, K^+)$ reaction & K1.8 & $\pi^+$ & completed \\
 E29 & Study of in medium mass modification for the $\phi$ meson using $\phi$ meson bound state in nucleus & K1.8BR & $\bar p$ & planned \\
 E31 & Spectroscopic study of hyperon resonances below $\bar K N$ threshold via the $(K^-,n)$ reaction on deuteron & K1.8BR & $K^-$ & completed \\
 E40 & Measurement of the cross sections of $\Sigma p$ scatterings & K1.8 & $\pi^{\pm}$ & ongoing \\
 E42 & Search for $H$-dibaryon with a large acceptance hyperon spectrometer & K1.8 & $K^-$ & forthcoming \\
 E45 & 3-body hadronic reactions for new aspects of baryon spectroscopy & K1.8 & $K^-$ & forthcoming \\
 E50 & Charmed baryon spectroscopy via the $(\pi^-, D^{*-})$ reaction & high-p & $\pi^-$ & planned \\
 E57 & Measurement of the strong interaction induced shift and width of the 1st state of kaonic deuterium at J-PARC & K1.8BR & $K^-$ & forthcoming \\
 E62 & Precision spectroscopy of kaonic helium 3 $3d \to 2p$ X-rays & K1.8BR & $K^-$ & completed \\
 E63 & Proposal of the 2nd stage of E13 experiment & K1.1 & $K^-$ & planned \\
 E70 & Proposal for the next E05 run with the $S$-$2S$ spectrometer & K1.8 & $K^-$ & forthcoming \\
 E72 & Search for a narrow $\Lambda^*$ resonance using the $p(K^-, \Lambda)\eta$ reaction with the hypTPC detector & K1.8BR & $K^-$ & forthcoming \\
 \hline
\end{tabular}
 \end{center}
\label{table:exp}
  \end{table}

\subsection{QCD Vacuum and Hadron Structure \label{subsec:QCD}}
Hadrons are the excitations of the QCD vacuum itself, which hold basic information regarding non-perturbative QCD. Precise spectroscopy of a nucleon resonance, {\it i.e.}, the spectroscopy of N*, is known to be a way to access that information. However, most of the nucleon resonances have a large width, so that there is no easy way to extract the pole position of the nucleon resonance. Therefore, partial wave analysis (PWA) to obtain the pole position of the resonance is mandatory. The study of nuclear resonances has a long history, and precise data for $\pi N$ interactions have already been accumulated. However, a high-quality data set, such as $\pi N \to \pi\pi N$, for the non-strangeness sector and $\pi N\to KY$, where $Y$ represents a baryon with a strange quark, are still missing. These measurements will be a key to accessing non-perturbative QCD vacuum structure via the spectroscopy of nucleon resonances. An experimental program to accumulate these missing data is now in preparation at the J-PARC K1.8 beamline, referred to as the J-PARC E45 experiment.

QCD allows for the existence of a hadron with non-meson and non-baryon configurations, the so-called exotic hadron, such as the tetra-quark ($qq\bar{q}\bar{q}$), pentaquark-baryon ($qqqq\bar{q}$), dibaryon ($qqqqqq$), and so forth. 
One of the critical missions for hadron physics is to clarify the existence of exotic hadrons and to identify their properties. 
The search for the pentaquark-baryon with strangeness$=+1$, {\it i.e.}, $\Theta^{+}$, which consists of $uudd\bar{s}$, was performed as the J-PARC E19 experiment. On the other hand, research to find one of the dibaryon states, the $H$-dibaryon, which consists of $uuddss$, is planned as the J-PARC E42 experiment.  

Another question needs to be answered with respect to understanding the effective degrees of freedom (DoF) to describe a hadron itself. For ordinary mesons and baryons, especially for the ground state hadrons, constituent quarks are known to be reasonable degrees of freedom to describe their properties, such as mass and spin-parity. However, especially for excited baryons or mesons, the correlation between a quark pair, {\it i.e.}, $qq$ inside the hadron, will be enhanced and those correlated quark pairs, which are referred to as a di-quark correlation, will be acting as DoF. One of the best ways to reveal the importance of a di-quark correlation is the spectroscopy of a baryon with a heavy quark, such as charmed baryons. An experiment to show the di-quark correlation in charmed baryons is planned at the J-PARC high momentum beamline as the J-PARC E50 experiment. 
 
Another significant issue for hadron physics is the origin of hadron mass, {\it i.e.}, to reveal the mechanism whereby mass is acquired from the QCD vacuum~\cite{Hayano:2008vn}. 
In the sector of light quarks, the underlying mechanism is accompanied by the spontaneous breaking of chiral symmetry. 
Efforts to access the information experimentally are in progress internationally. 
Partial restoration of the chiral symmetry in a nucleus is expected theoretically, and such partial restoration 
will lead to a reduction of the magnitude of the condensate, $|\langle \bar{q}q\rangle|$. 
QCD sum rules connect hadron spectral functions to such condensates. 
Therefore, spectral functions of mesons in the nucleus could be modified~\cite{Brown:1991kk,Hatsuda:1991ez,Klingl:1997kf,Gubler:2018ctz}. 
Precise measurement of the vector meson spectral functions, such as $\rho$, $\omega$ and $\phi$ mesons, is under preparation as the J-PARC E16 experiment.  

The other way to access information regarding the partial restoration of the chiral symmetry is to search for the meson-nuclear bound state, and determine its binding energy and decay width~\cite{Metag:2017yuh}.
The existence of the meson-nuclear bound state indicates a reduction of the meson mass in nuclear media. Experiments to search for the meson nuclear bound state, especially for $\omega$ and $\phi$ mesons, have been proposed as the J-PARC E26 and E29 experiments.

Details of obtained and expected results for each experiment are reviewed in the following sections.

\subsubsection{Search for the $\Theta^+$ penta-quark baryon \label{subsubsec:E19}}

In 2003, the $S=+1$  baryon was observed via the $\gamma n \to K^- K^+ n$ reaction at the SPring-8/LEPS experiment~\cite{Nakano:2003qx}. The results showed a peak structure in the $K^+$n invariant mass spectrum at 1535 MeV/$c^2$. The obtained width is surprisingly narrow; it is less than 25 MeV/$c^2$, which is consistent with the experimental resolution. The quantum number of the observed baryon can be identified as S=+1, and the width is much less than an ordinary nucleon resonance, which is of the order of 100 MeV/$c^2$. Thus the observed baryon is expected to be a candidate for an exotic hadron, {\it i.e.},  a pentaquark baryon with one anti-strange quark ($uudd\bar{s}$). The baryon was named $\Theta^+$, and it is expected to be one of the lightest anti-decuplet members of the set of baryons with four quarks and one anti-quark, which was predicted by D.~Diakono, V.~Petrov, and M.~Polyakov in 1997~\cite{Diakonov:1997mm}.  In 2009, the SPring-8/LEPS collaboration presented a new result for $\Theta^+$ via the $\gamma d \to K^+K^- pn$ reaction with high statistics~\cite{Nakano:2008ee}. They confirmed the previous result, namely, the existence of the peak structure in the $K^+ n$ invariant mass spectrum at the same position, $M_{K+n}=1535$ MeV/$c^2$. Intensive studies to confirm the existence of $\Theta^+$ have been performed in photoproduction at Jefferson Lab~\cite{Ireland:2007aa}, with proton--proton collisions at COSY~\cite{Abdel-Bary:2007}, and in high energy collider experiments~\cite{Bai:2004gk, Wang:2005fc}. However, no clear evidence has been reported. Results from 
KEK 12 GeV proton synchrotron (KEK-PS) E522 show a bump structure at 1530.6 $^{+2.2}_{-1.9}$(stat.)$^{+1.9}_{-1.3}$(syst.) MeV via the $\pi^-p\to K^-X$ reaction, 
for an incident pion momentum of 1.92  GeV/$c$, even though the statistical significance for the measurement was only 2.5 -- 2.7$\sigma$~\cite{Miwa:2006if}. Thus it is clear that a new experiment with a hadron beam is required to conclude whether the $\Theta^+$ baryon exists.
 
\begin{figure}[htbp]
\centerline{\includegraphics[width=8cm,angle=-90]{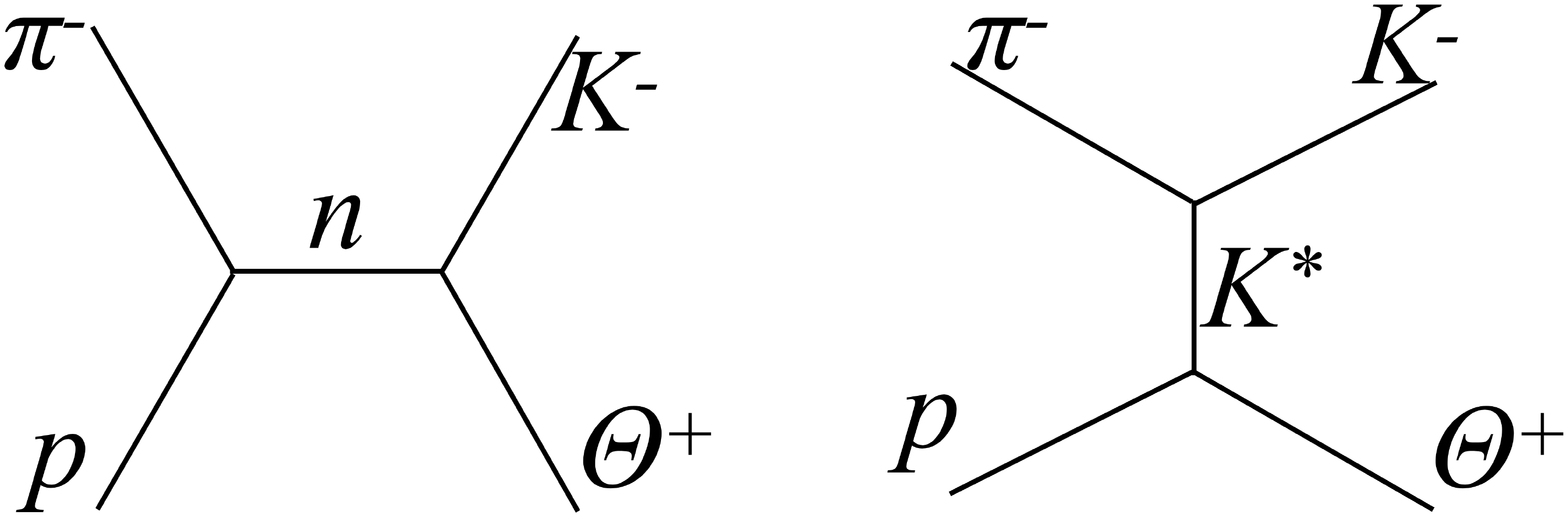}}
\caption{Feynman diagram for $\Theta^+$ production via $(\pi^-,K^-)$ reaction. }
\label{fig:penta_prod}
\end{figure}

Figure~\ref{fig:penta_prod}  shows diagrams for $\Theta^+$ production via the ($\pi^-,K^-$) reaction. 
To identify $\Theta^+$ production in the J-PARC E19 experiment, missing mass spectroscopy was performed by high precision analysis of the incident $\pi^-$ and out-going $K^-$ momenta using the high-resolution beamline spectrometer and the superconducting kaon spectrometer (SKS) placed on the K1.8 beamline, respectively. The experiment collected data for two different incident $\pi$ momenta settings, 1.92 GeV/$c$ and 2.01 GeV/$c$, to investigate the excitation function of $\Theta^+$ production for this production channel.  The achieved missing mass resolution for $\Theta^+$ production was found to be 1.72 MeV and 2.13 (FWHM), for incident pion momenta of 1.92 GeV/$c$ and 2.01 GeV/$c$, respectively.

\begin{figure}[htbp] 
\centerline{\includegraphics[width=17cm]{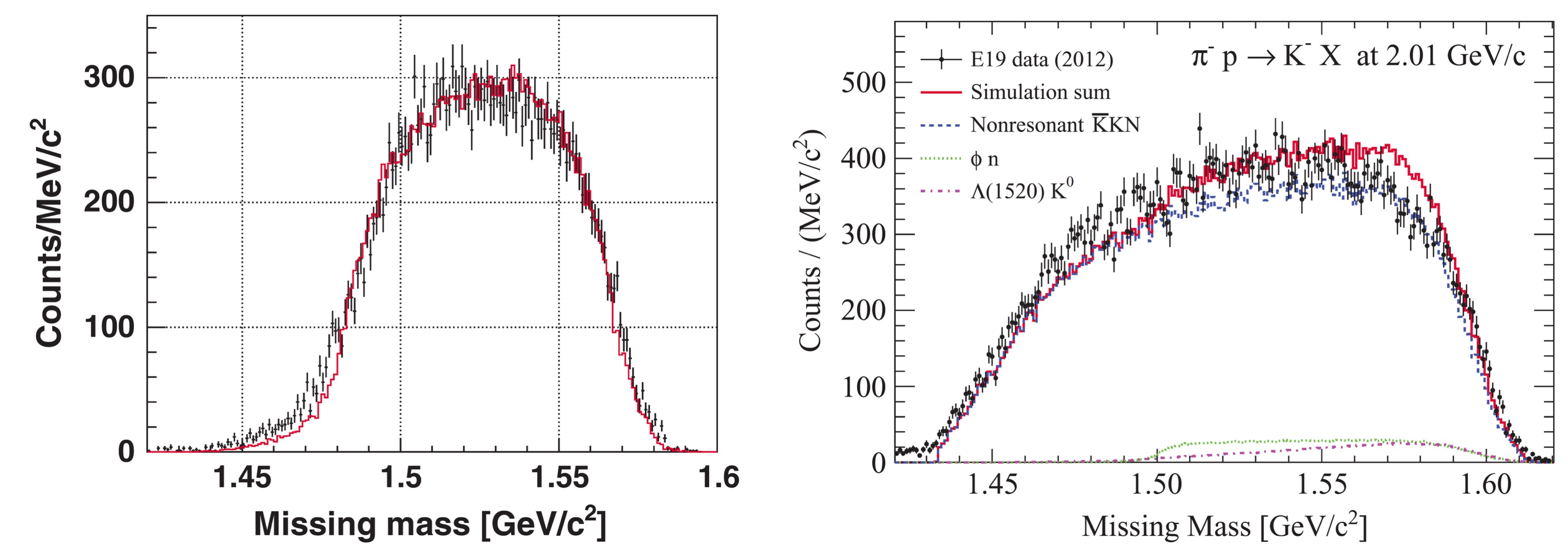}}
\caption{Missing mass spectra of $\pi^-p\rightarrow K^-X$ reaction at 1.92 GeV/$c$ (left) and 2.01 GeV/$c$ (right). The red lines show simulation results for both figures. No clear peak structure is  seen around the mass of $\Theta^+$, M $\sim$ 1530~MeV/$c^2$. The figures are taken from Ref.~\cite{Shirotori:2012ka} (left) and Ref.~\cite{Moritsu:2014bht} (right).}
\label{fig:E19_results}
\end{figure}

Figure~\ref{fig:E19_results} shows the obtained results from the E19 experiment~\cite{Shirotori:2012ka, Moritsu:2014bht}. The right and left figures show the missing mass spectra reconstructed via the $p(\pi^-,K^-)$ reaction with incident pion momenta of 1.92 GeV/$c$ and 2.01 GeV/$c$, respectively. No clear peak structure was observed around the expected $\Theta^+$ mass region for either momentum setting.     

\begin{figure}[htbp] 
\centerline{\includegraphics[width=14cm,angle=-90]{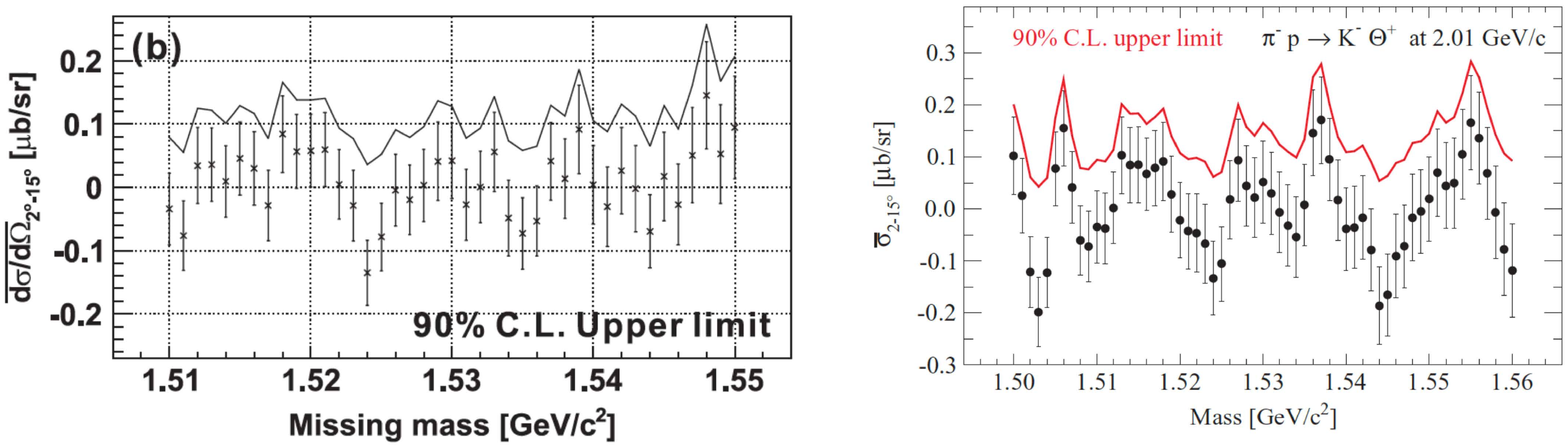}}
\caption{Production cross section of $\Theta^+$ as a function of its mass for incident pion beam momenta of 1.92 GeV/$c$ (Left) and 2.01 GeV/$c$ (right). The line indicates the upper limit of the production cross section at the 90\% confidence level. The figures are taken from Ref.~\cite{Shirotori:2012ka} (left) and Ref.~\cite{Moritsu:2014bht} (right).}
\label{fig:E19_results2}
\end{figure}

The E19 experiment also determined the upper limit of the production cross section of $\Theta^+$ for this reaction, as summarized in Figure~\ref{fig:E19_results2}.  The obtained upper limit is 0.28 $\mu$b/sr at 90\% confidence level (CL) in the mass region 1.510 to 1.550 GeV for both 1.92 and 2.01 GeV/$c$, which is an order of magnitude lower than the previous E522 result of 2.9 $\mu$b/sr. 
The decay width at the 90\% CL upper limit was also evaluated and derived to be less than 0.36 and 1.9 MeV for the assumed spin-parities of the $\Theta^+$ of $1/2^+$ and $1/2^-$, respectively, by combining the theoretical calculations using the effective Lagrangian. 

\subsubsection{Search for the $H$ dibaryon \label{subsubsec:E42}}

In 1977, R.L.~Jaffe predicted the existence of a hadron with six quarks including two strange quarks, as a compact object~\cite{Jaffe:1976yi}. The object was named $H$-dibaryon.
 
Recently, lattice QCD calculations have shown strong evidence for the existence of $H$-dibaryon near and above the double $\Lambda$ threshold region~\cite{Beane:2010hg, Inoue:2010es}. 
Thus, a new experimental program with high statistics and high precision is necessary. 
Many experimental efforts to find the $H$-dibaryon have been performed, including $K^-$ induced fixed-target experiments and collider experiments.
However, no positive results have been reported to date apart from the KEK-PS E224~\cite{Ahn:2000fu} and E522~\cite{Yoon:2007aq} experiments.  The E224/E522 experiments measured the invariant mass spectrum of two $\Lambda$s in the ($K^-, K^+$) reaction on a carbon target and found a peak structure near the double $\Lambda$ threshold.  However, the significance of the signal is very poor due to low statistics. Therefore, the experiment could not conclude the existence of $H$-dibaryon from the data. 

\begin{figure}[htbp]
\centerline{\includegraphics[width=14cm]{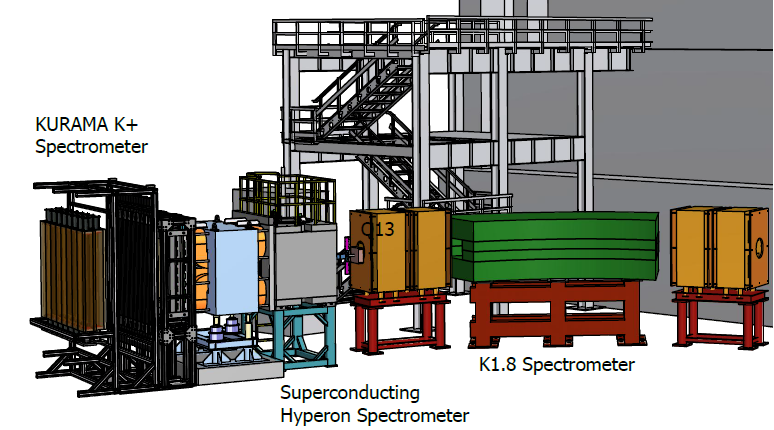}}
\caption{Schematic view of the E42 detector setup. The Hyperon Spectrometer consists of a superconducting Helmholtz-type magnet and a time projection chamber (HypTPC) with a KURAMA spectrometer~\cite{Ahn:2018qak,Kim:2019fsi}.}
\label{fig:E42_spectrometer}
\end{figure}

A new experiment to search for the $H$-dibaryon is under preparation as the J-PARC E42 experiment~\cite{E42-proposal}. A schematic view of the E42 spectrometer (Hyperon Spectrometer) is shown in Fig.~\ref{fig:E42_spectrometer}.
The experiment will be performed at the K1.8 beamline with a newly installed superconducting Helmholtz magnet with a large time projection chamber (HypTPC).   
A high intensity $K^-$ beam transported by the K1.8 beamline spectrometer is focused on a diamond target where two $\Lambda$s will be produced via the ($K^-, K^+$) reaction at 1.8~GeV/$c$. The outgoing $K^+$ will be detected and identified by the scattered particle KURAMA spectrometer. The KURAMA spectrometer consists of a large dipole magnet together with tracking chambers and a time-of-flight (ToF) wall to determine the particle species. Some fraction of the interaction may produce $H$-dibaryons, which immediately decay into two $\Lambda$s. The Hyperon Spectrometer will reconstruct those two $\Lambda$s by detecting its decaying particles, {\it i.e.}, two $\pi^-$s and two protons.

The expected double $\Lambda$ invariant mass resolution is ~1.5 MeV/$c^2$ at the $\Xi N$ mass threshold. 
10,000 $\Lambda\Lambda$ events will be accumulated over a data gathering period of 33 days, together with 1440 $H$-dibaryon signals. These statistics are about 120 times higher than in the previous E522 experiment. 

\subsubsection{Detailed investigations of nucleon resonances \label{subsubsec:E45}}

Due to the non-perturbative nature of the QCD in the low energy regime, the QCD vacuum structure shows a complex aspect. It is known that a baryon is an excitation of the QCD vacuum. Thus, one way to reveal the QCD vacuum structure is to investigate the baryon spectrum, especially for excited state baryons.

Spectroscopic studies of baryons, especially for nucleon resonances,  have gathered considerable data on $\gamma N \rightarrow \pi N$ and $\pi N \rightarrow \pi N$, and intensive analyses of the data have been performed.  A clear description of the nucleon resonances has been developed, though as yet no precise details on nucleon resonances with masses more than 1.6 GeV/$c^2$, which is above the $N\pi\pi$ threshold, have been obtained  because of the lack of data available. 
Therefore, high statistics and high-quality data are required to improve our knowledge of  nucleon resonances for $M>1.6$ GeV/$c^2$.

The J-PARC E45 experiment~\cite{E45-proposal} will accumulate high statistics and high-quality data on $\pi N \rightarrow \pi\pi N$ and $\pi N \rightarrow KY$ reactions. The experiment will be performed at the K1.8 beamline using the E42 spectrometer (Hyperon Spectrometer) with 
a large acceptance. This will allow a partial wave analysis to extract detailed resonance parameters. 

\subsubsection{Hyperon resonance near the $\Lambda\eta$ threshold \label{subsubsec:E72}}

A new hyperon resonance has been observed in the invariant mass of the $p K^-$ spectrum~\cite{Tanida:2019cif} in the Dalitz plot analysis of $\Lambda_c \to p K^- \pi^+$~\cite{Yang:2015ytm}.
The mass of the resonance was found to be 1663 MeV, and the width of the resonance was extremely narrow, about 10 MeV. Because no such narrow resonance has been observed to date~\cite{Tanabashi:2018oca}, the newly found resonance might be a candidate for an exotic $\Lambda^*$ resonance. 

On the other hand, based on old data presented by the Crystal Ball collaboration for the differential cross section for the $K^-p\to\Lambda\eta$ reaction~\cite{Starostin:2001zz}, two theoretical groups have indicated the existence of a new $\Lambda^*$ resonance around this mass region. The ANL-Osaka-KEK group claimed, based on a partial wave analysis, the existence of a narrow $\Lambda^*$ resonance at 1671$^{+2}_{-8}$ MeV with $J^{P}=3/2^+$~\cite{Kamano:2014zba, Kamano:2015hxa}. Unfortunately, a lack of information regarding the polarization observables in the data of the Crystal Ball collaboration means that the new $\Lambda^*$ cannot be confirmed unambiguously. 
On the other hand, B.C.~Liu and J.J.~Xie concluded, based on their reaction model approach, the presence of a narrow $\Lambda^*$ resonance at 1669 MeV with $J^{P}=3/2^-$~\cite{Liu:2011sw,Liu:2012ge}. Therefore, the properties of the resonance are still undetermined. 

The existence of the $\Lambda^*$, which strongly coupled with $\eta \Lambda$, needs to be identified together with its spin and parity to answer the question of whether new narrow $\Lambda^*$ resonance exists or not. 
The E72 experiment aims to search for the $\Lambda^*$ resonance with $J=3/2$ near the $\Lambda\eta$ threshold via the $p(K^-,\Lambda)\eta$ reaction at around 735 MeV/$c$~\cite{E72-proposal}. A detailed shape study of the differential cross section measurements in a narrow center-of-mass region around 1669 MeV will allow us to determine the spin of the resonance. Polarization measurements can also be used for the parity determination.  
The experiment will be performed at the K1.8BR or K1.1 beamline with a large solid angle detector, HypTPC. The experiment is under preparation.

\subsubsection{Charmed baryon spectroscopy \label{subsubsec:E50}}
Understanding the effective degree of freedom to describe a hadron is a fundamental question in hadron physics. Previously, a model based on the ``constituent quark'' was used to describe the properties of a ground state hadron, such as the mass and magnetic moment.
However, the model fails to reproduce excited state hadrons. A possible approach to address this discrepancy is to introduce a new degree of freedom, correlated di-quarks, inside the excited hadron. 
Interactions between quarks, especially the color-magnetic interaction inside a hadron, can be written as follows:
\begin{equation}
V_{CMI} \sim \frac{\alpha_s}{m_i m_j}(\lambda_i \cdot \lambda_j)(\vec{\sigma_i}\cdot\vec{\sigma_j}), 
\end{equation}
where $\lambda$ and $\sigma$ are the color and spin of the quarks, respectively.

\begin{figure}[htbp]
\centerline{\includegraphics[width=12cm]{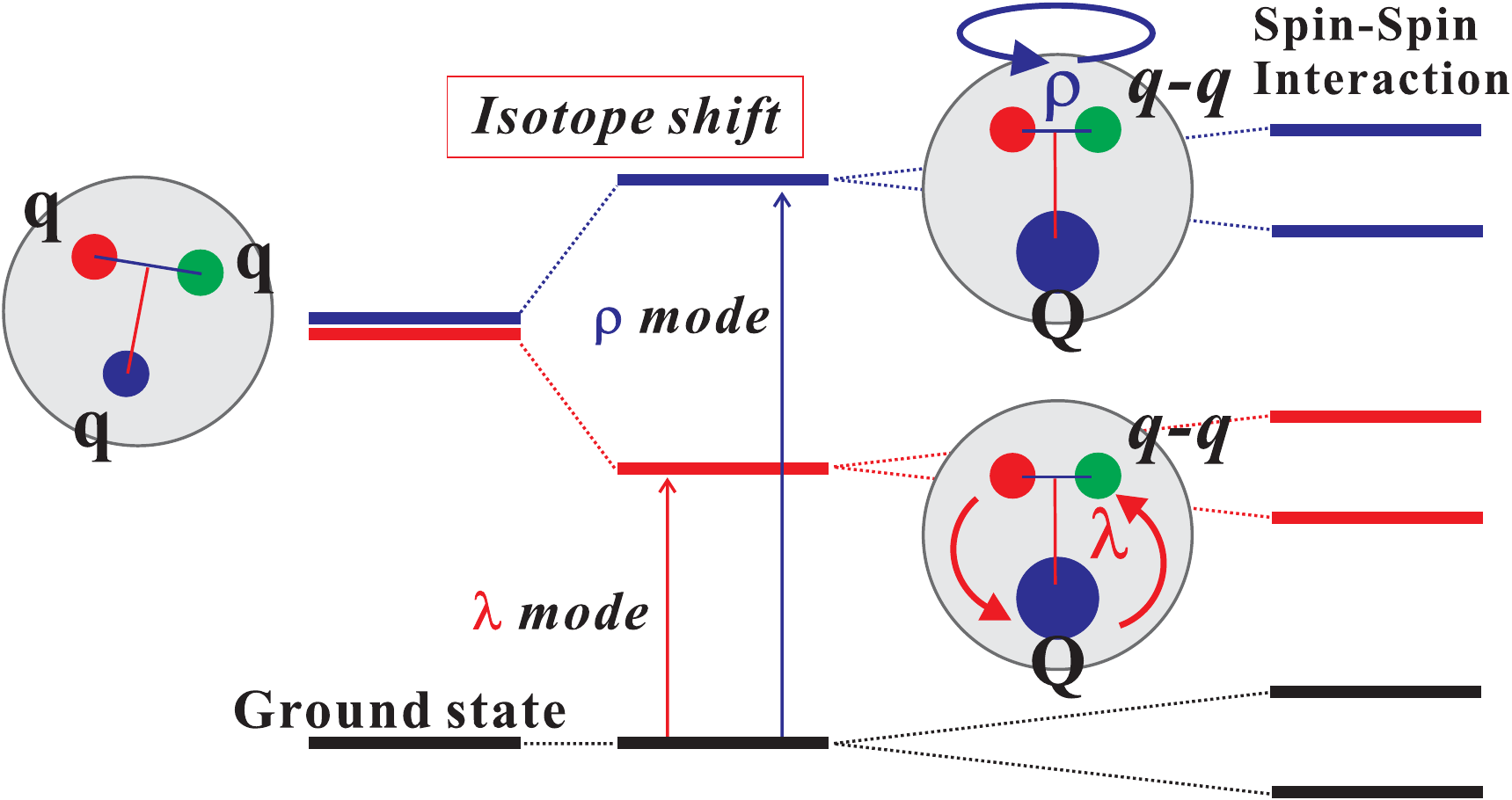}}
\caption{Schematic view of the $\rho$ and $\lambda$ excitations in $N^*$ and $Y^*_c$. Degeneracy of the energy level for the $\rho$ and $\lambda$ mode excitations is expected in nucleon resonances ($N^*$), while those levels are expected to be separate in excited charmed baryons. The figure is taken from Ref.~\cite{Shirotori:2015eqa}.}
\label{fig:E50_illust}
\end{figure}

\begin{figure}[htbp]
\centerline{\includegraphics[width=12cm]{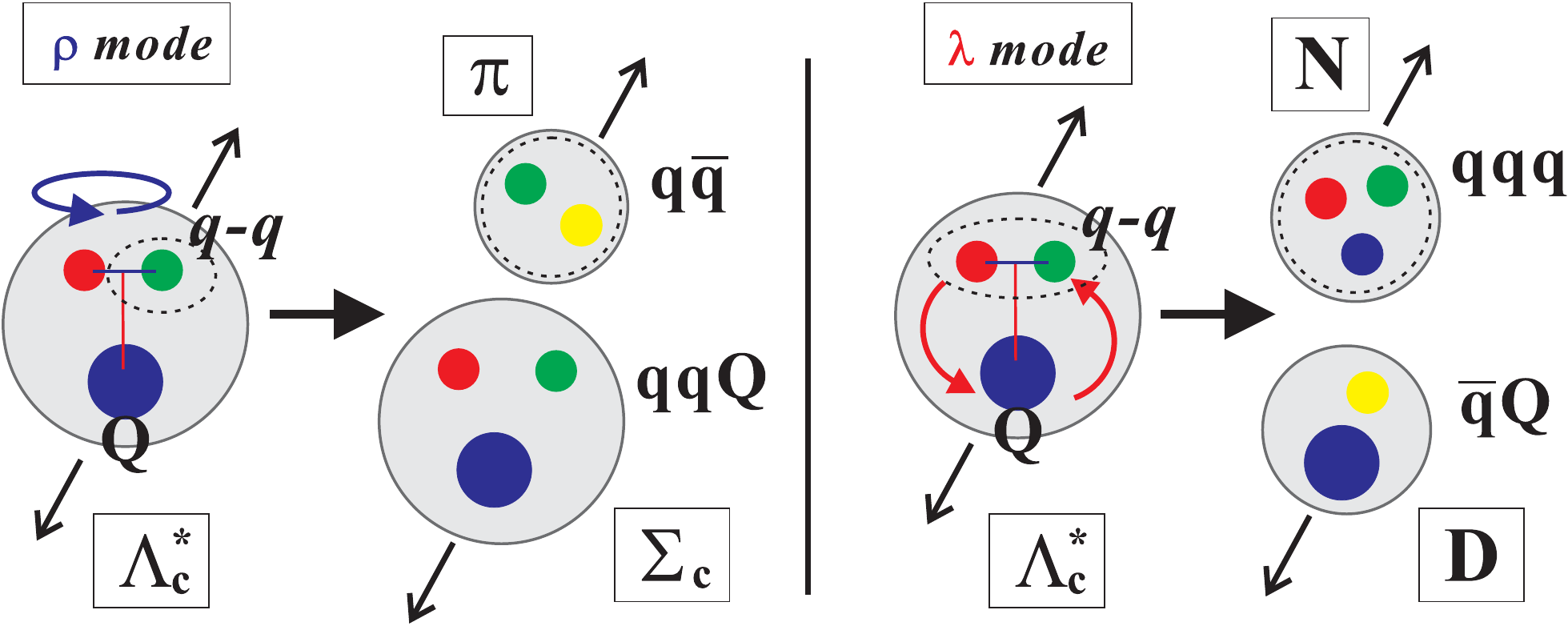}}
\caption{Expected decay channel for  $Y^*_c$. A charmed baryon with a $\rho$ mode excitation will favorably decay to a charmed baryon and a light meson, while a decay with a nucleon and charmed meson will be enhanced in the $\lambda$ mode excited charmed baryon. The figure is taken from Ref.~\cite{Shirotori:2015eqa}.} 
\label{fig:E50_decay}
\end{figure}

The equation shows that the strength of the interaction between two quarks is proportional to the inverse of the quarks' masses.   
For a baryon, we use the Jacobi coordinate system to solve the three-body equation. One coordinate is $\lambda$ and the other is $\rho$. 
Figure~\ref{fig:E50_illust} illustrates this coordinate systems. 
For an excited nucleon, $N^*$, where the three constituent quarks (written as qqq) have almost the same masses,  we cannot distinguish the $\rho$ and $\lambda$ directions.
However, let us introduce a heavy-quark Q in a baryon, such as a charm quark or even a strange quark, to form a ``qqQ'' system. In such a system, the color-magnetic interaction for ``qq'' is naturally stronger than the interaction for ``qQ.''  Therefore, a strong ``qq'' correlation appears inside a baryon. The next question is how the effect of the di-quark correlation appears in the observables. The answer is illustrated in Figure~\ref{fig:E50_illust}.  As we mentioned, for a $N^*$(qqq),  it is not possible to separate the $\lambda$ and $\rho$ coordinates. Thus, the excitation of the $\rho$ and $\lambda$ directions cannot be distinguished and thus the energy level of the excitation in the $\rho$ and $\lambda$ directions is degenerate. On the other hand, for a baryon with one heavy quark, the $\rho$ and $\lambda$ coordinates can be defined. Let us assume the coordinate between ``qq" is $\rho$  and ``Q--qq" is $\lambda$.  In this case, an excitation of the $\lambda$ direction, which corresponds to a rotation of two light quarks around one heavy quark, is easier than a $\rho$ mode excitation, where two light quarks are rotated. This leads to a separation of the $\rho$ mode and the $\lambda$ mode in baryon mass spectrum. 
Moreover, as shown in Figure~\ref{fig:E50_decay}, an excited baryon in a $\rho$ or $\lambda$ mode is expected to have favored decay channels. For example, light-meson ($\pi$) -- heavy-baryon ($Y_c$) decay will be dominant for a $\rho$ mode excited charmed baryon. On the other hand,
a heavy-meson (D) and light-baryon (N)  decay mode will be favored for $\lambda$ mode excited baryons.
In addition, a systematic comparison between charmed baryon (baryon with one charm quark) and hyperon (baryon with one strange quark) mass spectra, including the decay modes, will confirm whether the di-quark correlation in hadrons exists.  

\begin{figure}[htbp]
\centerline{\includegraphics[width=8cm]{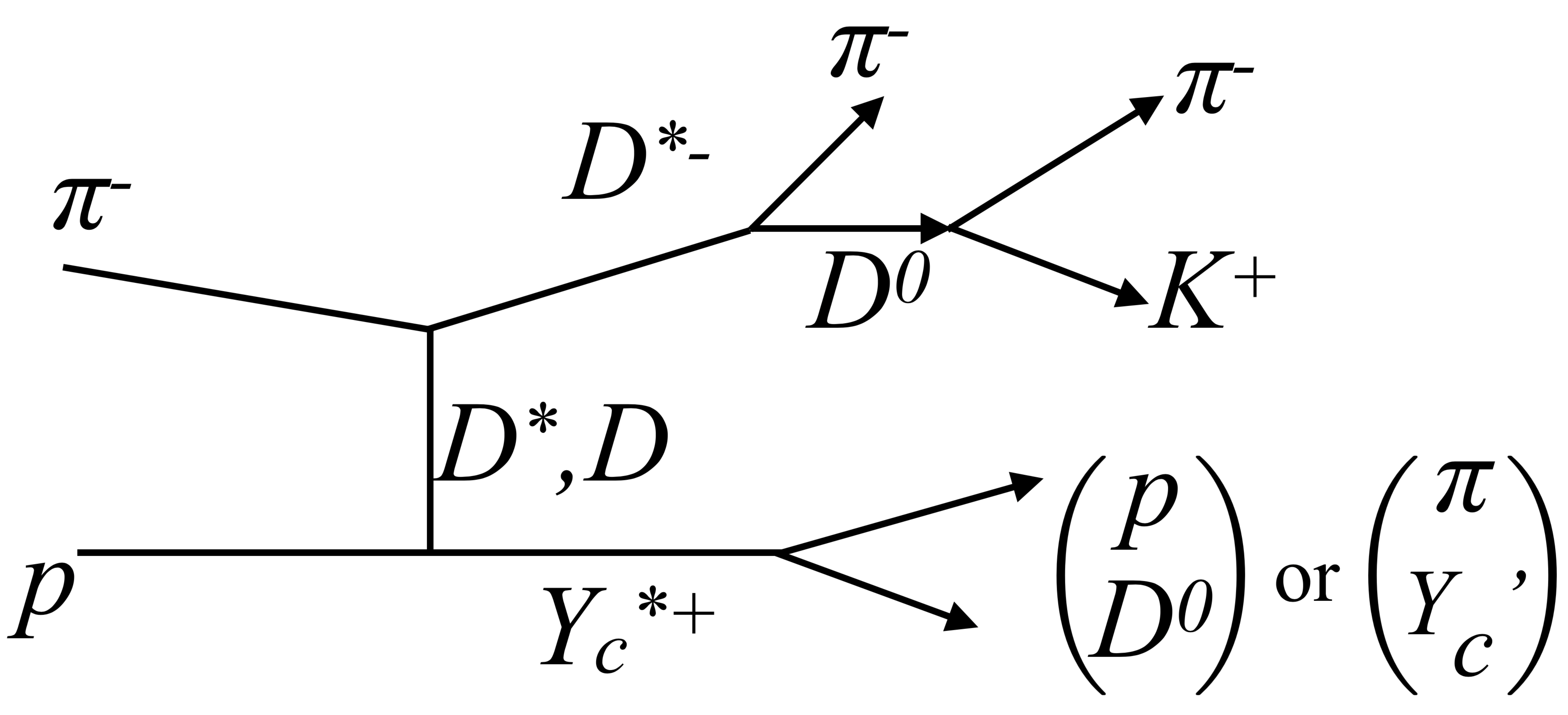}}
\caption{Feynman diagram for charmed baryon production via the $(\pi^-,D^{*-})$ reaction. The production of a charmed baryon can be identified via missing mass analysis by detecting forward going $D^*$ mesons.}
\label{fig:E50prod}
\end{figure}

The main goal of the J-PARC E50 experiment is to obtain evidence for di-quark correlation in baryons via charmed baryon spectroscopy~\cite{E50-proposal}. 
The E50 experiment focuses on charmed baryon production with t-channel $D^*$ exchange in the $p(\pi^-, D^{*-})Y_c^*$ reaction up to 20 GeV/$c$, where $Y_c^*$ denotes an excited charmed baryon. 
A typical production diagram is shown in Fig.~\ref{fig:E50prod}.   

\begin{figure}[htbp]
\centerline{\includegraphics[width=13cm]{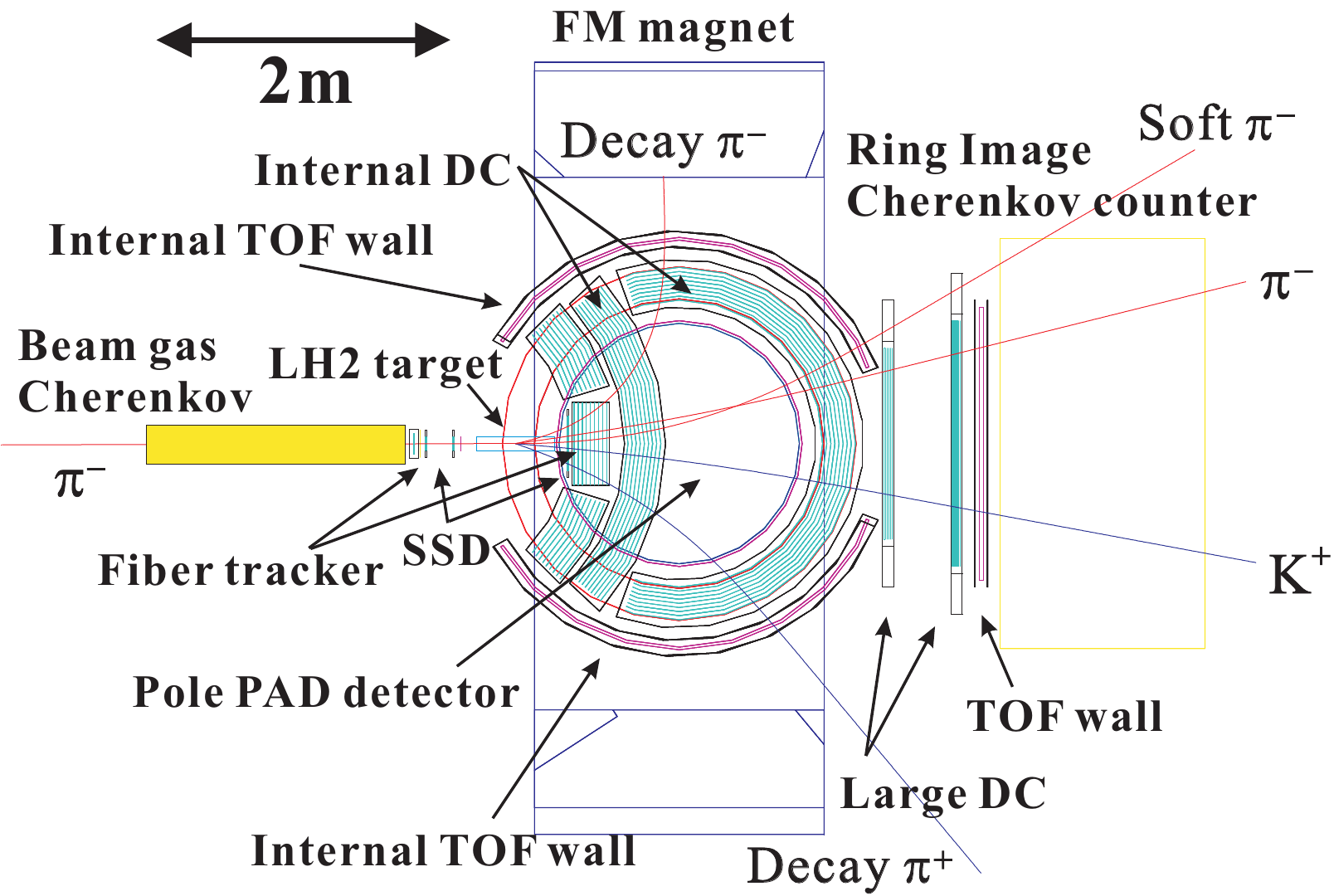}}
\caption{Schematic view of the E50 spectrometer. The figure is taken from Ref.~\cite{Shirotori:2015eqa}.}
\label{fig:E50spectrometer}
\end{figure}

The key to the experiment is a high momentum resolution beamline spectrometer, and a large acceptance and high precision decay particle spectrometer.  
A schematic view of the decay particle spectrometer is shown in Fig.~\ref{fig:E50spectrometer}. 
The spectrometer consists of a large-gap dipole magnet, wire chambers, and high-resolution ToF detectors to identify outgoing $D^*$ mesons and decay products of charmed baryons. 

\begin{figure}[htbp]
\centerline{\includegraphics[width=8cm,angle=-90]{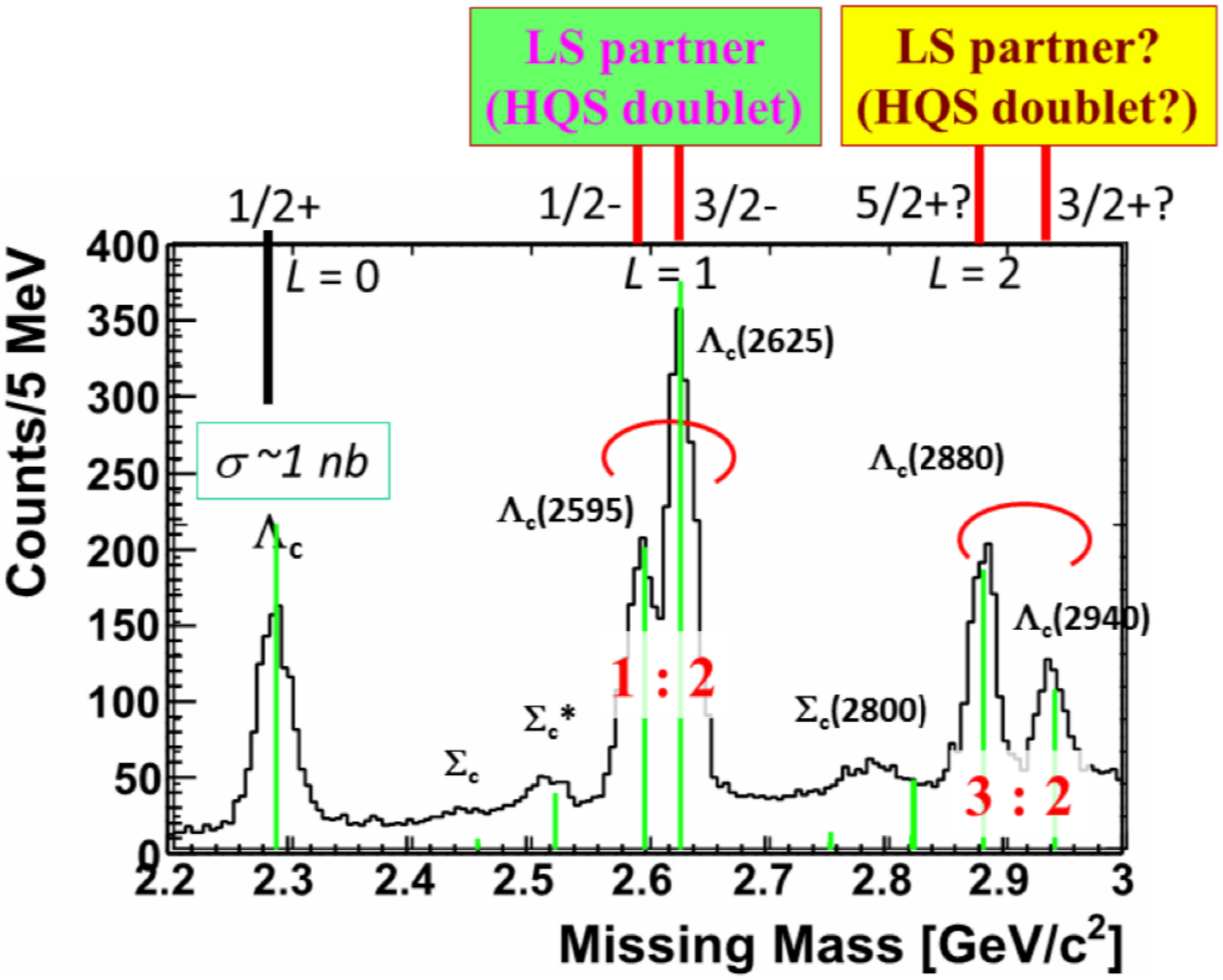}}
\caption{Expected missing mass spectrum simulated with the known charmed baryon states. The detector resolutions are included for this simulation.
The production cross section for a charmed baryon is  assumed to be $\sigma$($\Lambda^+_c$) =1~nb, which can be evaluated with a theoretical calculation~\cite{Kim:2015omh}. 
Clear spin double states, $\Lambda_c(2595)$ and $\Lambda_c(2625)$, and $\Lambda_c(2880)$ and $\Lambda_c(2940)$, are seen. 
For this simulation, the spin-parity of $\Lambda_c(2940)$ is assumed to be $3/2^+$. The figure is taken from Ref.~\cite{Noumi:2017sdz}.}
\label{fig:E50_spec}
\end{figure}

Figure~\ref{fig:E50_spec} shows the expected charmed baryon spectrum for one hundred days of beam time. If the strong di-quark correlation exists as expected, a two-peak structure which corresponds to the spin doublets (the $\rho$ and $\lambda$ mode excited charmed baryons) will appear in the spectrum~\cite{Noumi:2017sdz}. 

On the other hand, we can access the structure of an excited hyperon and explore the existence of the di-quark correlation in it by using $K^-$ as beam particles instead of $\pi^-$.  Due to the relatively high production cross section of excited hyperons compared with charmed baryons, hyperon spectroscopy will be an essential upgrade to the E50 experiment. We may find evidence for the existence of di-quark correlation by hyperon spectroscopy, which can be confirmed by charmed baryon spectroscopy. 

The beamline construction was completed during 2019, and a secondary target will be installed at the branch point of the high-momentum beamline to realize high-momentum $\pi^-$ and $K^-$ beams in the HEF.
The detector construction for charmed baryon spectroscopy is underway.

\subsubsection{Precise measurement of the spectral function of vector mesons via di-electrons \label{subsubsec:E16}}
QCD sum rules establish a link between hadron masses and spontaneous chiral symmetry breaking 
in the QCD vacuum.
One way to confirm this is to investigate meson properties in nuclear matter, 
where the symmetry is expected to be partially restored.
The vector mesons, $\rho$, $\omega$ and $\phi$, have attracted attention as powerful probes to investigate the spectral function in nuclear matter for two reasons. The first is that vector mesons have di-lepton decay channels, $e^+e^-$ and $\mu^+ \mu^-$, which do not cause spectral shape distortion via the strong interaction . The second is 
the narrow natural width of the order of 10 MeV, which makes it relatively easy to identify. 

To date, the KEK-PS E325 experiment has discovered a significant modification of the spectral function of $\rho$, $\omega$ and $\phi$  mesons in a nucleus via di-electron spectrum~\cite{Naruki:2005kd,Muto:2005za}. On the other hand, the CLAS experiment at Jefferson Lab observed width broadening of the vector meson spectral function in heavier target nuclei (Fe, Ti) without shifting of the pole position~\cite{Wood:2008ee}. 
The difference between the E325 experiment and the CLAS experiment may be due to the production mechanism of the vector mesons,   
namely hadron-production and photoproduction.   
Unfortunately, both experiments have only a limited amount of data. Therefore, new high statistics data are required to make detailed investigations on the spectral functions of vector mesons in nuclear matter.

\begin{figure}[htbp] 
\centerline{\includegraphics[width=16cm]{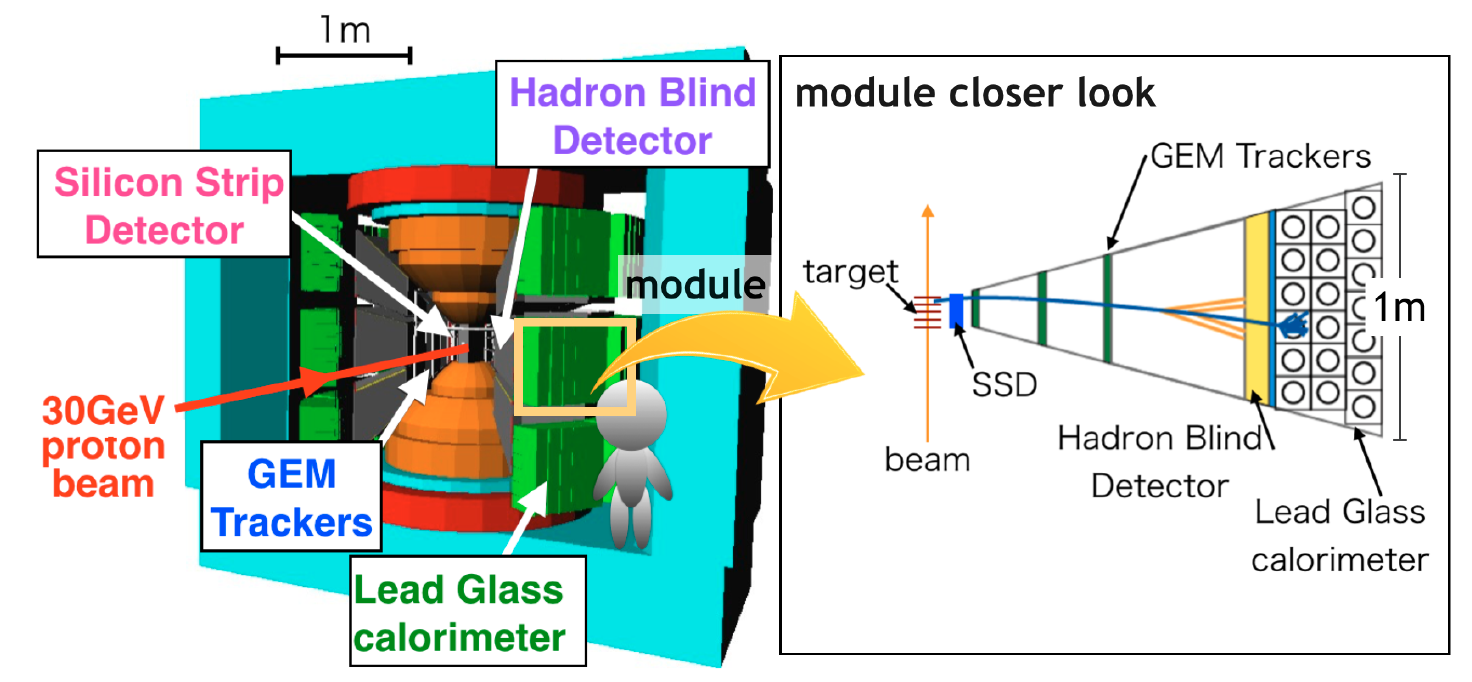}}
\caption{Schematic view of the E16 spectrometer. The figure is taken from Ref.~\cite{E16AshikagaQNP2018}}. 
\label{fig:E16_spectrometer}
\end{figure}

The J-PARC E16 experiment is the successor to the previous E325 experiment, with a significant detector upgrade incorporating acceptance for a high-intensity beam but maintaining a high mass resolution~\cite{E16-proposal}.  The main goal of the E16 experiment is to accumulate a hundred times more statistics than the previous E325 experiment. The new data will allow access to the spectral functions of vector mesons as a function of their momenta~\cite{Lee:1997zta,Kim:2019ybi}. This will subsequently allow measurements of the dispersion relation of vector mesons in nuclear matter. Figure~\ref{fig:E16_spectrometer} shows a schematic view of the E16 spectrometer. 
The spectrometer consists of four major subsystems. The first is a high precision and high rate tracking detector with a gas electron multiplier (GEM)~\cite{Sauli:1997qp}, with a low material budget to reduce the external radiative effect and multiple scattering
of electrons and positrons, which distort the reconstructed mass spectrum. 
The GEM tracker is used to compensate for the expected background hit rate of up to 5 kHz/mm$^2$~\cite{Komatsu:2013kua}. 
The second system is a hadron blind detector (HBD), which detects the electrons using Cherenkov light in CF$_4$ gas,
and does not respond to pions~\cite{Aoki:2011zza,Kanno:2016eaa}. The third is a lead-glass calorimeter (LG) for identifying electrons together with HBD signals.  The pion misidentification probabilities of the HBD and LG are estimated to be 0.6\% and 5\%, respectively~\cite{Komatsu:2017zkt}.
Finally, a layer of silicon strip detector (SSD) is introduced to reduce the possible effects of accidental hits due to the 10-MHz interaction rate at the target.

The phase one experiment, with part of the detector, will be performed in 2020, followed by experiments with the full detector setup in the near future.
  
\subsubsection{Search for vector-meson nuclear bound states \label{subsubsec:E29}}

As we discussed in the previous section, the mass of the hadron is strongly correlated with the expectation value of the chiral condensation, $\langle \bar{q}q \rangle$, in the environment where the hadron exists. The value of  $\langle \bar{q}q \rangle$ is expected to be reduced as a function of the nuclear matter density. Thus, the mass of a hadron, such as a vector meson, is expected to be reduced inside the nucleus. The mass reduction can be considered to indicate the existence of an attractive force between the vector meson and nuclear matter, raising the issue as to whether vector meson nuclear-bound states exist.

\begin{figure}[htbp]
\centerline{\includegraphics[width=10cm,angle=-90]{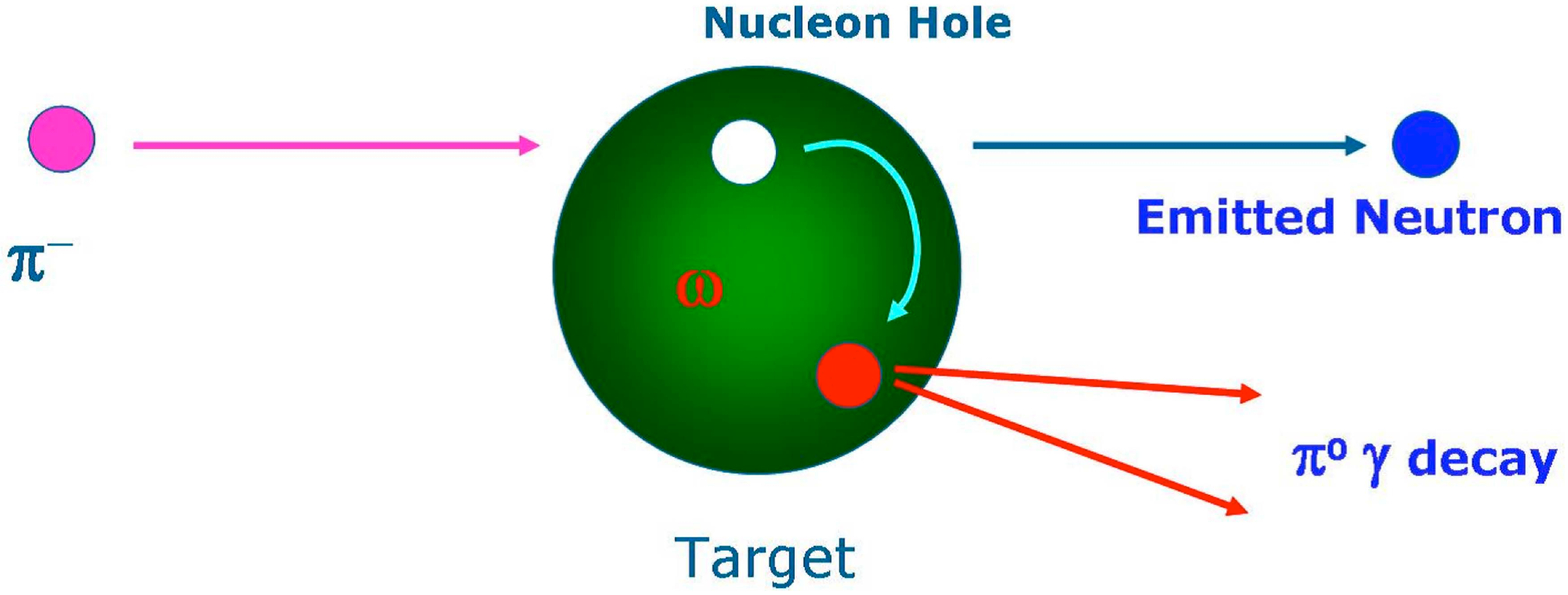}}
\caption{Schematic view of the production mechanism for the $\omega$ meson nucleus bound state.}
\label{fig:E26_process}
\end{figure}

The J-PARC E26 experiment is planned to be performed at the J-PARC K1.8 beamline, in an effort to search for the $\omega$ meson bound state~\cite{E26-proposal}.  Figure~\ref{fig:E26_process}  shows a schematic view of the measurement. 
The $\omega$ meson will be produced at a proton in the nucleus via the $p(\pi^-,n)\omega$ reaction. Neutron emission in the beam direction ({\it i.e.}, $0^\circ$) will be required for this experiment. In this situation, the direction of the produced $\omega$ meson will be the backward direction in the center-of-mass frame, {\it i.e.}, a low momentum $\omega$ meson in the laboratory frame will be produced. The typical momentum produced in this kinematics is $\sim$ 100 MeV/$c$. If the $\omega$--nucleus interaction is strong enough, a $\omega$ meson nuclear-bound state will be formed. Soon after the $\omega$-meson bound state is formed, some fraction of $\omega$, which remains inside the nuclear matter, will decay via the $\omega\to\pi^0\gamma$ mode. The invariant mass of $\pi^0\gamma$ from the $\omega$ meson will contain information on the $\omega$ mass in the nucleus, which is related to chiral condensation in the nucleus.
The E26 experiment can
identify the meson nucleus bound state via missing mass analysis and measurements of the spectral function of the $\omega$ meson via the invariant mass reconstruction by its decay products, simultaneously. 

The J-PARC E29 experiment will search for $\phi$ meson bound nuclei~\cite{E29-proposal}. The E29 experiment will focus on the $\bar{p}p\rightarrow\phi\phi$ reaction as a source for producing $\phi$ mesons in nuclei.
An advantage of using this reaction as an elementary process is that  double $\phi$ meson production is the dominant channel for the double $\bar{s}s$ pair production reaction around the production threshold, where the corresponding $\bar{p}$ beam momentum will be $\sim$ 1.0 GeV/$c$~\cite{Evangelista:1998zg}. Thus, detecting $\phi$ mesons in the reaction automatically assures that another $\phi$ meson has been produced. Moreover, if we require the production direction of one $\phi$ meson to be in the incoming beam direction, the momentum of the other $\phi$ in the laboratory frame will be $\sim$ 200 MeV/$c$. This momentum is comparable with the Fermi momentum in the nucleus. Therefore, the probability of a $\phi$ meson sticking in the nucleus will be enhanced if the interaction between $\phi$ meson and nucleus is sufficiently strong.  

\begin{figure}[htbp]
\centerline{\includegraphics[width=10cm,angle=-90]{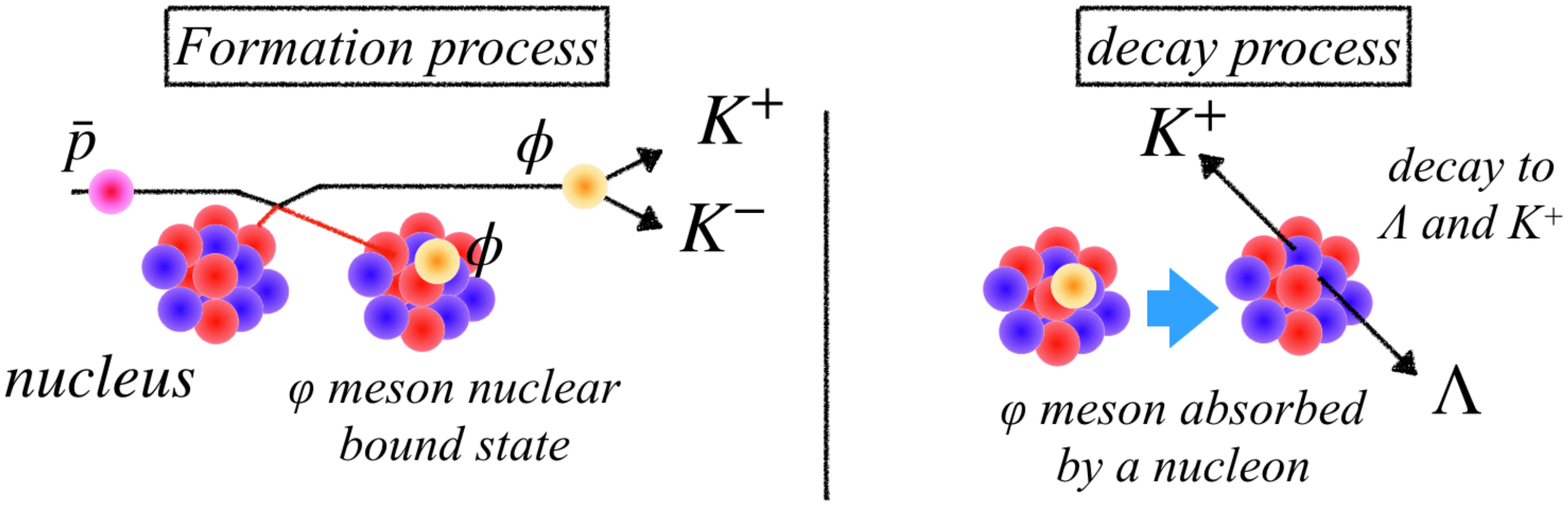}}
\caption{Schematic view of the production mechanism for the $\phi$ meson nucleus bound state.}
\label{fig:E29_process}
\end{figure}

Figure~\ref{fig:E29_process} shows a schematic view of the process of searching for the $\phi$ meson bound nucleus. 
The experiment is planned to be performed at the K1.8BR beamline.
A high intensity $\bar{p}$ beam will impact a carbon target and the $\bar{p}p\rightarrow\phi\phi$ reaction will occur in the nucleus. A $\phi$ meson will be ejected in the forward direction and a residual nucleus will capture the other $\phi$ meson by a strong interaction, forming a $\phi$ meson nuclear-bound state.
Soon after the $\phi$ meson nuclear-bound state formation, the nucleus will absorb the bound $\phi$ meson via the $\phi p\rightarrow K^+\Lambda$ reaction in the nucleus.

Formation of a $\phi$-meson bound nucleus can be identified by missing mass analysis via the forward emitted $\phi$ meson. 
In addition, the existence of a $\bar{s}s$ pair in the residual nucleus is assured by the detection of $\Lambda$ from the target. As discussed above, the double $\phi$ meson production is dominant in double strange quark pair production. Thus the required $\Lambda$  is already good evidence for the existence of a $\phi$ meson in the nucleus.

\subsection{Antikaon-Nucleon Interaction \label{subsec:MB_int}}
The meson-baryon interactions close to the mass thresholds provide
crucial information on the interplay between spontaneous and explicit
chiral symmetry braking.
Among meson-baryon interactions, the antikaon-nucleon ($\bar K N$)
interaction is an important probe to understand this important aspect of
low-energy QCD.
The $\bar K N$ interaction in the $I = 0$ channel is known to be strongly
attractive, which has been revealed from extensive measurements of
anti-kaonic hydrogen atoms~\cite{Iwasaki:1997wf,Beer:2005qi,Bazzi:2011zj}
and low-energy $\bar K N$ scattering~\cite{Martin:1980qe}.

From the point of view of this strong interaction between the antikaon
and nucleus at the low-energy limit, kaonic atoms in which an
electron is replaced by a $K^-$ meson have been well studied in precise
experiments of characteristic X-ray spectroscopy.
The effects of strong interaction can be measured as a level shift from
the binding energy calculated by only the electromagnetic interaction,
and the broadening width due to absorption of the kaon by the nucleus.
Information on the $\bar K$-nucleus potential can be extracted 
using the comprehensive data set of the shifts and widths obtained by
the kaonic atom measurements from lithium (Li) to uranium (U) with
density dependent optical potentials.
However, the obtained $\bar K$-nucleus potential still remains
controversial; the two major theoretical approaches with
phenomenological models and chiral unitary models have provided
conflicting results.
With a view to clarify this controversial situation, in the J-PARC E62 experiment, 
high-precision spectroscopy of the isotope shift between the
kaonic-$^3$He and kaonic-$^4$He $2p$ states was performed using a superconducting
transition-edge-sensor (TES) microcalorimeter with resolution one order of 
magnitude better than conventional semi-conductor
detectors.

The discrepancy of the calculated $\bar K$-nucleus potential is closely
related to the different approach to calculations of the most essential
$\bar K N$ interaction.
Extensive efforts have revealed the $\bar K N$ interaction to be
strongly attractive in the $I=0$ channel, and now the
$K^-p$ scattering amplitude in the low energy region has been precisely
obtained from the results of SIDDHARTA at DA$\Phi$NE~\cite{Bazzi:2011zj}
and from theoretical calculations based on this
measurement~\cite{Ikeda:2011pi,Ikeda:2012au,Cieply:2011nq}.
To determine the isospin dependent $\bar K N$ scattering length, which
is the most important but missing information in the $\bar K N$
interaction field, kaonic deuterium X-ray measurements have now been launched
as the J-PARC E57 experiment.

\begin{figure}[htbp]
\centerline{\includegraphics[width=10cm]{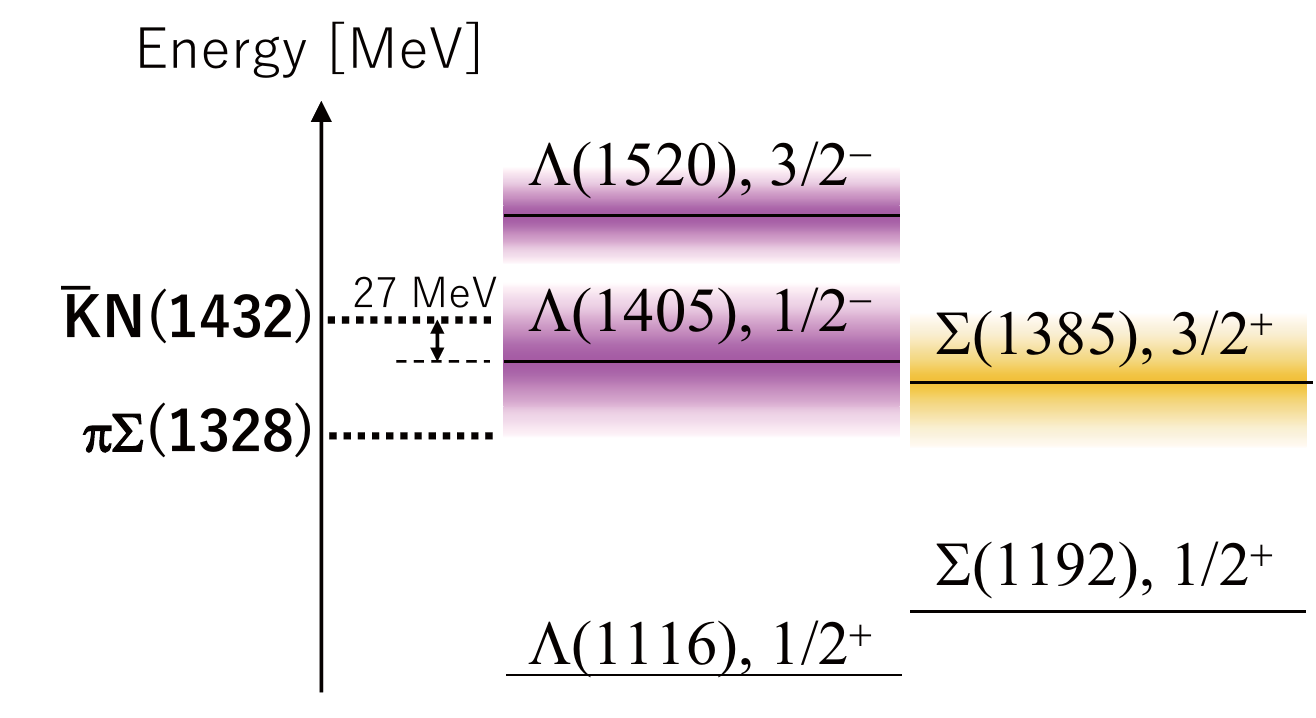}}
\caption{
Level scheme of $Y^*$ resonances around the $\pi\Sigma$ and the $\bar KN$
mass thresholds.
}
\label{fig:L1405}
\end{figure}

However, due to the presence of the $\Lambda(1405)$ state located just
below the $\bar K+N$ mass threshold as shown in Fig.~\ref{fig:L1405}, theoretical investigations of the
$\bar K N$ interaction are very complicated and thus difficult.
The $\Lambda(1405)$ state is still an unclear state because it cannot be
described by simple constituent quark models as an ordinary three-quark
state~\cite{Isgur:1978xj}; therefore, there is a long-standing discussion on
the interpretation of the $\Lambda(1405)$ state, such as a meson-baryon
($\bar K N$) quasi-bound state~\cite{Dalitz:1967fp} or other
exotic states such as a pentaquark baryon~\cite{Brau:2004ie,Helminen:2000jb}.
Among the interpretations of the $\Lambda(1405)$ state, the meson-baryon
scenario is widely supported because the $\Lambda(1405)$ state can be
naturally described as a quasi-bound $\bar K N$ state in the $\bar K N$-$\pi \Sigma$ coupled-channel system.
In such models, energy independent phenomenological models are constructed to reproduce
the $\Lambda(1405)$ mass pole with a single pole structure, which show the mass pole of around 1405 MeV with deep $\bar K N$
potential~\cite{Dalitz:1991sq,Akaishi:2002bg}. 
On the other hand, the chiral unitary approaches, which are based on a chiral
effective Lagrangian and are thus energy
dependent, have given shallow potential and predicted a double pole
structure of $\Lambda(1405)$; $\pi\Sigma$ and $\bar K N$ coupled
channels appear in lower and higher poles, respectively, and
consequently, the spectrum shape has a peak around 1420
MeV~\cite{Oller:2000fj,Jido:2003cb,Hyodo:2007jq,Hyodo:2011ur}.
The ``{\it $\Lambda(1405)$ problem}'' is one of the most important issues
to be solved in the field of hadron physics, because the meson-baryon
molecule picture of the $\Lambda(1405)$ state can also be applied to
other candidates of hadronic molecular states, such as the recently observed
$XYZ$~\cite{Choi:2003ue,Aubert:2005rm,Choi:2007wga} and $P_c$
states~\cite{Aaij:2015tga,Aaij:2019vzc}.
At J-PARC, the E31 experiment was performed to reveal the
$\Lambda(1405)$ line shape by using and focusing on the most promising
channel of $\bar K N \to \pi \Sigma$.

The possible existence of kaon-nuclear quasi-bound states has
been widely discussed, based on the concept of the $\Lambda(1405)$ state
as the $\bar K N$ quasi-bound state.
The kaon-nuclear quasi-bound states have been well studied theoretically with
various frameworks and the existence of these states is claimed today.
Among the kaonic nuclear states, the $\bar K NN$ system (symbolically
denoted as ``$K^-pp$'') has attracted strong interest in both 
theoretical and experimental studies because it is the lightest
predicted $S = -1$ $\bar K$ nucleus.
Many theoretical calculations have been conducted based on few-body
calculations using the $\bar K NN$-$\pi \Sigma N$-$\pi \Lambda N$ coupled
formalism; however, predictions of the binding energy and width
are still widely divergent, so that
there are as yet many uncertainties of the $\bar K N$ interaction below
the mass threshold of $\bar K + N$.
Experimental investigations of the kaon-nuclear quasi-bound states have also
been performed over the last few decades.
Despite these extensive efforts, only a small amount of experimental
information has become available; however, it is still insufficient to
discriminate between a variety of conflicting interpretations.
To pin down the strength of the $\bar K N$ interaction below the
threshold and to clarify the controversial situation of the kaon-nuclear
quasi-bound states, the J-PARC E27 and E15 experiments were conducted to
search for the ``$K^-pp$'' state via different reactions using the
world's highest-intensity $\pi^-$ and $K^-$ beams, respectively.

\subsubsection{ Measurement of kaonic-deuterium $2p \to 1s$ X-rays \label{subsubsec:E57}}

The $K^-p$ interaction at around the threshold energy has been studied by
precise kaonic hydrogen X-ray spectroscopy measurements of the $2p
\to 1s$ transition (K$_\alpha$). 
The strong-interaction effects can be derived via the shift $(\epsilon)$
and the width $(\Gamma)$ of the atomic levels relative to the
electromagnetic values, which are caused by the strong interaction
between kaons and protons.
The strong interaction shift $\epsilon_{1s}$ is defined as
\begin{equation}
 \epsilon_{1s} = E^{measured}_{2p \to 1s} - E^{EM}_{2p \to 1s},
\end{equation}
where $E^{measured}_{2p \to 1s}$ and $E^{EM}_{2p \to 1s}$ are the measured $2p
\to 1s$ transition energy and that calculated with only the
electromagnetic interaction, respectively.
The obtained $\epsilon_{1s}$ and $\Gamma_{1s}$ are used to derive the complex
$S$-wave $K^-p$ scattering length $a_{K^-p}$ with the
Deser-type formula by taking into account the isospin-breaking
corrections at the next-to-leading
order~\cite{Deser:1954vq,Trueman:1961zza,Meissner:2004jr}:
\begin{equation}
 \epsilon_{1s} + i\frac{\Gamma_{1s}}{2} = 2 \alpha^{3} \mu_{r}^2
  a_{K^-p}\times\left\{ 1 - 2 \alpha \mu_{r} ({\rm ln} \alpha
		   -1) a_{K^-p} \right\},
\end{equation}
where $\alpha$ and $\mu_r$ denote the fine-structure constant and
the reduced mass of the $K^-p$ system, $m_K M_p / (m_K + M_p)$, respectively.
The most precise values of the strong-interaction shift and width of the
kaonic hydrogen $1s$ state were obtained by the SIDDHARTA
experiment~\cite{Bazzi:2011zj}.

The obtained scattering length from the SIDDHARTA results is
\begin{equation}
 a_{K^-p} = (-0.65\pm0.10) + i (0.81\pm0.15)~{\rm fm}.
\end{equation}
Together with the total cross sections of the $K^-p$ scattering
amplitude and the threshold branching ratios, the complex $S$-wave $K^-p
\to K^-p$ scattering amplitude can be evaluated with strong constraint
at the mass threshold, and then extrapolated to the sub-threshold
region~\cite{Ikeda:2011pi,Ikeda:2012au,Cieply:2011nq}.

The $K^-p$ scattering amplitude has thus been determined, especially at
around and above the threshold.
However, the information on the isospin dependent $\bar K N$ scattering
length is still missing.
Due to isospin conservation, only the averaged scattering length
in the isospin $I=0$ and $I=1$ ($a_0$ and $a_1$) channels is obtained
from the kaonic hydrogen measurement:
\begin{equation}
 a_{K^-p} = \frac{1}{2}(a_0+a_1).
\end{equation}
Measurement of the kaonic deuterium $1s$ state is required
to determine the individual isoscaler $a_0$ and isovector $a_1$ scattering
lengths.
The complex $S$-wave $K^-d$ scattering length $a_{K^-d}$, can be obtained
from the shift and the width of the $2p \to 1s$ transition using a
similar formula to the $K^-p$ case~\cite{Meissner:2006gx}:
\begin{equation}
 \epsilon_{1s} + i\frac{\Gamma_{1s}}{2} = 2 \alpha^{3} \mu_{r}^2
  a_{K^-d}\times\left\{ 1 - 2 \alpha \mu_{r} ({\rm ln} \alpha
		   -1) a_{K^-d} \right\},
\end{equation}
where $\mu_r$ denotes the reduced mass of the kaon-deuteron system.
Information on $a_0$ and $a_1$ can then be obtained from the
combination of $a_{K^-p}$ and $a_{K^-d}$:
\begin{eqnarray}
 a_{K^-n} &=& a_1,\\
 a_{K^-d} &=& \frac{4(m_N+m_K)}{2m_N+m_K}Q+C \label{eq:a_Kd},
\end{eqnarray}
where
\begin{equation}
 Q = \frac{1}{2}(a_{K^-p}+a_{K^-n}) = \frac{1}{4}(a_0+3a_1).
\end{equation}
The first and second terms of $a_{K^-d}$ in Eq.~\ref{eq:a_Kd} represent
the lowest-order impulse approximation of the $K^-N$ scattering in the
$K^-d$ system, {\it i.e.}, $K^-$ scattering with each nucleon in
deuterium, and higher-order corrections such as the $K^-d$
three-body interaction, respectively.

There are many theoretical calculations on the $K^-d$ scattering length
that give consistent values of the shift and width, as summarized
in Table~\ref{table:Kd}.
However, no experimental results have yet been
obtained due to the difficulty of the measurement, which is caused by large
absorption in the $2p$
state~\cite{Koike:1995sa,Jensen:2004,Raeisi:2009,Faber:2010iw}.
To date, only the SIDDHARTA group has shown an exploratory measurement on
the X-rays from kaonic deuterium, and the upper limits for the yield
of the $K$-series transitions were reported at a liquid deuterium
density of 1.5\% (13.9 times the STP density): total and $K_{\alpha}$ yields of
$Y(K_{tot}) < 0.0143$ and $Y(K_{\alpha}) < 0.0039$ (90\% CL),
respectively~\cite{Bazzi:2013vft}.
The yield is one order of magnitude smaller than the kaonic-hydrogen
yield, which is known to be $\sim$ 0.01 for 
$K_{\alpha}$~\cite{Iwasaki:1997wf,Bazzi:2016rlt}.

\begin{table}[htbp]
 \caption{
 Calculated $K^-d$ scattering length $a_{K^-d}$, and
 corresponding experimental observables, $\epsilon_{1s}$ and
 $\Gamma_{1s}$.
 The values are taken from Ref.~\cite{Curceanu:2019uph}.
 }
 \begin{center}
  \begin{tabular}{cccc}
   \hline
   $a_{K^-d}$ (fm) & $\epsilon_{1s}$ (eV) &  $\Gamma_{1s}/2$ (eV) & Reference \\
   \hline
   -1.66 + $i$1.28 & -884 & 665 & \cite{Meissner:2006gx} \\
   -1.42 + $i$1.09 & -769 & 674 & \cite{Gal:2006cw} \\
   -1.46 + $i$1.08 & -779 & 650 & \cite{Doring:2011xc} \\
   -1.48 + $i$1.22 & -787 & 505 & \cite{Shevchenko:2012np} \\
   -1.58 + $i$1.37 & -887 & 757 & \cite{Mizutani:2012gy} \\
   -1.42 + $i$1.60 & -670 & 508 & \cite{Hoshino:2017mty} \\
   \hline
  \end{tabular}
  \label{table:Kd}
 \end{center}
\end{table}

\begin{figure}[htbp]
\centerline{\includegraphics[width=15cm]{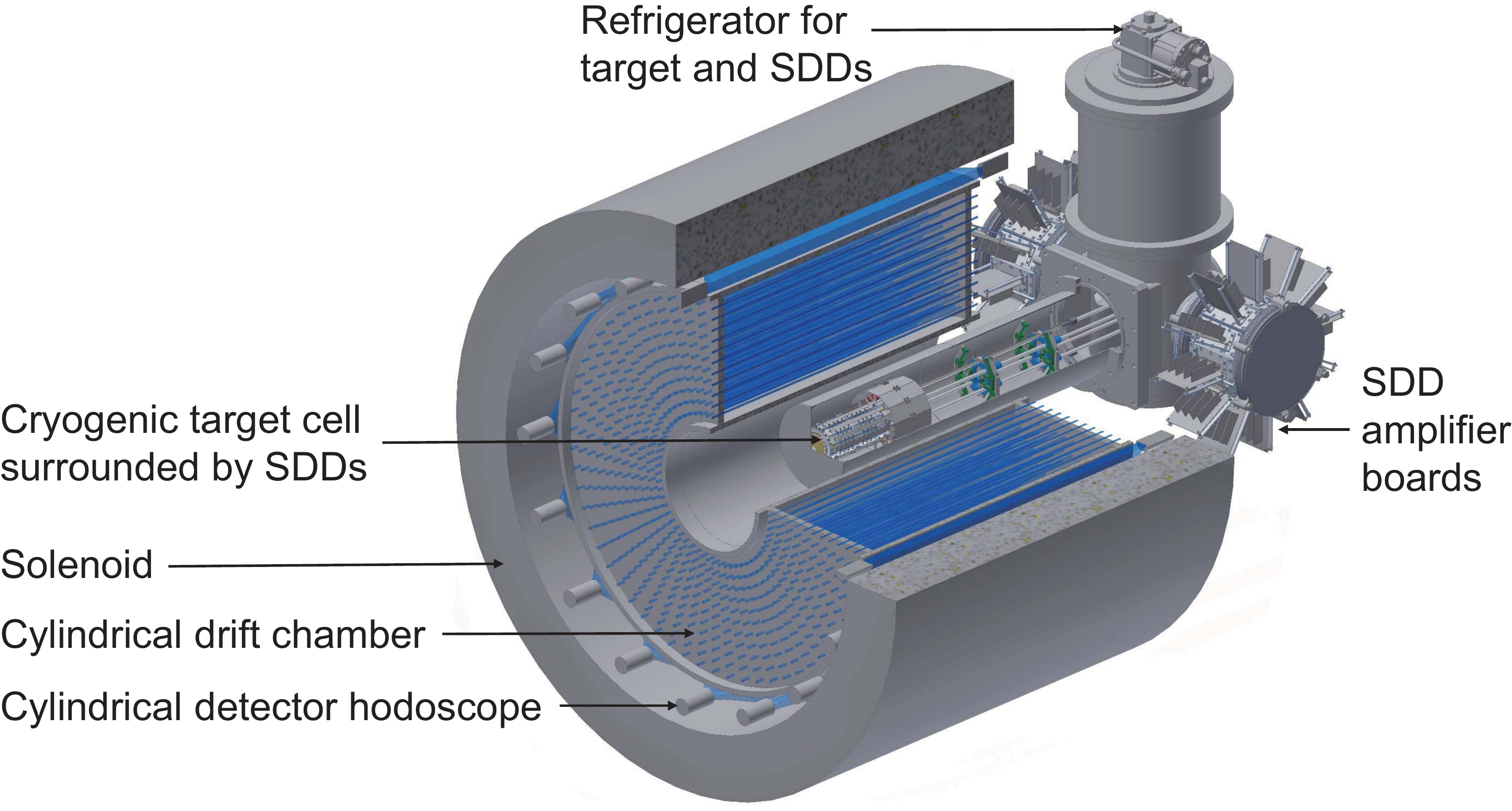}}
\caption{
Schematic illustration of the CDS layout for the E57 setup.
}
\label{fig:E57}
\end{figure}

The J-PARC E57 experiment at the K1.8BR beamline has been proposed to measure the shift and width of
the kaonic-deuterium $1s$ state with an accuracy of 60~eV and 140~eV,
respectively~\cite{E57-proposal}.
The experiment uses a gaseous deuterium target at a density of 4\% of the
liquid deuterium density (30 K with 0.35 MPa, $\sim$ 30 times the STP
density), where a $K_{\alpha}$ yield of $\sim$ 0.1\% is
expected~\cite{Koike:1995sa,Jensen:2004}.
To efficiently measure low-yield X-rays with a large width,
a large solid angle is covered with many arrays of silicon drift
detectors (SDDs)~\cite{Gatti:1984uu,Lechner:1996,Gatti:2005dw}.
In addition, a cylindrical detector system (CDS) is used to improve the
S/N ratio by the removal of charged-particle hits on the SDDs and
selection of the reaction vertex.
Charged decay particles from the target are detected by the CDS, which
consists of a solenoid magnet, a cylindrical wire drift chamber (CDC),
and a cylindrical detector hodoscope (CDH) with a solid angle coverage
of 59\%~\cite{Agari:2012gj}.
Figure~\ref{fig:E57} shows a schematic illustration of the CDS for the E57
setup.
Tracking information of charged particles is obtained from the 
CDC, which operates in a solenoidal magnetic field of 0.7~T, and particle
identification is performed using ToF together with a
beamline trigger counter.

In the experiment, negatively-charged kaons incident with a momentum of
$\sim$ 700 MeV/c are degraded in carbon and copper degraders and are
finally stopped inside the gaseous deuterium target.
The kaonic deuterium atoms are produced via the stopped-$K^-$ reaction
in the deuterium target, of which the target cell is cylindrical with a length of
19 cm and a 6 cm diameter.
The $2p \to 1s$ X-ray transition of the kaonic deuterium atom is
measured with SDDs surrounding the target cell to give a total
area of 246 cm$^2$.
Each monolithic SDD array that has been developed for the study of
kaonic deuterium at DA$\Phi$NF (SIDDHARTA-2) and J-PARC
(E57) has 8 square cells with a total active
area of 5.12 cm$^2$~\cite{Curceanu:2019uph}, and in total 48 SDDs are used for the experiment.
The read-out of the SDD is based on a CMOS charge sensitive preamplifier
(CUBE)~\cite{Quaglia:2015}.
The performance of the SDDs has been confirmed to achieve an energy
resolution of 130~eV at 6~keV with the temperature at 120 K.
The cryogenic target cell surrounded by the SDDs is placed in the center
of the CDS, as shown in Fig.~\ref{fig:E57}.

Following a pilot experiment with a hydrogen target in the near future, the
kaonic deuterium measurement will be performed.

\subsubsection{ Precise measurement of kaonic-helium $3d \to 2p$ X-rays \label{subsubsec:E62}}

Kaonic atoms with $Z \geq 2$ have been studied in terms of nuclear
medium effects for the strong interaction at the threshold
energy to derive information on the possible appearance of kaon
condensation and the evolution of strangeness in high-density stars, such
as neutron stars.
Many experimental measurements of kaonic atom X-rays have been
performed with various nuclear targets from helium (He) to uranium (U),
and the $\bar K$-nucleus potential has been theoretically obtained using
the density dependent optical potentials.
The depth of the real-part potential obtained with the phenomenologically
well known $t\rho$ potential is typically 180
MeV~\cite{Friedman:1993cu,Friedman:1994hx,Batty:1997zp,Friedman:1998xa}.
Such substantial attractive potentials have led to the possible existence of
`deeply' bound kaonic nuclear systems, which will be discussed in
Sec.~\ref{subsubsec:E27_E15}.
The potential has also been constructed from the
effective $\bar K N$ interaction obtained from a coupled channel chiral
unitary approach to the low-energy $\bar K N$ data.
The depth of the chiral based potential is typically 50 MeV~\cite{Ramos:1999ku,Hirenzaki:2000da,Cieply:2001yg}, which is
shallower than the potential obtained with the phenomenological
models.
However, the existing data is not sufficient to discriminate between
the conflicting interpretations; both of the calculated results using
the different potentials agree well with the experimental data within
the uncertainties.

In such a situation, a possible breakthrough has been specifically
pointed out;
a high-precision measurement of the isotope shift
between the kaonic-$^3$He and kaonic-$^4$He $2p$ states could resolve
the question of whether the potential is deep or shallow.
Table~\ref{table:KHe} shows preliminary results for the calculated
strong-interaction shifts of the kaonic-$^3$He and kaonic-$^4$He $2p$
states~\cite{Yamagata:2015}.
The difference in the isotope shift between the two models has been
predicted to be $\sim$ 0.6 eV.

\begin{table}[htbp]
 \caption{
 Preliminary results for the calculated shifts of the 
 kaonic-$^3$He and $^4$He $2p$ states~\cite{Yamagata:2015}.
 The two different optical potentials of the phenomenological
 model~\cite{Mares:2006vk} and the chiral unitary
 model~\cite{Ramos:1999ku} are assumed, where a Gaussian expansion method
 is used for the charge density distributions of $^3$He and
 $^4$He~\cite{Hiyama:2015}.
 }
 \begin{center}
  \begin{tabular}{ccc}
   \hline
   & Phenomenological &  Chiral unitary \\
   & [$V_{optical} \sim -180 + i73$ MeV] & [$V_{optical} \sim -40 + i55$ MeV] \\
   \hline
   $K^-$ $^4$He & -0.41 eV & -0.09 eV \\
   $K^-$ $^3$He & +0.23 eV & -0.10 eV \\
   Isotope shift ($K^-$ $^4$He - $K^-$ $^3$He) & -0.64 eV & +0.01 eV \\
   \hline
  \end{tabular}
  \label{table:KHe}
 \end{center}
\end{table}

However, the most precise experimental measurement was achieved by
the KEK-PS E570 experiment using a liquid helium target and
SDDs~\cite{Okada:2007ky}:
\begin{equation}
 \epsilon_{2p}(K^- {^4}{\rm He}) = +2\pm2 ({\rm stat.})\pm2 ({\rm
  syst.})~{\rm eV}.
\end{equation}
The SIDDHARTA experiment also measured the isospin dependence of the
level shift by the strong interaction using gaseous helium targets and
SDDs~\cite{Bazzi:2009zz,Bazzi:2010zt}:
\begin{eqnarray}
 \epsilon_{2p}(K^- {^4}{\rm He}) = + 5\pm3 ({\rm stat.})\pm4 ({\rm
  syst.})~{\rm eV},\\
 \epsilon_{2p}(K^- {^3}{\rm He}) = - 2\pm2 ({\rm stat.})\pm4 ({\rm
  syst.})~{\rm eV}.
\end{eqnarray}
Therefore, the experimental accuracy obtained to date has been an order of magnitude
worse compared to the expected shifts obtained by theoretical calculations.
The required precision of the kaonic-helium isotope measurements is in
the order of $\sim$ 0.2 eV, which also makes it possible to determine the
sign of the level shift for each isotope.

To realize high-resolution and high-accuracy measurements of 
kaonic-helium atoms X-rays, the J-PARC E62 experiment utilized a
superconducting transition-edge-sensor (TES) microcalorimeter.
The TES is a highly sensitive thermal sensor that measures energy
deposition by measurement of the increase in the resistance of a
superconducting material biased within the sharp phase transition
between the normal and superconducting
phases~\cite{Irwin:2005,Doriese:2017}.
The energy resolution of the TES microcalorimeter is
$\sim$ 5 eV (FWHM) at 6 keV, which enables determination of the level shift to
as good as 0.2 eV, which is one order of magnitude better than the precision of
a semiconductor detector such as the SDD.
Recent technological advances in the multiplexed readout of a TES
multi-pixel array has enabled the use of the TES detector in measurements of
kaonic-helium X-rays with reasonably large acceptance.

The experiment employed a 240 pixel TES array (effective area of 23
mm$^2$) designed for hard X-ray measurements developed by the National
Institute of Standards and Technology (NIST).
As a demonstration of the feasibility of the TES-based detector to perform
hadronic atom X-ray measurements, a precedence experiment was carried out
at the PSI $\pi$M1 beamline in 2014, with the same type of TES
spectrometer as that used in the E62 experiment~\cite{Tatsuno:2017,Okada:2016slo}.
In this experiment, the {\it in situ} energy calibration method was
demonstrated by shining characteristic X-rays on the TES, excited by an
X-ray tube source during data acquisition.
The FWHM energy resolution obtained was 6.8 eV at 6.4 keV (Fe $K_{\alpha
1}$), and the uncertainty of the absolute energy calibration was achieved
to 0.1 eV under a high-rate hadron beam condition of 1.45 MHz, which
matches the E62 goal of the kaonic-helium measurement.

\begin{figure}[htbp]
\centerline{\includegraphics[width=15cm]{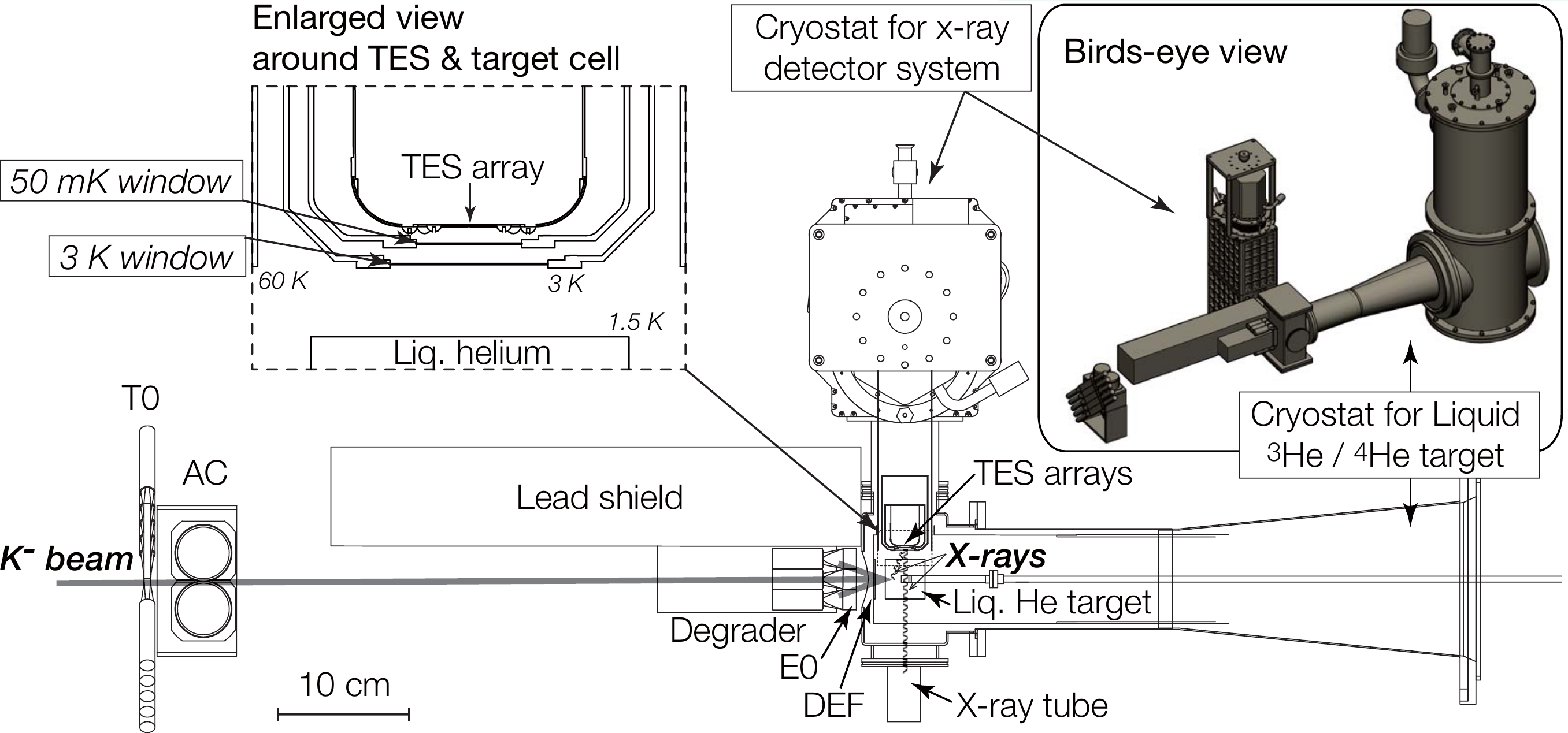}}
\caption{
Schematic diagram of the E62 experimental setup around the experimental
target.
The insets show a birds-eye view and an enlarged view around the TES and
the helium target cell.
The figure is taken from the E62 proposal~\cite{E62}.
}
\label{fig:E62}
\end{figure}

After a commissioning experiment at the secondary K1.8BR beamline in
2016~\cite{Hashimoto:2017}, the E62 experiment was performed in 2018.
Figure~\ref{fig:E62} shows the experimental setup.
Incident kaons extracted with a momentum of 900 MeV/$c$ were counted with
beamline counters, degraded, and stopped inside the
liquid-helium target.
X-rays emitted from the kaonic-helium atoms were detected by the TES
spectrometer. 
An X-ray generator and secondary target metals were also installed to
perform the absolute energy calibration for every single readout
channel, as demonstrated in the precedence experiment.

During the $\sim$ 20 days beam time, more than 100 X-ray counts from both
kaonic-$^3$He and $^4$He $3d \to 2p$ transitions were accumulated~\cite{Hashimoto:QNP2018}.
The energy resolutions achieved were $\sim$ 7 eV (FWHM) at 6 keV.
The expected statistical precisions of the kaonic-helium X-ray energies
were 0.35 and 0.2 eV for $K^-$ $^3$He and $K^-$ $^4$He, respectively,
with a systematic uncertainty of 0.1 eV.

\subsubsection{ Investigation of the $\Lambda(1405)$ line shape \label{subsubsec:E31}}

Extensive measurements of the $K^-p$ scatterings has also led to a
theoretical prediction of the existence of the $\Lambda(1405)$ state
below the $K^- + p$ mass threshold~\cite{Dalitz:1959dn}.
After the first observation of the $\Lambda(1405)$ state using a
hydrogen bubble chamber~\cite{Alston:1961zzd}, the $\Lambda(1405)$
state has been well established today and is listed in the table of the
Particle Data Group as a four-star state~\cite{Tanabashi:2018oca}.
The $\Lambda(1405)$ state is known to have strangeness $S=-1$ and
isospin $I=0$, which is located slightly below the $K^- + p$ mass
threshold (Fig.~\ref{fig:L1405}) and decays into the $\pi\Sigma$
channels with a 100\% branching ratio.
The quantum number $J^P = 1/2^-$ was recently derived by the CLAS
collaboration at Jefferson Lab (JLab) using the reaction $\gamma p \to K^+
\Lambda(1405)$~\cite{Moriya:2014kpv}.

It is well known that there is a difficulty in describing the
$\Lambda(1405)$ state as an ordinary three-quark state in simple
constituent quark models because its mass is lower than any other excited
spin $1/2$ baryons~\cite{Isgur:1978xj}.
Thus, the $\Lambda(1405)$ state has been widely interpreted as a dynamically
generated resonance through a $\bar K N$-$\pi \Sigma$ coupled channel
system~\cite{Dalitz:1967fp}.
A recent lattice QCD calculation also strongly suggested that the
structure of the $\Lambda(1405)$ state is dominated by a bound $\bar K N$
component~\cite{Hall:2014uca}.

During the past decade, many new experimental results on the
$\Lambda(1405)$ state have been reported; photoproduction at the
LEPS (SPring-8)~\cite{Niiyama:2008rt} and the CLAS
(JLab)~\cite{Moriya:2013eb}, electroproduction at the CLAS
(JLab)~\cite{Lu:2013nza}, and 
proton-proton collision at the ANKE (COSY)~\cite{Zychor:2007gf} and the
HADES (GSI)~\cite{Agakishiev:2012xk}. 
Using these precise $\pi\Sigma$ spectra, detailed theoretical studies
on the $\Lambda(1405)$ line shape have also been performed.
However, despite many such experimental attempts and theoretical
analyses, the most fundamental unsettled question still remains:
whether the $\Lambda(1405)$ state is located at 1405 MeV/$c^2$
or at 1420 MeV/$c^2$, which correspond to the respective single- or double-pole nature
of the $\Lambda(1405)$ state.

The phenomenological models are based on theoretical fits of the
$\pi\Sigma$ invariant mass spectra with a single pole energy independent structure, which shows the $\Lambda(1405)$ mass pole is around
1405 MeV with a deep $\bar K N$
potential~\cite{Dalitz:1991sq,Akaishi:2002bg}.
In contrast, the chiral unitary approaches are based on an effective
Lagrangian of the chiral perturbation theory, and are thus energy dependent.
The potential is constructed by combining the experimental data of
the low-energy $\bar K N$ scatterings and the kaonic hydrogen, the
result of which gives a shallow potential.
In the chiral unitary approaches, the $\Lambda(1405)$ state is predicted to
have a double pole structure, and consequently the line shape has 
a single-like peak structure around 1420 MeV; the first and main pole
coupled to the $\bar K N$ channel is located near the $\bar K N$ mass
threshold ($\sim$ 1432 MeV) with a relatively small width, and the
second pole coupled to the $\pi\Sigma$ channel appears near the $\pi
\Sigma$ threshold ($\sim$ 1328 MeV) with a large
width~\cite{Oller:2000fj,Jido:2003cb,Hyodo:2007jq,Hyodo:2011ur}.

The Review of Particle Physics by the Particle Data
Group (PDG)~\cite{Tanabashi:2018oca} has adopted the values obtained by the
phenomenological analyses based on the single-pole nature of
$\Lambda(1405)$; a mass of $1405.1^{+1.3}_{-1.0}$ MeV/$c^2$ and a width
of $50.5 \pm 2.0$ MeV/$c^2$.
These values were obtained from the theoretical fits to the $\pi\Sigma$
invariant mass spectra from the reaction $K^-p \to \Sigma^+ \pi^-
\pi^+ \pi^-$ at 4.2 GeV/$c$ measured by a hydrogen bubble chamber
experiment~\cite{Hemingway:1984pz,Dalitz:1991sq}, the stopped $K^-$
reaction in $^4$He measured by a helium bubble chamber
experiment~\cite{Riley:1975rg,Esmaili:2009iq}, and the reaction $pp \to
\Sigma^{\pm} \pi^{\mp} K^+ p$ at 3.5 GeV from the HADES
experiment~\cite{Agakishiev:2012xk,Hassanvand:2012dn}.

On the other hand, the precise $\pi\Sigma$ spectra obtained from the
$\gamma$-induced reaction at the CLAS experiment
were also theoretically
analyzed using the chiral motivated models, {\it i.e.} based on the
double-pole nature of
$\Lambda(1405)$~\cite{Roca:2013av,Nakamura:2013boa,Mai:2014xna}.
The CLAS data have been well reproduced by the theoretical models,
the obtained pole positions of which are summarized in the review {\it
``Pole structure of the $\Lambda(1405) region$ (U.-G.~Meissner and
T.~Hyodo, 2015)''}~\cite{Tanabashi:2018oca}.
However, a recent phenomenological analysis of the CLAS data has also
extracted the pole of the $\Lambda(1405)$ state based on a single pole model,
and claimed that the pole is consistent with the PDG
value~\cite{Hassanvand:2017iif}.
Also, a recent partial wave analysis has shown the good description
of experimental data with a single pole model having the pole of $(1422
\pm 3, -i(21\pm3))$ MeV~\cite{Anisovich:2019exw}; photoproduction
data from the CLAS experiment and $K^-$ induced reaction data obtained
with the Crystal Ball multiphoton spectrometer at
BNL~\cite{Prakhov:2004an} and old bubble chambers were utilized.

Therefore the situation is still controversial in terms of the
theoretical analysis of the $\Lambda(1405)$ line shape.
To clarify which scenario is valid, decomposition of the
$\Lambda(1405)$ state coupled to $\bar K N$ is of essential
importance; $\Lambda(1405)$ lies below the $\bar K N$
threshold and thus has no decay channel into $\bar K N$; therefore, 
investigation of the $\bar K N$ collision process in a virtual state is required.

The J-PARC E31 experiment aims to exclusively show the $\Lambda(1405)$ line
shape via the $\bar K N \to \pi \Sigma$ channels using the $(K^-,n)$
reaction on a deuterium target.
$\Lambda(1405)$ production initiated by virtual $\bar K N$
scattering is theoretically expected to be enhanced in the $K^- d \to
\Lambda(1405) n$ process~\cite{Jido:2009jf}; therefore, this process is
important to investigate the $\Lambda(1405)$ properties via the
sub-threshold $\bar K N \to \pi \Sigma$ channels.
In the experiment, the missing-mass spectra of the $d(K^-,n)$
reaction are measured in coincidence with decay particles from
the $\Lambda(1405)$ state.
To decompose the isospin amplitudes of $I = 0, 1$ and their interference
term in the $\pi \Sigma$ spectrum, all of the $\pi^+ \Sigma^-$, $\pi^-
\Sigma^+$, and $\pi^0 \Sigma^0$ final states are identified.
An exclusive measurement is realized by the detection of a neutron from the 
$(K^-,n)$ reaction and two charged particles from the $\pi\Sigma$
decays, where the missing neutral particle(s) is identified by the missing
mass of the reaction:
\begin{eqnarray}
 \Lambda(1405) &\to& \pi^+ \Sigma^- \to \pi^+ \pi^- n ~ (33\%), \nonumber \\
 &\to& \pi^- \Sigma^+ \to \pi^+ \pi^- n ~ (16\%) ~/~ \pi^0 \pi^- p ~ (17\%), \nonumber \\
 &\to& \pi^0 \Sigma^0 \to \gamma \pi^0 \pi^- p ~ (21\%),
\label{eq:L1405}
\end{eqnarray}
where the numbers in parentheses show the branching ratio in each reaction with the
assumption of a branching ratio of 33.3\% for each decay channel of 
$\Lambda(1405) \to \pi^{\pm0} \Sigma^{\mp0}$.
The $\Sigma(1385)^0$ resonance ($I=1$) is a significant
background to isolate the $\Lambda(1405)$ state ($I=0$) in the
$\pi^{\pm}\Sigma^{\mp}$ channels:
\begin{eqnarray}
 \Sigma(1385)^0 &\to& \pi^+ \Sigma^- \to \pi^+ \pi^- n ~ (6\%), \nonumber \\
 &\to& \pi^- \Sigma^+ \to \pi^+ \pi^- n ~ (3\%) ~/~ \pi^0 \pi^- p ~ (3\%), \nonumber \\
 &\to& \pi^0 \Lambda \to \pi^0 \pi^- p ~ (56\%).
\end{eqnarray}

The measurement was performed using the K1.8BR spectrometer.
The incident $K^-$ momentum of 1.0 GeV/$c$ was selected to maximize the
$(K^-,n)$ reaction rate~\cite{Kishimoto:1999yj}.
The scattered neutrons from the $(K^-,n)$ reaction were measured by a
neutron counter (NC) located $\sim$ 15~m downstream from the target
position, with a missing-mass resolution of $\sim$ 10 MeV/$c^2$
$(\sigma)$ for the region of interest.
The decay charged particles associated with the reaction were detected
by the CDS surrounding a liquid deuterium
target system.
To detect a backward boosted proton from the sequential decay of the
$\Lambda(1405) \to \pi^0\Sigma^0 \to \gamma \pi^0 \pi^- p $ decay, the
E31 experiment used backward-proton detectors installed just upstream
of the target system in the CDS~\cite{Agari:2012gj}.

\begin{figure}[htbp]
\centerline{\includegraphics[width=10cm]{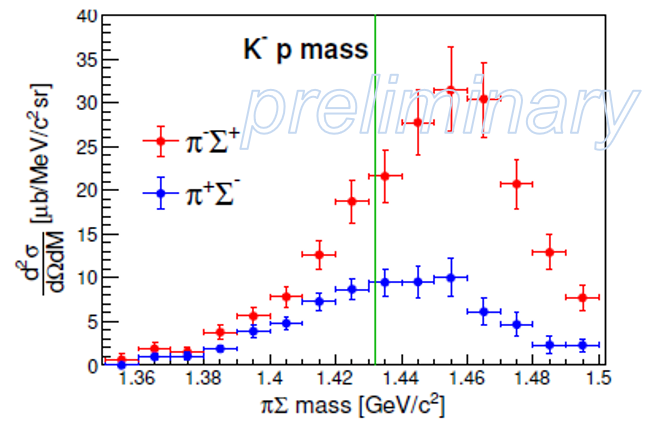}}
\caption{
Results of the $d(K^-,n)\pi^{\pm}\Sigma^{\mp}$ missing-mass
spectra obtained with the E31-1st data.
Reproduced from Ref.~\cite{Asano}, with the permission of AIP Publishing.
}
\label{fig:E31_1}
\end{figure}

In 2016 and 2018, the first (E31-1st) and second (E31-2nd) experiments
were conducted with respectively $1.5\times10^{10}$ and $3.9\times10^{10}$ kaons
on a deuteron target.
Figure~\ref{fig:E31_1} shows results of the
$d(K^-,n)\pi^{\pm}\Sigma^{\mp}$ missing-mass spectra~\cite{Asano}.
In the analysis, a neutron from the $\Sigma^{\pm}$ decay was identified
using the missing-mass of $d(K^-,n\pi^+\pi^-)X$ as $X=n$, and the
$\Sigma^{\pm}$ were then obtained from the missing-mass of
$d(K^-,n\pi^{\mp})X$ as $X=\Sigma^{\pm}$.
The ratio of the $\pi^+\Sigma^-$ to $\pi^+\Sigma^-$
states was evaluated with a Monte Carlo template fitting in each 10
MeV/$c^2$ bin.

\begin{figure}[htbp]
\centerline{\includegraphics[width=10cm]{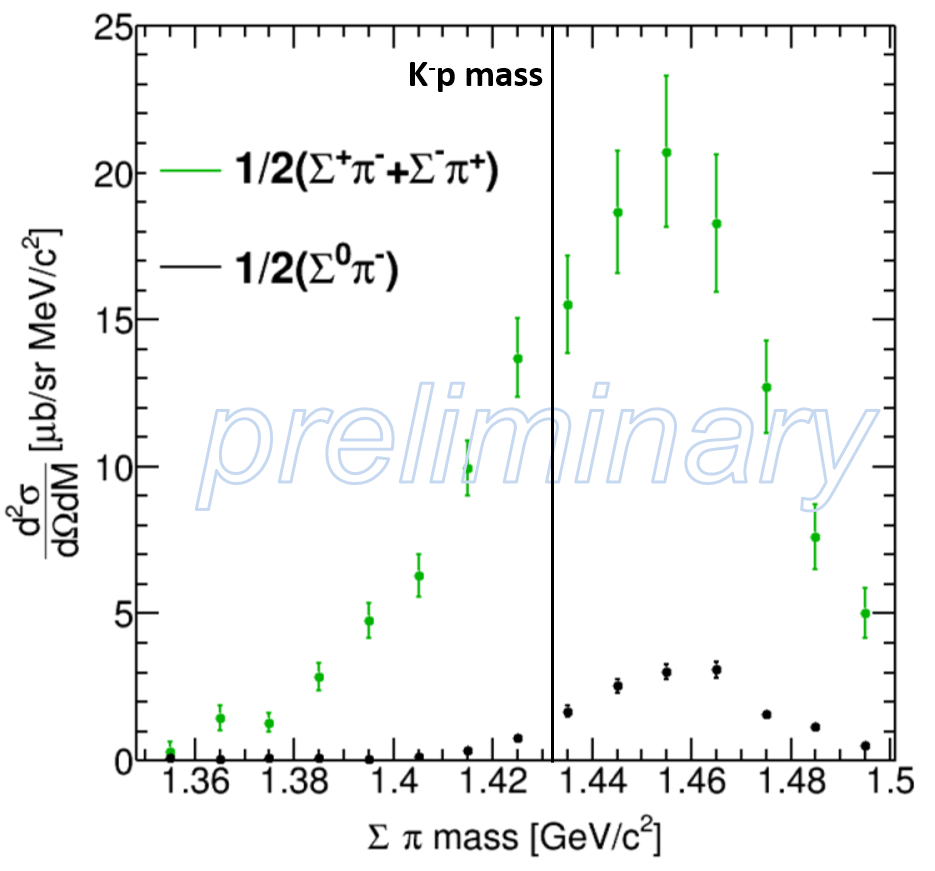}}
\caption{
Result of the average of the $d(K^-,n)\pi^{\pm}\Sigma^{\mp}$
missing-mass spectra obtained with the E31-1st data.
The $d(K^-,p)\pi^{-}\Sigma^{0}$ missing-mass spectrum divided by two is
also shown.
Reproduced from Ref.~\cite{Asano}, with the permission of AIP Publishing.
}
\label{fig:E31_2}
\end{figure}

The difference of the spectral shape and strength between the
$\pi^+\Sigma^-$ and $\pi^-\Sigma^+$ spectra is clearly evident in
Fig.~\ref{fig:E31_1}.
This is strong evidence of the interference between the $I = 0$ and 
$I = 1$ channels in the $\pi^{\pm}\Sigma^{\mp}$ states, which can
be given as~\cite{Nacher:1998mi}:
\begin{eqnarray}
 \frac{d\sigma}{d\Omega}(\pi^+\Sigma^-) &\propto& \frac{1}{3}|f_{I=0}|^2 +
  \frac{1}{2}|f_{I=1}|^2 + \frac{\sqrt{6}}{3}{\rm Re}(f_{I=0}f_{I=1}^*), \\
 \frac{d\sigma}{d\Omega}(\pi^-\Sigma^+) &\propto& \frac{1}{3}|f_{I=0}|^2 +
  \frac{1}{2}|f_{I=1}|^2 - \frac{\sqrt{6}}{3}{\rm Re}(f_{I=0}f_{I=1}^*), \\
  \frac{d\sigma}{d\Omega}(\pi^0\Sigma^0) &\propto& \frac{1}{3}|f_{I=0}|^2,
\end{eqnarray}
where $f_{I=0}$ and $f_{I=1}$ denote the $I = 0$ and $I = 1$
amplitudes, respectively.
To cancel out the interference between the $I=0$ and $I=1$
components, the $\pi^{\pm}\Sigma^{\mp}$ spectra are averaged, as shown in
Fig.~\ref{fig:E31_2}.
In addition, the pure $I=1$ component of the $d(K^- p)\pi^-\Sigma^0$
reaction was obtained with the same data set using a forward proton ToF
counter array located alongside the NC.
The missing-mass spectrum of $d(K^-, p)\pi^-\Sigma^0$ is plotted in
Fig.~\ref{fig:E31_2} together with the averaged $\pi^{\pm}\Sigma^{\mp}$
spectra.
Under an assumption of the similarity of the $d(K^-,n)$ and $d(K^-,p)$
reaction mechanisms, these two amplitudes can be written as:
\begin{eqnarray}
 \frac{1}{2}\left(\frac{d\sigma}{d\Omega}(\pi^+\Sigma^-) +
	     \frac{d\sigma}{d\Omega}(\pi^-\Sigma^+)\right)
 &\propto& \frac{1}{3}|f_{I=0}|^2 +  \frac{1}{2}|f_{I=1}|^2, \\
 \frac{1}{2}\frac{d\sigma}{d\Omega}(\pi^-\Sigma^0) &\propto&
 \frac{1}{2}|f_{I=1}|^2.
\end{eqnarray}
Therefore, the results strongly indicate that the amplitude of the $I=0$
component is dominated at the forward direction, in particular, below
the $\bar K N$ threshold.

\begin{figure}[htbp]
\centerline{\includegraphics[width=10cm]{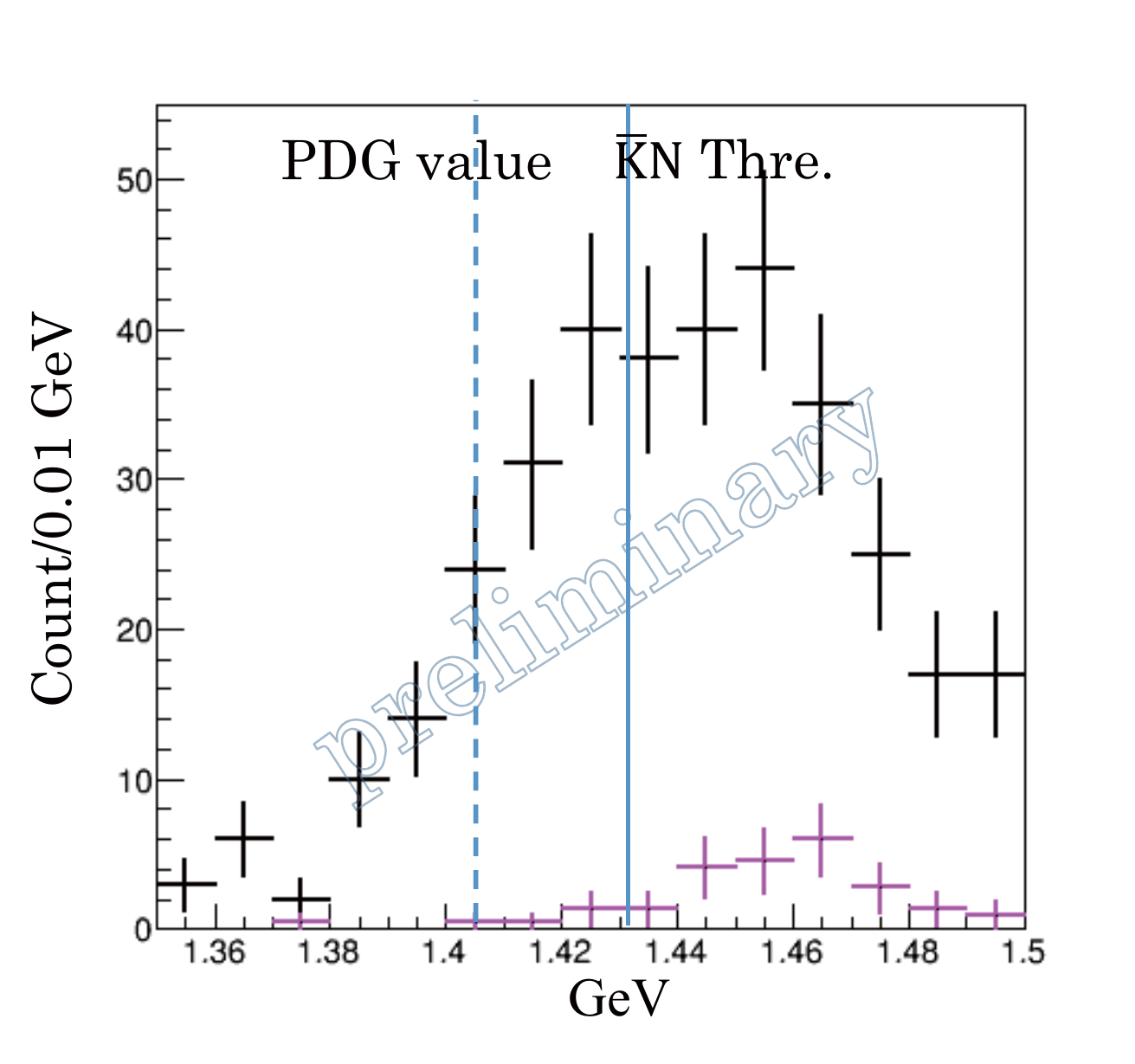}}
\caption{
Result of the $d(K^-,n)\pi^0\Sigma^0$ missing-mass spectrum
without the acceptance correction.
The background contribution (magenta) from the $K^-d \to
 \pi^-\Sigma^+n / \pi^0\Lambda n \to \pi^0\pi^-pn$ reactions is also plotted.
The figure is taken from Ref.~\cite{Kawasaki}.
}
\label{fig:E31_4}
\end{figure}

Figure~\ref{fig:E31_4} shows the result of the $d(K^-,n)\pi^0\Sigma^0$
missing-mass spectrum without the acceptance correction.
In the analysis, a negative pion and a backward-going proton in the decay
chain of $\pi^0\Sigma^0 \to \gamma\pi^0\Lambda \to \gamma\pi^0\pi^-p$
were identified by the CDS and the backward-proton detectors,
respectively.
The $\Lambda \to \pi^- p$ decay was then identified in the $\pi^-p$
invariant mass spectrum, and the $K^-d \to \pi^0\Sigma^0n$ event was
selected using the missing-mass spectrum of $d(K^-,n\Lambda)X$ as
$X=\gamma\pi^0$.
In the figure, the background from the $K^-d \to \pi^-\Sigma^+n /
\pi^0\Lambda n \to \pi^0\pi^-pn$ reactions is also plotted.
The significant yield below the $\bar K N$ threshold is evident, where
the $\pi^-\Sigma^+ / \pi^0\Lambda$ background is negligibly small.

The results obtained can be compared with theoretical
calculations, taking into account the kinematics of the E31 experiment;
the $(K^-,n)$ measurement at $0^{\circ}$ with a kaon momentum of 1.0
GeV/$c$.
In Ref.~\cite{Kamano:2016djv}, a model of the $K^- d \to \pi \Sigma N$
reactions was developed with the off-shell amplitudes of the $\bar
K N \to \bar K N$ and $\bar K N \to \pi Y$ reactions generated from a
dynamical coupled-channels (DCC) model that was constructed by fitting
the existing data of the $K^-p \to \bar K N, \pi\Sigma, \pi\Lambda,
\eta\Lambda$, and $K\Xi$ reactions~\cite{Kamano:2014zba}.
The model indicates the appearance of the two-pole structure of the
$\Lambda(1405)$ state, similar to the structure obtained
in the chiral unitary models.
Furthermore, in Ref.~\cite{Miyagawa:2018xge}, the reaction $K^- d \to \pi
\Sigma N$ has been studied within a Faddeev-type approach using the
$\bar K N \to \bar K N$ amplitude obtained by a partial wave
analysis~\cite{Zhang:2013cua} and chiral unitary models of the $\bar K N
\to \pi \Sigma$
interaction~\cite{Oset:2001cn,Cieply:2011nq,Ohnishi:2015iaq}.
Both of the calculated $K^- d \to \pi \Sigma N$ spectra in
Ref.~\cite{Kamano:2016djv} and Ref.~\cite{Miyagawa:2018xge} show
reasonable agreement with the E31 results.
Therefore, the E31 results show significant potential to reveal the details of the
$\bar K N \to \pi \Sigma$ amplitude and to provide substantial
constraints on the line shape of the $\Lambda(1405)$ state in conjunction with
various theoretical calculations.

\subsubsection{ Search for kaonic nuclei \label{subsubsec:E27_E15}}
The $\Lambda(1405)$ state can be inherently interpreted as the $\bar K
N$ quasi-bound state due to the strongly attractive $\bar K N$
interaction in the $I=0$ channel; therefore, the existence of the $\bar
K$-nuclear quasi-bound state has been widely
discussed~\cite{Nogami:1963xqa,Akaishi:2002bg,Yamazaki:2002uh} and
various interesting phenomena have been suggested.
The most significant interest of these exotic states is that they could form
high-density cold nuclear matter, such as the inner core of neutron
stars, where the chiral symmetry is expected to be
restored~\cite{Dote:2002db,Dote:2003ac}.
Therefore, investigation of the $\bar K$-nuclear quasi-bound states is
expected to provide new insights, not only on meson-baryon interactions
in the low-energy QCD, but also on the change of the meson properties in
nuclear media.

The lightest predicted $S = -1$ $\bar K$ nucleus is the $\bar K NN$
system (symbolically denoted as ``$K^-pp$''), which is expected to be a
$[\bar K \otimes {NN}_{I=0, S=0} ]_{I=1/2}$ with $J^{P} = 0^-$ state.
The existence of the $\bar K NN$ quasi-bound state is supported by many
theoretical predictions based on few-body calculations using the
$\bar K NN$-$\pi \Sigma N$-$\pi \Lambda N$ coupled formalism.
However, the calculated properties, such as the binding energy
($B$) and the width ($\Gamma$), are scattered due to the large 
dependence of the model on the $\bar K N$ interaction.
Table~\ref{table:Kpp} summarizes the calculated binding energies and widths
obtained from various theoretical works.
Calculations based on the energy independent models (phenomenological
models) show rather large binding energies of $\sim$ 40--90
MeV~\cite{Yamazaki:2002uh,Shevchenko:2006xy,Shevchenko:2007ke,Ikeda:2007nz,Ikeda:2008ub,Wycech:2008wf,Revai:2014twa,Dote:2017veg,Ohnishi:2017uni},
whereas this becomes as small as $\sim$ 10--50 MeV with the energy dependent models
(chiral unitary
models)~\cite{Dote:2008in,Dote:2008hw,Ikeda:2010tk,Barnea:2012qa,Bayar:2012hn,Revai:2014twa,Ohnishi:2017uni,Dote:2017wkk}.
With regard to the widths, the predicted values are widely spread
($\sim$ 20--110 MeV) and independent of the $\bar K N$ interaction models.
In the framework of the chiral model, a double pole structure of the
$\bar K NN$ has also been proposed in relation to the
$\Lambda(1405)$ state~\cite{Ikeda:2010tk,Dote:2014via}.

\begin{table}[htbp]
 \caption{Calculated binding energy ($B$) and width ($\Gamma$) of the
 $\bar K NN$ quasi-bound state from various theoretical works.}
 \begin{center}
  {\tabcolsep = 1.5mm
  \begin{tabular}{c ccc c ccc c c}
   \hline
   &
   \multicolumn{9}{c}{Energy independent $\bar K N$ interactions} \\
   &
   \multicolumn{9}{c}{(Phenomenological model)} \\ 
   \cline{2-10}
   &
   \multicolumn{3}{c}{\multirow{2}{*}{Variational}} & &
   \multicolumn{3}{c}{\multirow{2}{*}{Faddeev}} & &
   Complex- \\
   & & & & & & & & & scaling \\
   \cline{2-4}
   \cline{6-8}
   \cline{10-10}
   B (MeV) & 48 & 40--80 & 49 & &
   50--70 & 60--95 & 47--53 & &
   51 \\
   $\Gamma$ (MeV) & 61 & 40--85 & 62 & &
   90--110 & 45--80 & 50--65 & &
   32 \\
   Ref. &
   \cite{Yamazaki:2002uh} & \cite{Wycech:2008wf} & \cite{Ohnishi:2017uni} & &
   \cite{Shevchenko:2006xy,Shevchenko:2007ke} &
   \cite{Ikeda:2007nz,Ikeda:2008ub} & \cite{Revai:2014twa} & &
   \cite{Dote:2017veg} \\

   \hline 
   & & & & & & & & & \\
   &
   \multicolumn{9}{c}{Energy dependent $\bar K N$ interactions} \\
   &
   \multicolumn{9}{c}{(Chiral unitary model)} \\ 
   \cline{2-10}
   &
   \multicolumn{3}{c}{\multirow{2}{*}{Variational}} & &
   \multicolumn{3}{c}{\multirow{2}{*}{Faddeev}} & &
   Complex- \\
   & & & & & & & & & scaling \\
   \cline{2-4}
   \cline{6-8}
   \cline{10-10}
   B (MeV) &
   17--23 & 16 & 26--28 & &
   9--16 & 15--30 & 32 & &
   14-50 \\
   $\Gamma$ (MeV) &
   40--70 & 41 & 31--59 & &
   34--46 & 75--80 & 49 & &
   16--38 \\
   Ref. &
   \cite{Dote:2008in,Dote:2008hw} & \cite{Barnea:2012qa} & \cite{Ohnishi:2017uni} & &
   \cite{Ikeda:2010tk} & \cite{Bayar:2012hn} & \cite{Revai:2014twa} & &
   \cite{Dote:2017wkk} \\

   \hline
  \end{tabular}
  }
  \label{table:Kpp}
 \end{center}
\end{table}

There are several reports on the experimental observation of a
``$K^-pp$'' candidate with a binding energy of around
100 MeV, which were measured with non-mesonic decay branches of
$\Lambda p$ and/or $\Sigma^0 p$ in different reactions.
The FINUDA collaboration measured back-to-back $\Lambda p$ pairs in
stopped $K^-$ reactions on light nuclear
targets~\cite{Agnello:2005qj}, and the DISTO collaboration analyzed the 
$pp \to \Lambda p K^+$ channel at a proton energy of 2.85
GeV~\cite{Yamazaki:2010mu}.
However, the obtained spectra can also be understood without the
inclusion of a bound state;
the AMADEUS collaboration reported that the spectra in the stopped $K^-$
reactions can be well described by $K^-$ multi-nucleon absorption
processes without the need for the ``$K^-pp$''
component~\cite{Doce:2015ust,DelGrande:2018sbv}, and the $K^+ \Lambda p$
final state in the $p+p$ collision can be explained with resonant and
non-resonant intermediate states that decay into a $K^+ \Lambda$ pair, as reported by the
HADES collaboration~\cite{Agakishiev:2014dha}.
The LEPS collaboration also reported only upper limits on the cross
section for the ``$K^-pp$'' bound state in $\gamma$-induced
reactions~\cite{Tokiyasu:2013mwa}.

Thus, the experimental situation to search for the ``$K^-pp$'' bound state 
was also controversial.
To clarify whether or not the ``$K^-pp$'' bound state exists, the key to
the experimental search is to adopt a simple reaction and to measure it
exclusively.
Simple reactions, such as in-flight $\bar K$ induced reactions with
light target nuclei, would make the ``$K^-pp$'' production mechanism
clear.
Exclusive measurement is thus crucial to distinguish a small and broad
signal from largely and widely distributed quasi-free backgrounds.

At J-PARC, two experimental searches for the ``$K^-pp$'' bound state
were conducted.
The J-PARC E27 experiment aimed to search for the ``$K^-pp$'' bound state in the
$\pi^+ + d \to K^+ + X$ reaction at 1.69 GeV/$c$.
In this reaction, a possible ``$K^-pp$'' production
mechanism has been suggested to produce the ``$K^-pp$'' state via the
$\Lambda(1405) + p$ collision as a doorway process; the off-shell
$\Lambda(1405)$ (`$\Lambda(1405)$') is produced by the $\pi^+ +n \to K^+
+$ `$\Lambda(1405)$' reaction followed by the `$\Lambda(1405)$' $+ p \to
$ ``$K^-pp$'' reaction~\cite{Yamazaki:2007cs}.
However, due to the large recoil momentum of the $\Lambda(1405)$ state in the
reaction, a small sticking probability of the `$\Lambda(1405)$' $+ p
\to$ ``$K^-pp$'' reaction is expected.

\begin{figure}[htbp]
 \begin{center}
  \centerline{\includegraphics[width=10cm]{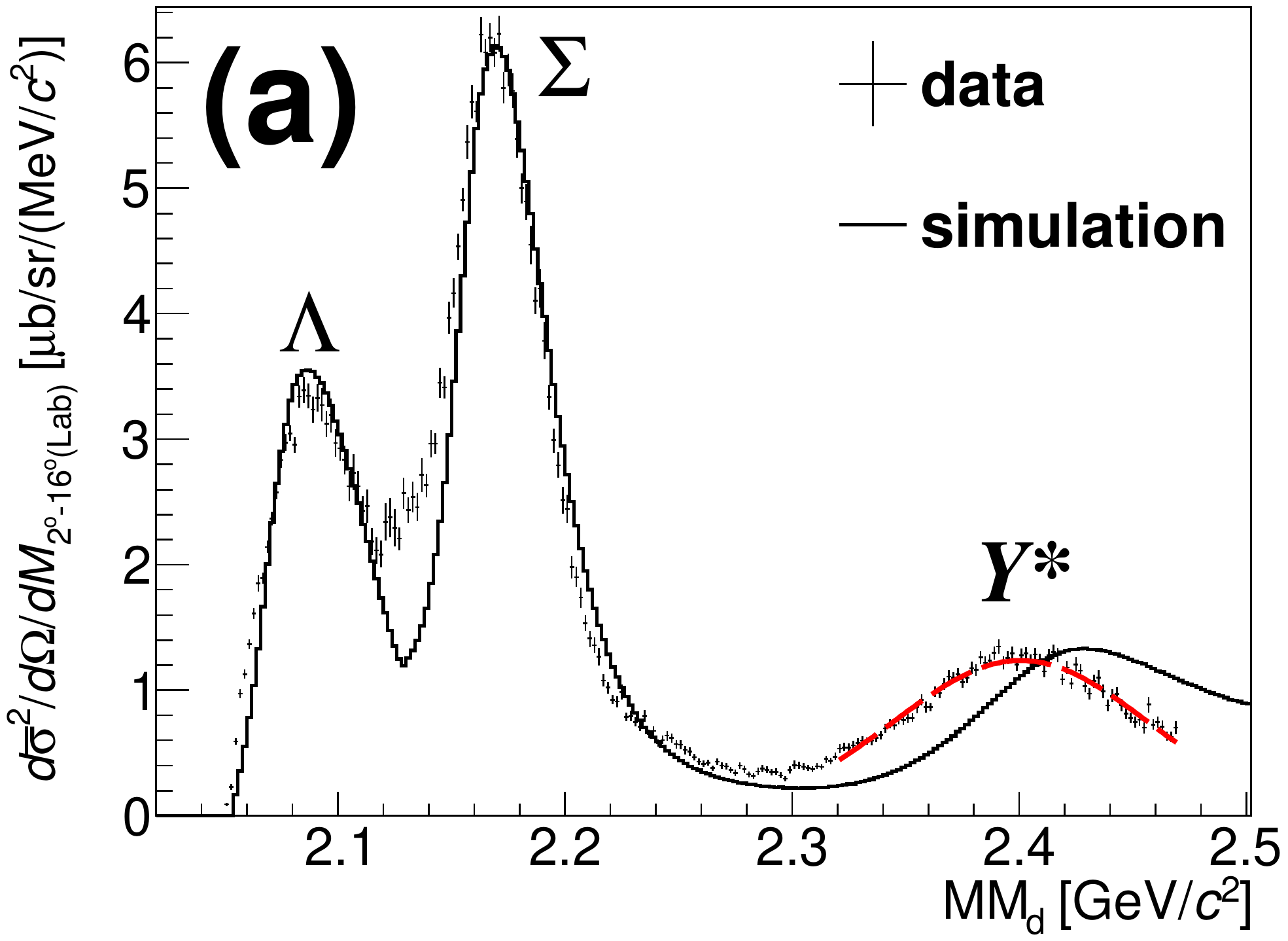}}
  \caption{
  Inclusive missing-mass spectrum of the $d(\pi^+, K^+)X$ reaction
  measured in the J-PARC E27 experiment.
  The crosses show the data and the solid line shows a simulated spectrum
  assuming quasi-free kinematics.
  The figure is taken from Ref.~\cite{Ichikawa:2014rva}.
  }
  \label{fig:E27_1}
 \end{center}
\end{figure}

The E27 experiment was conducted at the K1.8
beamline in 2012 with $3.3\times10^{11}$
incident pions. The details of the experimental setup
are given in Ref.~\cite{Ichikawa:2014rva}.
The incident $\pi^+$ and outgoing $K^+$ momenta were reconstructed
with the K1.8 beamline spectrometer and the SKS covering the scattering
angle between $2^{\circ}$ and $16^{\circ}$ in the laboratory system,
respectively.
Figure~\ref{fig:E27_1} shows the inclusive missing-mass spectrum of the
$d(\pi^+, K^+)$ reaction at 1.69 GeV/$c$, where quasi-free $\Lambda$,
quasi-free $\Sigma^+$ and $\Sigma^0$, and quasi-free $Y^* (Y^* =
\Lambda(1405)/\Sigma(1385)^+/\Sigma(1385^0))$ production is
clearly evident.
Although the overall structure of the spectrum is well reproduced
by a full Monte-Carlo simulation, as shown by the solid line, 
a difference is observed at the $\Sigma N$ mass threshold of 2.13
GeV/$c^2$, which is originated from the cusp effect.
Furthermore, the peak position of the quasi-free $Y^*$ production is shifted by approximately
30 MeV/$c^2$ toward low-mass side, which may indicate that a strong attractive
$Y^*N$ interaction exists ($\Lambda(1405)N$ and/or
$\Sigma(1385)N$)~\cite{Ichikawa:2014rva}.
Note that the LEPS group reported a similar inclusive spectrum for the
$d(\gamma, K^+\pi^-)X$ reactions in the 1.5--2.4 GeV photon energy region,
although no significant shift was observed in the $Y^*N$
region~\cite{Tokiyasu:2013mwa}.
This puzzling ``shift'' should be clarified theoretically by taking into
account the final state interaction.
Focusing below the $K^-+p+p$ mass threshold of 2.37 GeV/$c^2$, a
possible signal of the ``$K^-pp$'' not readily visible in the inclusive
spectrum.

\begin{figure}[htbp]
 \begin{center}
  \centerline{\includegraphics[width=10cm]{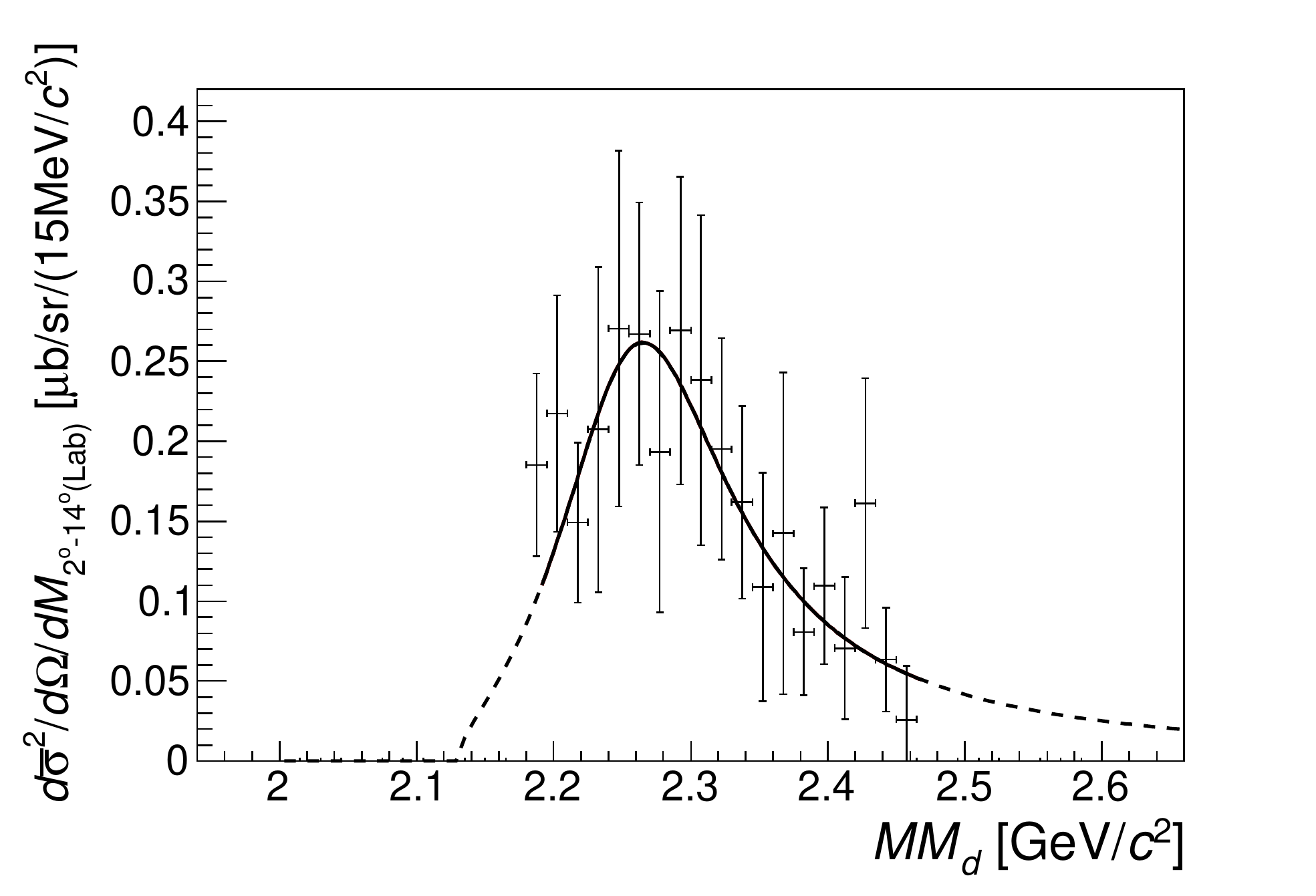}}
  \caption{
  Exclusive missing-mass spectrum of the $d(\pi^+, K^+)X$ reaction with 
  the $X \to \Sigma^0 p$ identification measured in the J-PARC E27
  experiment.
  The spectrum is fitted with a relativistic Breit-Wigner function.
  The figure is taken from Ref.~\cite{Ichikawa:2014ydh}.
  }
  \label{fig:E27_2}
 \end{center}
\end{figure}

To reconstruct the expected ``$K^-pp$'' $\to \Lambda/\Sigma^0 + p$ decays
exclusively, a range counter array (RCA) was installed surrounding the
target system.
By requiring high-momentum proton(s) over 250 MeV/$c$ at large emission
angles ($39^\circ < \theta_{lab}<122^\circ$) with an RCA, a broad
enhancement in the $d(\pi^+, K^+)X$ missing-mass spectrum was
observed around 2.27 GeV/$c^2$~\cite{Ichikawa:2014ydh}.
The decay branches of $\Lambda p (\to \pi^-pp)$, $\Sigma^0 p (\to
\gamma\pi^- pp)$, and $\pi Y N (\to \pi\pi pp)$ were decomposed using the
$d(\pi^+, K^+pp)X$ missing-mass spectra.
Figure~\ref{fig:E27_2} shows the missing-mass spectrum of the $d(\pi^+,
K^+)$ reaction for the $\Sigma^0 p$ decay-branch events, where the
spectrum is fitted with a relativistic Breit-Wigner function.
The obtained binding energy and width are
$95^{+18}_{-17}(stat.)^{+30}_{-21}(syst.)$ MeV and
$162^{+87}_{-45}(stat.)^{+66}_{-78}(syst.)$ MeV,
respectively~\cite{Ichikawa:2014ydh},
both of which are significantly larger than those of the theoretical
calculations for the ``$K^-pp$'' bound state.
The obtained production cross section $d \sigma / d \Omega_{\Sigma^0
p-decay}$ is $3.0\pm0.31 (stat.)^{+0.7}_{-1.1}(syst.)$ $\mu$b/sr.
The cross section of the peak structure in the $\Lambda p$ decay branch was
also evaluated from the fitting by assuming the same distribution of the
structure in the missing-mass spectrum of $d(\pi^+, K^+)X$.
The obtained branching fraction between the $\Sigma^0 p$ and $\Lambda p$
decay modes is $\Gamma_{\Lambda p} / \Gamma_{\Sigma^0
p} = 0.92^{+0.16}_{-0.14}(stat.)^{+0.60}_{-0.42}(syst.)$, which is
consistent with the theoretically calculated ratio of
1.2~\cite{Sekihara:2009yk}.
Such a branching ratio gives important information to investigate the
properties of the observed structure.

A comparison with the $\Lambda(1405)$ cross section was
performed to reveal the production mechanism related to the
$\Lambda(1405) + p \to$ ``$K^-pp$'' doorway process.
By comparing with a past measurement of the $\Lambda(1405)$
production in the $p(\pi^-, K^0)$ reaction at 1.69
GeV/$c$~\cite{Thomas:1973uh}, a production ratio $(d \sigma / d
\Omega_{Yp-decay})$ / $(d \sigma / d \Omega_{\Lambda(1405)})$ of 7
$\sim$ 8\% was obtained~\cite{IchikawaD},
which is much larger than the theoretical expectation in
the order of 0.1--1.0\% evaluated with a naive coalescence
mechanism of $\Lambda(1405) + p \to$ ``$K^-pp$''~\cite{Yamazaki:2007cs}.

These discrepancies between the experimental and theoretical binding energies, widths, and production ratios
could be
explained not by the {\it usually} expected $\bar K NN$ bound
state with the quantum number of $(Y = 1, I = 1/2, J^P = 0^-)$, but by a $\pi
\Sigma N - \pi \Lambda N$ bound state with $(Y = 1, I = 3/2, J^P =
2^+)$~\cite{Garcilazo:2012rh}.
However, further measurement
with high statistics and a larger acceptance detector system is required to identify the origin of the structure.
One possible detector system to conduct further $(\pi^+,
K^+)$ experiment is the Hyperon Spectrometer.
A successor experiment to E27 with the new spectrometer system is now
under discussion.

The other experiment, J-PARC E15, measured the $K^- +$ $^3$He reaction at 1.0
GeV/$c$.
In the reaction, ``$K^-pp$'' can be produced via the $(K^-,n)$
reaction; a recoiled virtual kaon (`$\bar K$') generated by $K^-N
\to$ `$\bar K$' $n$ processes can be directly induced into 
two residual nucleons within the strong interaction range.
The momentum of the `$\bar K$' is described as the momentum difference
of the incident kaon and the outgoing neutron, $q = |\bm{p^{lab}_{K^-}} -
\bm{n^{lab}_n}|$.
When the backscattered `$\bar K$' is realized, $q$ is as small
as $\sim$ 200 MeV/$c$, which makes the probability for the formation of ``$K^-pp$'' large.

\begin{figure}[htbp]
 \begin{center}
 \begin{tabular}{c}
  \begin{minipage}{0.475\linewidth}
  \includegraphics[width=8cm]{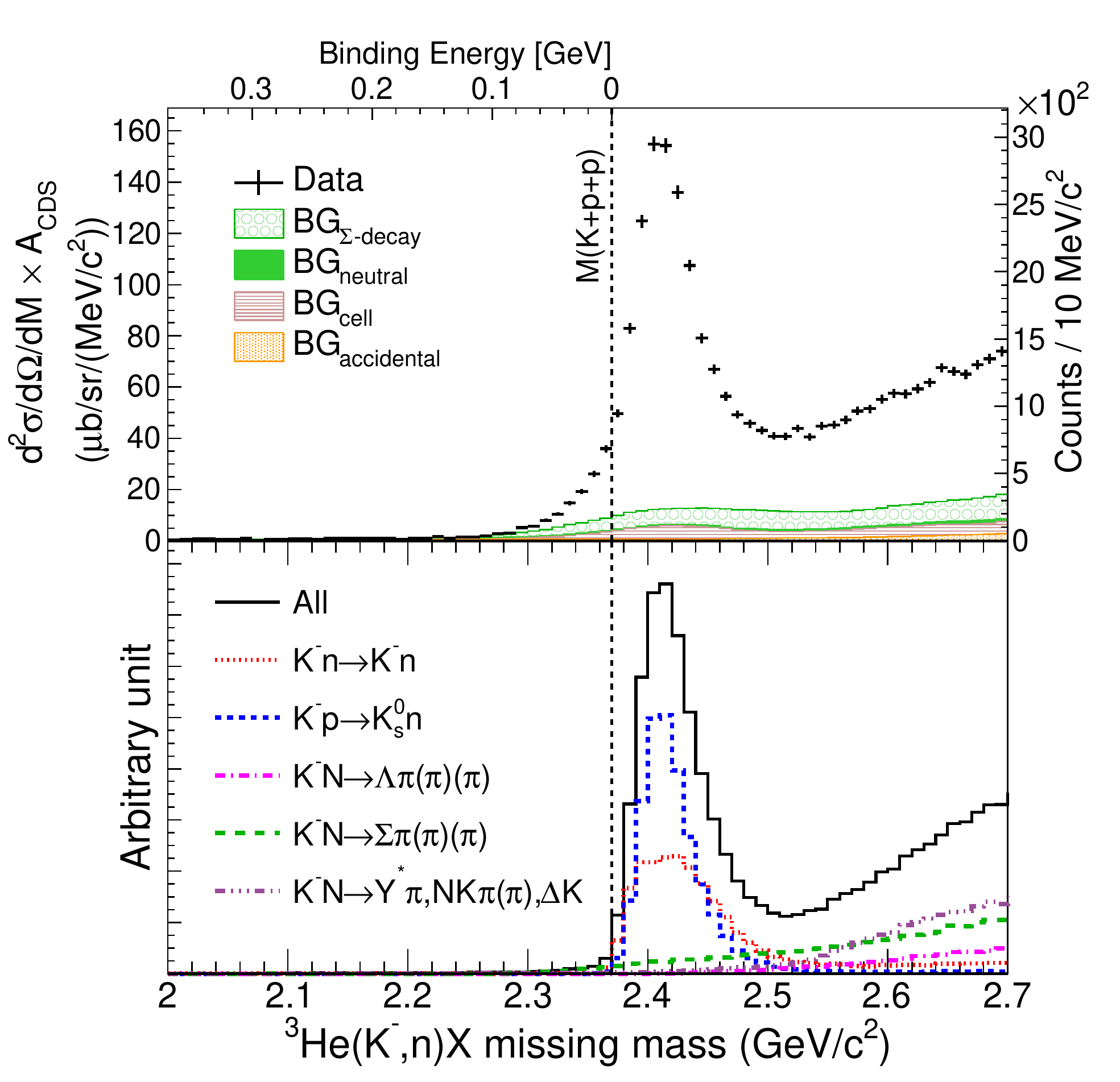}
  \caption{
  $^3$He$(K^-, n)X$ semi-inclusive missing-mass spectra obtained in the
   E15-1st experiment (top) and by simulation (bottom).
   The $K^- + p + p$ binding threshold (2.37 GeV/$c^2$) is shown as a
   dotted line.
   The figure is taken from Ref.~\cite{Hashimoto:2014cri}.
  }
  \label{fig:E15_1}
  \end{minipage}
  \hspace{5mm}
  \begin{minipage}{0.475\linewidth}
  \includegraphics[width=8cm]{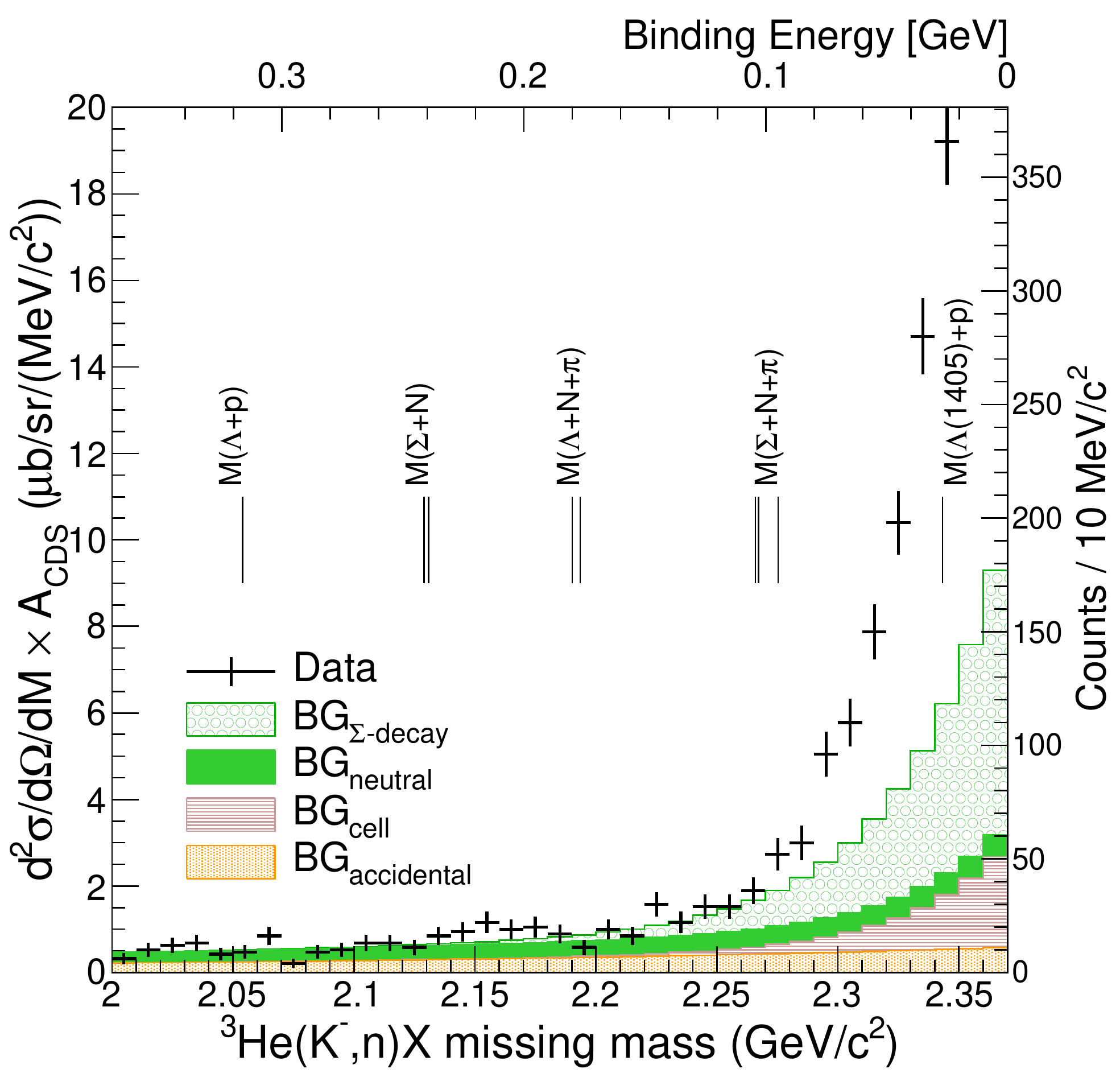}
  \caption{
  Close-up view of the $^3$He$(K^-, n)X$ semi-inclusive missing-mass 
  spectrum on the ``$K^-pp$'' bound region in
   Fig.~\ref{fig:E15_1} (top).
   The figure is taken from Ref.~\cite{Hashimoto:2014cri}.
  }
  \label{fig:E15_2}
\end{minipage}
\end{tabular}
 \end{center}
\end{figure}

The E15 experiment was performed at the K1.8BR beamline.
The first (E15-1st) and second (E15-2nd) physics runs were performed in
2013 with $5.3\times10^9$ kaons~\cite{Hashimoto:2014cri,Sada:2016nkb} and in 2015 with $32.5\times10^9$
kaons~\cite{Ajimura:2018iyx} on a liquid $^3$He target, respectively.
In the semi-inclusive $^3$He$(K^-, n)X$ analysis, a forward-going
neutron produced from the $^3$He$(K^-,n)$ reaction was detected by the
forward NC array at $\theta^{lab}_n = 0^{\circ}$, where at
least one charged track in the CDS was required to reconstruct the
reaction vertex.
Figure~\ref{fig:E15_1} shows the obtained $^3$He$(K^-,n)X$ missing-mass
spectrum and a simulated spectrum~\cite{Hashimoto:2014cri}.
The quasi-elastic scattering ($K^- n \to K^- n$) and the
charge-exchange reaction ($K^- p \to K^0 n$) are clearly evident,
of which the cross sections were evaluated to be $\sim$ 6 and $\sim$ 11 mb/sr,
respectively.

In the $(K^-,n)$ spectrum, a definitive excess of the yield is observed
in the bound region, whereas the global spectrum in the unbound region
is well reproduced by elementary reactions.
Figure~\ref{fig:E15_2} shows a close-up view of the region below the
$K^- + p + p$ mass threshold.
The excess that reaches to $\sim$ 100 MeV below the
$K^-+p+p$ mass threshold cannot be explained by any simulation result
with the elementary reactions.
Therefore, the excess will be attributed not only to the
attractive $\bar K N$ interaction, but also to two-nucleon absorption
processes such as $K^- N \to$ `$\bar K$' $N$ followed by `$\bar K$' $+
NN \to Y^* N$ and/or `$\bar K$' $+ NN \to$ ``$K^-pp$'' reactions.
However, no significant peak structure was found in the deeply bound region around the binding energy of $\sim$ 100
MeV/$c^2$, where the
FINUDA, DISTO and E27 collaborations reported the peak structure in
different production reactions.
The upper limits of the deeply bound ``$K^-pp$'' state obtained are
30 -- 270 $\mu$b/sr for natural widths of 20 -- 100 MeV at a 95\% confidence
level~\cite{Hashimoto:2014cri}.
These values correspond to (0.5 -- 5)\% and (0.3 -- 3)\% of the quasi-elastic
and the charge-exchange reaction cross sections, respectively.

To kinematically discriminate the ``$K^-pp$'' signal from the
two-nucleon absorption processes, an exclusive $^3$He$(K^-,\Lambda p)n$
analysis was conducted.
Two protons and one negative pion were detected with the CDS, and a
missing neutron was identified by the missing-mass of
$^3$He$(K^-,pp\pi^-)X$ as $X=n$.
In the E15-1st experiment, a peak structure was identified below the mass
threshold in the $\Lambda p$ invariant mass spectrum, which is
concentrated in the low momentum-transfer region
of the $(K^-,n)$ reaction, as expected~\cite{Sada:2016nkb}.

\begin{figure}[htbp]
 \begin{center}
  \includegraphics[width=10cm]{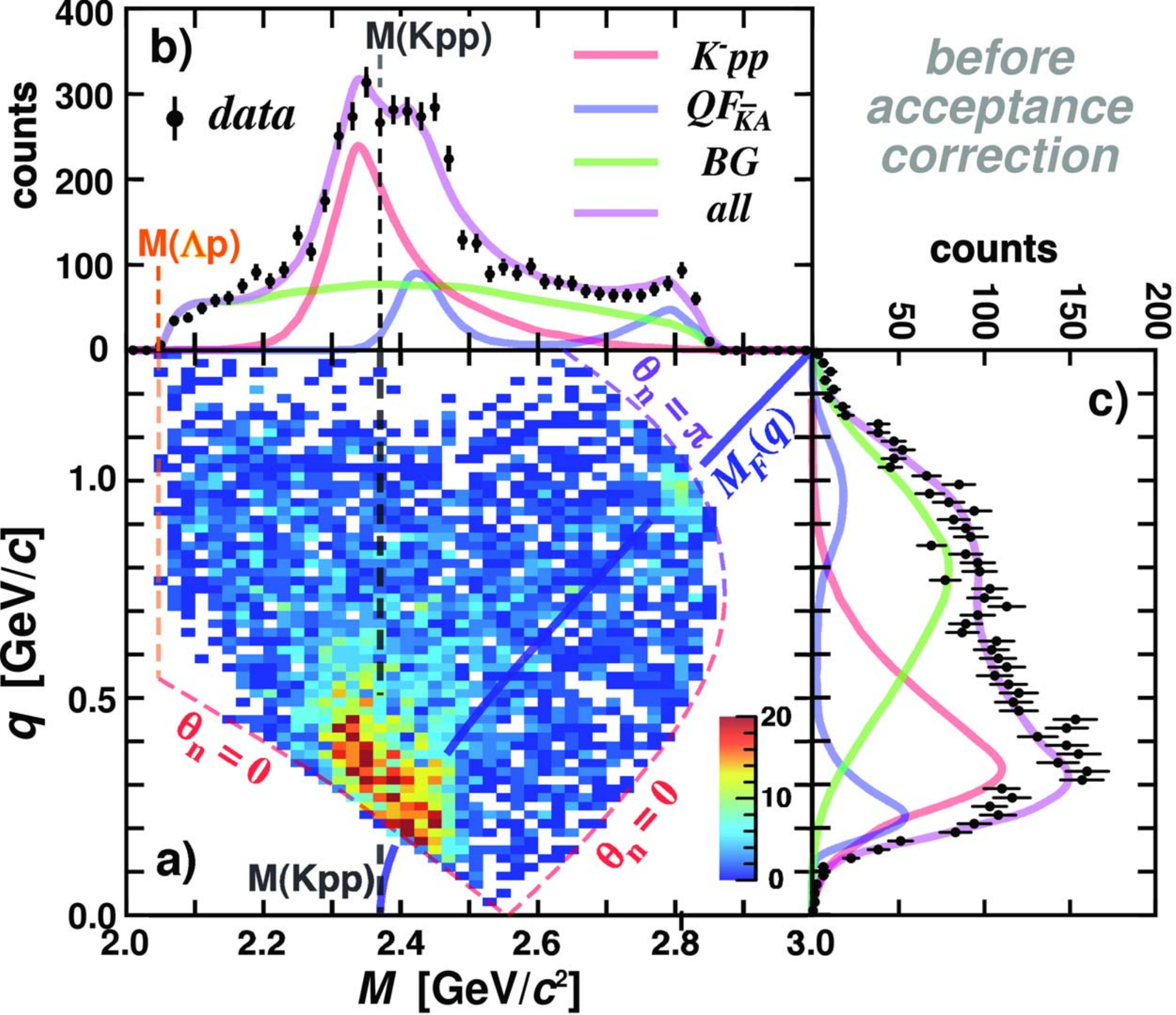}
  \caption{
  Event distribution on $M$ (= $IM(\Lambda p)$) and the momentum
  transfer $q$, for the $\Lambda pn$ final state (a).
  Histograms projected onto the $M$ axis (b) and $q$ axis (c).
  The fitting results with the simple model are also plotted as colored
  curves.
  The figure is taken from Ref.~\cite{Ajimura:2018iyx}.
  }
  \label{fig:E15_3}
 \end{center}
\end{figure}

The peak structure observed in the $\Lambda p$ spectrum was confirmed
with $\sim$ 30 times more data on the $\Lambda pn$ final
state by focusing on the $\Lambda pn$ event accumulation
in the E15-2nd experiment.
Figure~\ref{fig:E15_3} shows the event distribution of the momentum transfer
$q$, and the $\Lambda p$ invariant mass $IM(\Lambda p)$~\cite{Ajimura:2018iyx}, where the
reaction of $K^- +$ $^3$He $\to$ ``$K^-pp$'' $+n \to \Lambda + p + n$
can be uniquely described by these two parameters, $q$ and $IM(\Lambda
p)$.
Figure~\ref{fig:E15_3}(a) shows a strong event concentration near the
mass threshold of $K^-+p+p$ in the lower $q$ region, as
previously reported in Ref.~\cite{Sada:2016nkb}.
The structure near the threshold is composed of two
structures that cannot be represented as a single Breit-Wigner
function as assumed in the previous analysis.
The centroid of the structure just below the mass threshold does
not depend on $q$ within the statistical uncertainty.
This behavior is strong evidence of the existence of a bound state.
On the other hand, the distribution centroid above the mass threshold
is dependent on $q$, {\it i.e.}, the centroid shifts to heavier mass for 
larger $q$.
A natural interpretation of the structure above the threshold is
non-resonant absorption of the backward quasi-free `$\bar K$'
by the $NN$ spectator, where the `$\bar K$'
propagates as an on-shell particle; the $\Lambda p$ final state is
generated by the `$\bar K$' $+ NN \to \Lambda p$ conversion due to the
final state interaction.
The structure below the threshold cannot be reproduced by $Y^{(*)}$
productions in two nucleon absorption process followed by $\Lambda p$
conversion, which also have the $q$-dependence: $K^-$ $^3$He $\to
Y^{(*)} N N_{R} \to \Lambda p n$, where $N_R$ denotes a residual nucleon.

\begin{figure}[htbp]
 \begin{center}
  \includegraphics[width=8cm]{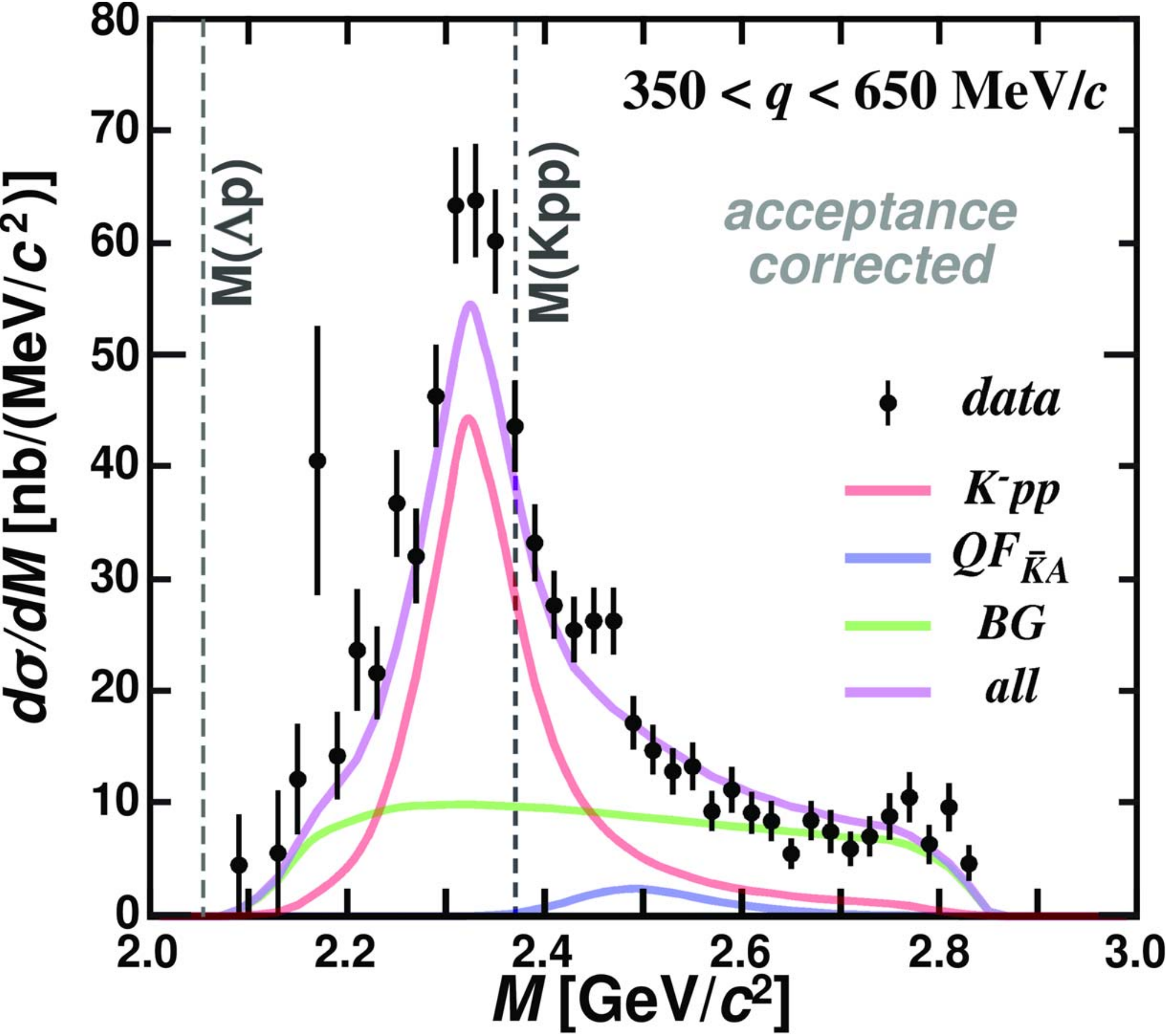}
  \caption{
  Efficiency and acceptance corrected $\Lambda p$ invariant mass in the
  region of $0.35 < q < 0.65$ GeV/$c$.
  The figure is taken from Ref.~\cite{Ajimura:2018iyx}.
  }
  \label{fig:E15_4}
 \end{center}
\end{figure}

The simplest model fitting that takes into account the bound state, the
quasi-free process, and a broad background reproduces the event
distribution, and the results are shown as
colored curves in Figs.~\ref{fig:E15_3}(b) and (c).
The details of the fitting are given in Ref.~\cite{Ajimura:2018iyx}.
Figure~\ref{fig:E15_4} shows the $\Lambda p$ invariant mass
spectrum corrected by the detector acceptance and the experimental
efficiency in the momentum transfer window of $350 < q < 650$ MeV/$c$,
where the signal from the bound state and the quasi-free contribution
are clearly separated; the interference between the bound state formation
and the quasi-free process will occur near the kinematical boundary.
The yields of other processes are largely suppressed in
contrast to the bound state, and the quasi-free distribution is also
clearly separated from the peak of the bound state.
The fitting result provides the Breit-Wigner pole position at $M = 2324
\pm3 (stat.) ^{+6}_{-3} (syst.)$ MeV/$c^2$, ({\it i.e.,} a binding energy of
$B = 47\pm3 (stat.) ^{+6}_{-3} (syst.)$ MeV/$c^2$) with a width of
$\Gamma = 115\pm7 (stat.) ^{+10}_{-20} (syst.)$ MeV/$c^2$, and the
$S$-wave Gaussian form factor parameter $Q = 381\pm14 (stat.) ^{+57}_{0}
(syst.)$ MeV/$c^2$~\cite{Ajimura:2018iyx}.
The $q$-integrated ``$K^-pp$'' formation yield going to the $\Lambda
p$ decay channel evaluated is $\sigma_{Kpp} \cdot Br_{\Lambda p} =
11.8 \pm 0.4 (stat.) ^{+0.2}_{-1.7}(syst.)$ $\mu$b.

\begin{figure}[htbp]
 \begin{center}
  \includegraphics[width=8cm]{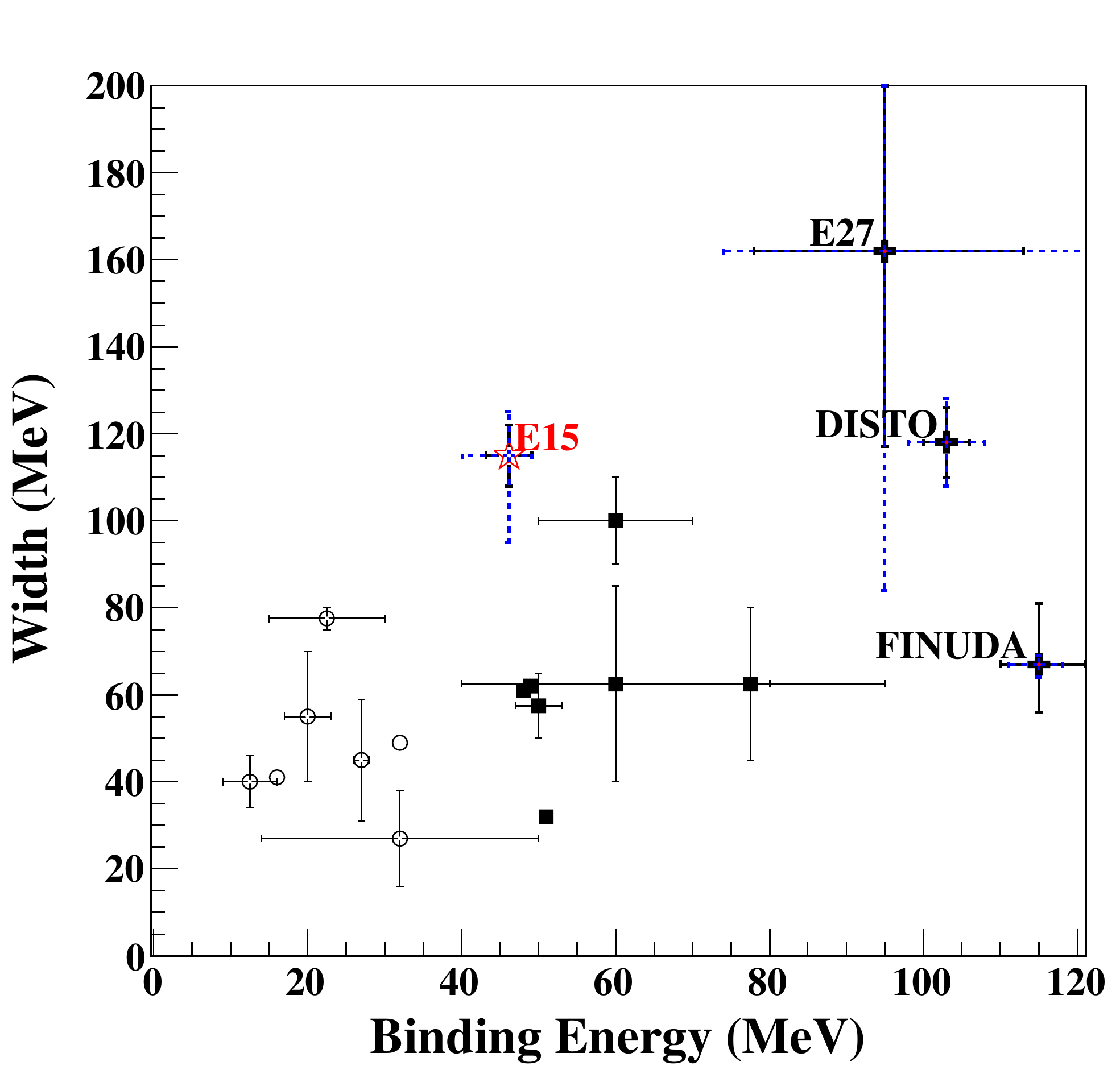}
  \caption{
  Summary of binding energies and widths for the ``$K^-pp$'' bound state.
  The open star shows the values obtained in the E15
  experiment~\cite{Ajimura:2018iyx}, and
  closed crosses show those obtained in the FINUDA~\cite{Agnello:2005qj},
  DISTO~\cite{Yamazaki:2010mu}, and E27~\cite{Ichikawa:2014ydh}
  experiments.
  Closed squares and open circles denote those from theoretical
  calculations based on energy
  independent (phenomenological)~\cite{Yamazaki:2002uh,Shevchenko:2006xy,Shevchenko:2007ke,Ikeda:2007nz,Ikeda:2008ub,Wycech:2008wf,Revai:2014twa,Dote:2017veg,Ohnishi:2017uni}
  and energy
  dependent (chiral unitary)~\cite{Dote:2008in,Dote:2008hw,Ikeda:2010tk,Barnea:2012qa,Bayar:2012hn,Revai:2014twa,Ohnishi:2017uni,Dote:2017wkk}
  $\bar K N$ interaction models, respectively.
  }
  \label{fig:Kpp}
 \end{center}
\end{figure}

In Ref.~\cite{Ajimura:2018iyx}, the E15 collaboration has
concluded that ``the simplest and natural interpretation is a kaonic
nuclear bound state ``$K^-pp$''''.
The binding energy and width obtained in the experiments are
summarized in Fig.~\ref{fig:Kpp} together with those taken from the
theoretical calculations of the ``$K^-pp$''.
The binding energy of $\sim$ 50 MeV obtained in the E15 experiment is
deeper than those obtained by the chiral motivated calculations, while
rather consistent with those obtained by the phenomenological based
calculations.
The width of $\sim$ 100 MeV is wide, meaning very absorptive.
It should be similar to that of $\Lambda(1405) \to \pi \Sigma$ ($\Gamma
\sim$ 50 MeV) if the ``$K^-pp$'' decays like `$\Lambda(1405)$' $+$
`$p$' $\to \pi \Sigma p$, as theoretical calculations taken into
account only mesonic decay channels show. 
Thus the observed large width indicates that the non-mesonic
$YN$ channels would be the major decay mode of the ``$K^-pp$''.
The observed values of the large form factor of $\sim$ 400 MeV/$c$ and the
large binding energy of $\sim$ 50 MeV imply the formation of a
compact system, although the static form-factor parameter of the ``$K^-pp$''
should be studied more quantitatively.
The structure below the mass threshold in the $\Lambda p$ spectrum
obtained has also been 
theoretically interpreted as the $\bar K NN$ quasi-bound
system~\cite{Sekihara:2016vyd,Sekihara:QNP2018}, in
which the experimental spectrum can be reproduced with the $\bar
K NN$ quasi-bound system and the quasi-free processes based on 
theoretical treatment of the $^3$He$(K^-,\Lambda p)n$ rection.

To confirm that the observed structure is the theoretically predicted
$\bar K NN$ quasi-bound state, the question whether or not the
``$K^-p$'' bound state is also produced in the same $K^- +$ $^3$He
reaction must be answered.
The $\Lambda(1405)$ state is theoretically considered as a quasi-bound
state of $\bar K N$ in the $I = 0$ channel, as discussed in Sec~\ref{subsubsec:E31}.
Therefore, not only the ``$K^-pp$'' bound state should be observed, but also the
$\Lambda(1405)$ state in the same $K^- +$ $^3$He reaction.
Furthermore, from a theoretical perspective, the $\bar K NN$ is a
resonant state in the $\bar K NN$-$\pi \Sigma N$-$\pi \Lambda N$
coupled-channel system, and the mesonic $\pi \Sigma N$ decays are
expected to be dominant compared to the non-mesonic $Y N$
decays~\cite{Ohnishi:2013rix}.
Measurements of the mesonic $\pi\Sigma N$ channels originated from the
``$K^-pp$'' bound state and the $\Lambda(1405) p$ final state are therefore
necessary to obtain further information on the kaonic nuclei.

For this purpose, analysis on the $K^-$ $^3$He $\to
\pi^{\pm}\Sigma^{\mp}pn$ reactions has been in progress.
$\pi^{\pm}\Sigma^{\mp}p \to \pi^+\pi^-np$ decays were detected with the
CDS, where {\it the neutron from the $\Sigma^{\pm}$ decays} was identified
using the CDH by requiring no charged track of the CDC in front of a CDH
hit and no hit in both neighboring sides of the CDH segment.
One neutron in the reaction was identified with the missing-mass method
of the $^3$He$(K^-, \pi^+\pi^-np)X$ with $X = n$.
With the E31 data set, {\it i.e.}, the $K^- + d$ reactions, a similar
analysis on $\Sigma^{\pm}$ detection using the CDS has been
conducted to measure the momentum transfer dependence of the
$\Lambda(1405)$ cross section and to discuss the reaction form factor.
In both analyses, the preliminary results clearly show the $\Lambda(1405)$
signal in the $\pi^{\pm}\Sigma^{\mp}$ invariant-mass spectra (see
Ref.~\cite{sakuma} for instance).

The system size dependence provides effective information to obtain further understanding of the kaonic nuclei.
As summarized in Table~\ref{tab:kaonic}, there are several theoretical
calculations of the mass number ($A$) dependence of the light kaonic
nuclei with the different $\bar K N$ interaction models.
In the calculations, the values are widely spread but all
the results show larger binding energies with larger $A$.
Therefore, a systematic study of the mass number dependence of the kaonic
nuclei will provide significant information on the $\bar K N$ interaction
below the mass threshold, and will be a good touchstone to establish the
few-body calculations including the three-body problem.
Form factor and decay-branching ratio measurements will also shed light
on the internal structure of the kaonic nuclei, which is predicted to be
compact and thus to form high density matter due to the strong
attraction of the $\bar K N$ interaction.
A few-body calculation recently
indicated that the $\bar K NNNN$ ($A=4$) system has the largest
central density of approximately 0.7~fm$^{-3}$ in light kaonic nuclei up to
the $\bar K NNNNNN$ ($A=6$) systems~\cite{Ohnishi:2017uni}.
The spin and parity measurements are also crucial to understand
whether or not the observed ``kaonic nuclei'' are the theoretically
predicted objects.
To these ends, a new experiment with a new large $4\pi$ spectrometer
equipped with neutral-particle detectors has been proposed to perform 
systematic measurements of light kaonic nuclei, from the most
fundamental ``$\bar K N$'' state ({\it i.e.},  the $\Lambda(1405)$ state) to
the ``$\bar K NNNN$'' state via the $(K^-,N)$ reactions~\cite{knucl_LOI}.

\begin{table}[htbp]
 \caption{Summary of theoretical calculations on light kaonic
 nuclear states.
 }
 \begin{center}
  \begin{tabular}{l cc cc cc cc} \hline
   &\multicolumn{2}{c}{Ref.~\cite{Akaishi:2002bg,Yamazaki:2002uh}}
   &\multicolumn{2}{c}{Ref.~\cite{Wycech:2008wf}}
   &\multicolumn{2}{c}{Ref.~\cite{Barnea:2012qa}}
   &\multicolumn{2}{c}{Ref.~\cite{Ohnishi:2017uni}}
   \\
   & $B$(MeV) & $\Gamma$(MeV) & $B$(MeV) & $\Gamma$(MeV)
   & $B$(MeV) & $\Gamma$(MeV) & $B$(MeV) & $\Gamma$(MeV)
   \\ \hline
   $K^-pp$ & 48 & 61 & \multirow{2}{*}{40--80} & \multirow{2}{*}{40--80} & \multirow{2}{*}{16} & \multirow{2}{*}{41} & 26--49 & 31--62
   \\
   $K^-pn$ & - & - &  &  &  &  & 25--48 & 32--62
   \\ \hline
   $K^-ppp$ & 97 & $\sim$24 & \multirow{3}{*}{90--150} & \multirow{3}{*}{20--30} & \multirow{3}{*}{19--29} & \multirow{3}{*}{31--33} & - & -
   \\
   $K^-ppn$ & 108 & 20 &  &  &  &  & - & -
   \\
   $K^-pnn$ & - & - &  &  &  &  & 45--73 & 26--79
   \\ \hline
   $K^-pppn$ & $\sim$105 & $\sim$26 & \multirow{3}{*}{120--220} & \multirow{3}{*}{10--30} & \multirow{3}{*}{-} & \multirow{3}{*}{-} & - & -
   \\
   $K^-ppnn$ & 86 & 34 &  &  &  &  & 68--85 & 28--87
   \\
   $K^-pnnn$ & - & - &  &  &  &  & 70--87 & 28--87
   \\
   \hline
  \end{tabular}
 \end{center}
 \label{tab:kaonic}
\end{table}

\subsection{Baryon-Baryon Interaction \label{subsec:BB_int}}
It is important to understand the interactions between baryons in a unified way, 
{\it i.e.}, $B_{8}B_{8}$ interaction under SU(3)$_{\rm f}$.
The $B_8B_8$ ineteractions are decomposed to the following
six multiplets by SU(3)$_{\rm f}$:
\begin{equation}
{\bf 8} \otimes {\bf 8} = {\bf 27} \oplus {\bf 8_s} \oplus {\bf 1}
\oplus \overline{{\bf 10}}
                 \oplus {\bf 10} \oplus {\bf 8_a}, \label{su3f-decomp}
\end{equation}
where ${\bf 27}$, ${\bf 8_s}$, and ${\bf 1}$ are symmetric under the exchange of
two baryons, and $\overline{\bf 10}$, ${\bf 10}$, and ${\bf 8_a}$ 
are asymmetric under the exchange of two baryons or more (basically two flavors of quarks).
Among these, ${\bf 27}$ and $\overline{\bf 10}$ include $NN$ interactions
and are well known, at least phenomenologically, from many scattering experiments.
Other multiplets are 
hyperon-nucleon ($YN$) and $YY$ interactions with newly introduced strangeness added.

According to theoretical $B_{8}B_{8}$ interaction models based on
the quark-gluon picture, the Pauli effect on quarks (quark Pauli effect) as well as
color-magnetic interactions between quarks are important to produce
a repulsive core~\cite{OkaYazaki1984QandN, Yazaki:1988hk, Shimizu:1992sf}.
In these models, the quark Pauli effect is strongly shown in ${\bf 8_s}$ 
(termed the Pauli forbidden state) and $\bf 10$ (almost Pauli forbidden state).
For example in 
${\bf 10}$, the $\Sigma^+ p$ channel, where four $u$ quarks exist, is expected to have a large repulsive core. 
In the SU(3)$_{\rm f}$ singlet (${\bf 1}$), which is a part of 
$\Lambda\Lambda$, $\Xi N$, and $\Sigma\Sigma$ interactions,
an attractive core rather than a repulsive core is expected to exist because the quark Pauli effect
does not act and color-magnetic interaction is attractive only in this channel.
As predicted by R.L.~Jaffe~\cite{Jaffe:1976yi}, the $H$-dibaryon, which consists of $uuddss$ quarks, 
is expected in this channel, although it has yet to be experimentally observed.
Thus, observation of the $H$-dibaryon would be direct evidence of the attractive core
and confirmation of a quark-based scenario.
Measurements of its mass and width will also provide important information on
the interaction.

It is difficult to realize hyperon beams and to conduct
hyperon scattering experiments at low energies
due to its production rate and short lifetime.
The scattering data obtained are very limited;
bubble chamber data in the 1970s~\cite{Dosch:1966, Engelmann:1966, Rubin:1967,
Alexander:1969cx, SechiZorn:1969hk, Charlton:1970bv, Kadyk:1971tc,  Eisele:1971mk}
and data using scintillating fiber or 
scintillating image detectors at KEK 12-GeV PS in 
the 1990s~\cite{Ahn:1997wa, Kondo:2000hn, Ahn:2005gb, Kadowaki:2002wf}.
Therefore, $YN$ and $YY$ interactions have been
understood by the construction of interaction models
based on the limited $YN$ scattering data
with the guide of SU(3)$_{\rm f}$ symmetry
and reference to the rich data on $NN$ scattering.
On the other hand, the structure of hypernuclei such as level energies
has been calculated using the constructed interactions
and compared with the experimental hypernuclear
spectroscopic data. 
Interaction models have been subsequently refined so as 
to reproduce experimental data on hypernuclei.
Therefore, the experimental studies on hypernuclei 
together with the development of accurate many-body calculations 
are essential to achieve an understanding of the $B_{8}B_{8}$ interaction.

The nuclear many-body effect is another subject of
hypernuclear physics at J-PARC.
In hypernuclei, large mixing effects due to
$\Lambda N$-$\Sigma N$ or $\Lambda\Lambda$-$\Xi N$ coupling 
are expected because the mass differences are small:
80 and 28 MeV for $\Lambda$-$\Sigma$ and
$\Lambda\Lambda$-$\Xi N$, respectively.
These mixing effects affect the energies of levels of hypernuclei 
and change the effective interaction
in a nucleus from that of a bare interaction.
Small changes of structures and interactions are interesting
from a nuclear physics perspective of the many-body problem.
On the other hand, understanding these changes is also very important to obtain
the bare $B_{8}B_{8}$ interaction from hypernuclear data.

In-medium hadron properties are of interest with respect to
the spontaneous breaking of chiral symmetry.
The hypernucleus provides one such platform for investigation.
A hyperon is a different particle from a nucleon; therefore,
the hyperon in the hypernucleus can be placed deep inside a nucleus
and be distinguishable from other nucleons in the hypernucleus.
If the hyperon properties in a hypernucleus can be measured
and compared with those in free space, then 
a possible change of the baryon properties in a nuclear medium can be investigated.

Detailed investigations on $\Lambda$ hypernuclei
with $S = -1$ have become possible with high intensity beams, especially that of $K^-$ at the HEF.
In $\gamma$-ray spectroscopy experiments,
hypernuclear $\gamma$-rays are measured
by Ge-detectors in coincidence with the production of hypernuclei 
identified via the $(K^-,\pi^-)$ reaction by magnetic spectrometers.
From such detailed investigations of $\Lambda$ hypernuclei,
information 
not only on $\Lambda N$ interactions including spin-dependence,
but also on the nuclear many-body effects could be obtained.
Full-scale investigations on the $S = -2$ systems, 
{\it i.e.}, double-$\Lambda$ hypernuclei, $\Xi$ hypernuclei, and 
$S = -2$ exotic systems have begun.
Studies on those systems produced and identified
by the $(K^-,K^+)$ reaction will provide important information
on $\Lambda\Lambda$,
$\Xi N$, and $\Xi N \rightarrow \Lambda\Lambda$ interactions,
and $H$-dibaryons in SU(3)$_{\rm f}$ singlet channel.

Recent developments of experimental techniques, such as
fine segmented detectors with high-rate abilities and
high-speed data acquisition (DAQ) systems, enable direct
hyperon scattering experiments to be performed.
Scattering events are identified kinematically 
without treatment of the image data as with a bubble chamber; therefore, 
high statistics data can be obtained in such modern experiments.
At J-PARC, the $\Sigma^{\pm}$-proton scattering experiment is currently in progress.

\subsubsection{Neutron-rich $\Lambda$ hypernuclei \label{subsubsec:E10}}

Neutron-rich $\Lambda$ hypernuclei are one of main research objects
of the $S = -1$ hypernuclear system at J-PARC.
For the $\Lambda$ hypernuclei, the structure change of the core nucleus
was already observed in the $^7_\Lambda$Li hypernucleus~\cite{Tanida:2000zs}.
Such a change results from the attractive $\Lambda N$ interaction
(glue-like role of $\Lambda$).
In addition to the glue-like role of $\Lambda$, the effect of the 
$\Lambda$-$\Sigma$ mixing
may be enhanced in neutron-rich $\Lambda$ hypernuclei.
Although $\Lambda$ and $\Sigma$ are not mixed in free space, 
$\Sigma$ may appear in nuclei as an intermediate state of
$\Lambda N$, $\Lambda NN$, etc., through the $\Sigma N$-$\Lambda N$
coupling.
However  $\Sigma$-mixing in the $\Lambda$ hypernuclear
state with zero isospin ($N=Z$) core nucleus is forbidden without
exciting the core nucleus, and thus the mixing is largely suppressed .
Conversely, in the case of a neutron-rich $\Lambda$ hypernucleus,
the mixing effect is expected to be enhanced because of  the large isospin value of
the core nucleus. 
Potential information in the neutron-rich environment
is closely related to an equation of states (EOS) for high-density neutron matter,
which could provide  a deeper understanding of the inside of neutron stars.
  
Neutron-rich $\Lambda$ hypernuclei can be produced by double
charge exchange (DCX) reactions such as the $(\pi^-, K^+)$ and
$(K^-, \pi^+)$. 
In these reactions, two protons are converted to
a $\Lambda$ and a neutron.
The $(e, e'K^+)$ reaction, in which a proton is 
converted to a $\Lambda$,  can also produce neutron-rich $\Lambda$ hypernuclei.
However the DCX reactions can produce $\Lambda$ hypernuclei
more far from $N=Z$ nuclei than the  $(e, e'K^+)$ reaction.
Reaction mechanism is also in discussion for the $\Lambda$ 
hypernuclear production via the DCX reaction.
Two different mechanisms have been proposed.
One is a two-step process that consists of two sequential reactions of
the $\Lambda$ production  and the charge exchange reaction;
$\pi^- + (pp)$ $\rightarrow$  $\pi^0 + (pn)$ $\rightarrow$ $K^+ + (\Lambda n)$
or
$\pi^- + (pp)$ $\rightarrow$ $K^0 + (\Lambda p)$ $\rightarrow$
$K^+ + (\Lambda n)$.
The other is a single-step process in which a $\Sigma^-$ admixture
in the $\Lambda$ hypernuclear state appears due to
$\Sigma^- p$-$\Lambda n$ coupling (the so-called $\Sigma$-doorway reaction).
In the former case, the production cross section is expected to peak at
1.05 GeV/{\it c} based on the  $\Lambda$ production process.
While it is expected to show the similar dependence with the $\Sigma$ production 
in the latter case.

 In the previous experiment at KEK (KEK-PS E521), $^{10}_{\,\,\Lambda}$Li was attempted to produce 
using the $^{10}$B$(\pi^-, K^+)$ reaction~\cite{Saha:2005zx}. 
Clear signal events were observed in the bound region, although its production
cross section was obtained to be very small of 11.3$\pm$0.3 nb/sr 
at 1.2 GeV/{\it c}.
This value is three-orders of magnitude smaller than  
that by the $(\pi^+, K^+)$ reaction ($\sim$ 10 $\mu$b/sr).
A comparison of the production cross sections at 1.05 and 1.2 GeV/{\it c},
suggested that the single-step process is favored which
is consistent with theoretical calculations~\cite{Harada:2008zs}.

\begin{figure}[htbp]
\centerline{\includegraphics[width=0.65\textwidth]{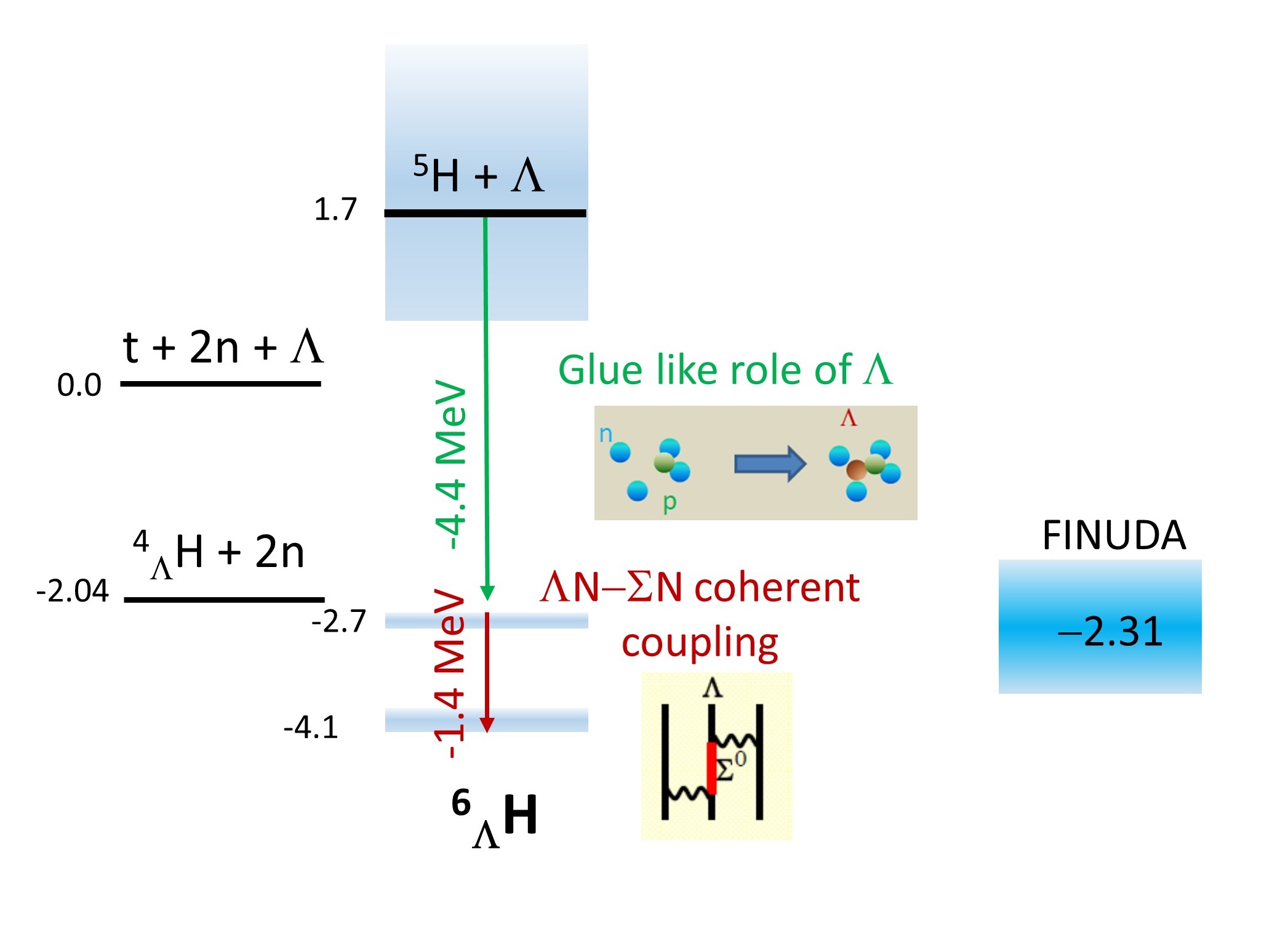}}
\caption{Expected level energy of $^6_\Lambda$H. 
Unbound $^5$H ground state becomes bound by adding a $\Lambda$ hyperon due
to glue-like role of $\Lambda$. Additional binding as a result of so-called ``coherent 
$\Sigma N$-$\Lambda N$ coupling'' is suggested 
by Khin Swe Myint and Y.~Akaishi~\cite{Myint:2003qf}.
Result of the FINUDA experiment~\cite{Agnello:2011xr} is also drawn.
}
\label{6LH-level}
\end{figure}

The J-PARC E10 experiment was planned to measure $^6_\Lambda$H and $^9_\Lambda$He
hypernuclei using $^6$Li and $^9$Be targets, respectively, 
at an incident $\pi^-$ momentum of 1.2 GeV/{\it c}.
The  $^6_\Lambda$H run was carried out in 2012 and 2013 
at the K1.8 beamline~\cite{Sugimura:2013zwg, Honda:2017hjf}.
Khin Swe Myint and  Y. Akaishi theoretically discussed the structure of
 $^6_\Lambda$H~\cite{Myint:2003qf}. 
As shown in Fig.~\ref{6LH-level}, unbound $^5$H ground state
reported as a broad resonance state of 1.7$\pm$0.3 MeV~\cite{Korsheninnikov:2001buf}
becomes to be bound by adding a $\Lambda$ due to the glue-like role of $\Lambda$.
Additional binding could occur by taking the effect of so-called 
``coherent $\Sigma N$-$\Lambda N$ coupling'',  leading to $B_\Lambda$=5.8 MeV
with respect to the $^5$H + $\Lambda$ system.
A rather small $B_\Lambda$=3.83$\pm$0.08$\pm$0.22 MeV was predicted by
A.~Gal and D.~J.~Millener based on shell-model calculations~\cite{Gal:2013ooa}.
On the other hand, E.~Hiyama {\it et al.} noted that the value of 
$B_\Lambda$ is significantly affected by the spatial size of the core nucleus, $^5$H; within the framework of the four-body cluster model that 
reproduce the energy and width of $^5$H, the ground state of $^6_\Lambda$H is obtained as a resonant state (0.87 MeV unbound above 
the $^4_\Lambda$H + 2$n$ threshold)~\cite{Hiyama:2013uqa}.
Therefore experimental confirmation of the existence
of $^6_\Lambda$H and determination of its energy is important.
Experimentally, the FINUDA group reported that 3 candidates of 
 $^6_\Lambda$H had been observed in coincidence of the $\pi^+$
from the $^6$Li$(K^-_{\rm stopped}, \pi^+)^6_\Lambda$H production
with the  $\pi^-$ from $^6_\Lambda$H $\rightarrow$ $^6$He + $\pi^-$ 
decay~\cite{Agnello:2011xr}. 
The obtained averaged mass of 5801.43 MeV 
($-2.31$ MeV with respect to the $t$+2$n$+$\Lambda$) 
indicates that the coherent $\Sigma N$-$\Lambda N$ coupling effect is small.

\begin{figure}[htbp]
\centerline{\includegraphics[width=0.65\textwidth]{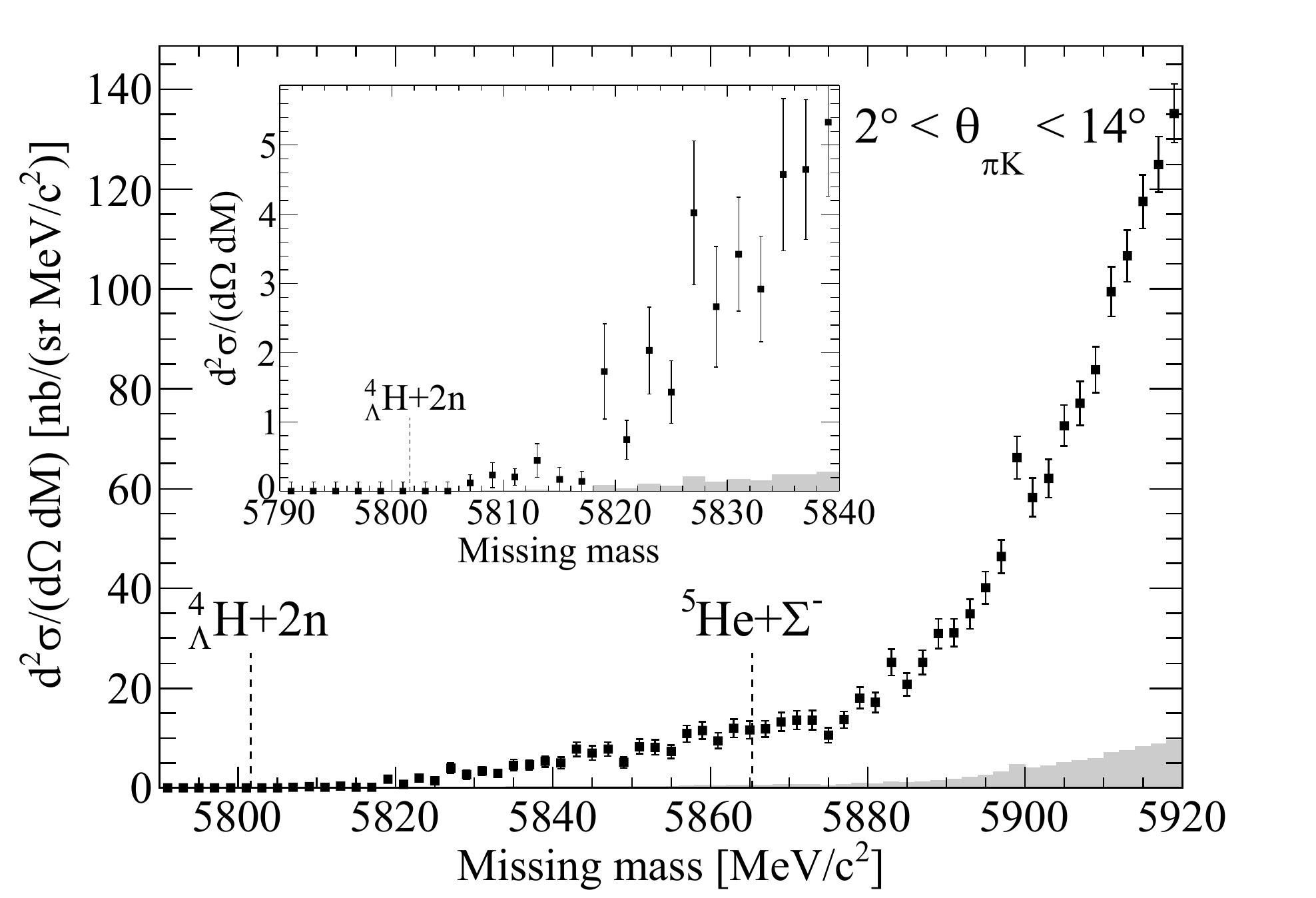}}
\caption{Missing-mass spectrum for the $^6$Li$(\pi^-, K^+)$ reaction
obtained from the improved analysis taken from Ref.~\cite{Honda:2017hjf}. 
No event was observed in the bound region or near the particle-bound threshold.
}
\label{e10-6LH-spec}
\end{figure}

In the experiment, 
a high-intensity ( 1.2--1.4 $\times$10$^7$/spill) 
$\pi^-$ beam of 1.2 GeV/{\it c} was
irradiated to the enriched $^6$Li (95.54\%) target 
with 3.5 g/cm$^2$ in thickness. 
The beam $\pi^-$'s were measured by the
K1.8 beam spectrometer, while the outgoing $K^+$'s
were identified and analyzed by the SKS 
(superconducting kaon spectrometer)~\cite{Takahashi:2012cka}.
The missing-mass resolution was estimated to be 3.2 MeV(FWHM)
from the  $^{12}$C$(\pi^+, K^+)^{12}_{\,\,\Lambda}$C spectrum
measured with a 3.6 g/cm$^2$ graphite target.
No peak was observed in the bound region of $\Lambda$ and 
an upper limit of 1.2 nb/sr (90\% CL) was reported 
in the initial paper~\cite{Sugimura:2013zwg}.
In the improved analysis described in Ref.~\cite{Honda:2017hjf},
no event was identified in the bound region
or the near threshold region as shown in Fig.~\ref{e10-6LH-spec},
and the upper limit was reduced to be 0.56 nb/sr (90\% CL).
Thus both bound and the resonance states seem not to
exist, which contradicts the FINUDA result.
Recently a new measurement on $^5$H resonance energy has been reported
to be 2.4$\pm$0.3 MeV~\cite{Wuosmaa:2017siq}, which is a larger value than
the previously reported one of 1.7$\pm$0.3 MeV~\cite{Korsheninnikov:2001buf}.
This may lead no particle bound state of $^6_\Lambda$H in a naive estimation.
Thus, the existence of $^6_\Lambda$H is still in discussion both experimentally
and theoretically.
Further studies on neutron-rich $\Lambda$ hypernuclei will be carried
out in future at the extended HEF.

\subsubsection{$\gamma$-ray spectroscopy of $\Lambda$ hypernuclei
for detailed studies on $\Lambda$ interactions \label{subsubsec:E13}}

$\gamma$-ray spectroscopy using Ge detectors
is one of  the most powerful methods to investigate fine structures
of nuclei with a few keV or better resolution,
and has revealed various many-body aspects 
of nuclei for many years.
The first successful application of this spectroscopy method to hypernuclei 
have been performed at KEK-PS in 1998~\cite{Tamura:2000ea}.
Since then, $p$-shell $\Lambda$ hypernuclei were intensively
studied by this method at KEK-PS~\cite{Tamura:2000ea, Tanida:2000zs, 
Miura:2005mh, Ma:2007zz, Hosomi:2015fma} and at
BNL-AGS~\cite{Akikawa:2002tm, Ukai:2004zz, Tamura:2004pz, Ukai:2006zp}
to obtain the spin-dependent $\Lambda N$ effective interaction.

The $\Lambda N$ effective interaction can be expressed as,
\begin{equation}
V_{\Lambda N}(r) = V_0(r) + V_\sigma(r) \mathbf{s}_N \cdot \mathbf{s}_\Lambda 
 + V_\Lambda(r)\mathbf{l}_{N\Lambda} \cdot \mathbf{s}_\Lambda 
 + V_N(r)\mathbf{l}_{N\Lambda} \cdot \mathbf{s}_N
 + V_T(r)[3(\mathbf{\sigma}_\Lambda \cdot \mathbf{\hat{r}})
 (\mathbf{\sigma}_N \cdot \mathbf{\hat{r}}) - 
 \mathbf{\sigma}_\Lambda \cdot \mathbf{\sigma}_N ],
\label{eqn:LNint}
\end{equation}
where, $\mathbf{s}_N$(=$\frac{1}{2}\mathbf{\sigma}_N$) and  
$\mathbf{s}_\Lambda$(=$\frac{1}{2}\mathbf{\sigma}_\Lambda$) are the spins 
of a nucleon and a $\Lambda$, respectively, and
$\mathbf{l}_{N\Lambda}$ and $\mathbf{r}$ are the relative
angular momentum and the coordinate between a nucleon and 
a $\Lambda$, respectively.
Level energies of low-laying states are primarily determined by
strengths of the spin-dependent terms of Eq.~\ref{eqn:LNint}.
In the case of $p$-shell $\Lambda$ hypernuclei with 
a $\Lambda$ in the $s$-orbit and valence nucleons in the $p$-orbit
in shell model picture,
these strengths are parametrized as $\Delta$ (spin-spin),
$S_\Lambda$ ($\Lambda$ spin-orbit), 
$S_N$ (nucleon spin-orbit), and $T$ (tensor).
These terms are defined as  the  radial integrals for the wave functions of a $\Lambda$ 
in the $s$-orbit ($s_\Lambda$) and a nucleon in the $p$-orbit ($p_N$).
From $\gamma$-ray data on $p$-shell $\Lambda$ hypernuclei, 
these parameters were determined to be
\[
\Delta = 0.43 (A=7)\  \mathrm{or}\  0.33 (A>10) \ \mathrm{MeV}, 
\ S_\Lambda = -0.01 \ \mathrm{MeV},
\ S_N = -0.4\  \mathrm{MeV},
\ \mathrm{and}
\ T = 0.03 \ \mathrm{MeV},
\]
although some inconsistent results were obtained for
$^{10}_{\,\,\Lambda}$B and $^{11}_{\,\,\Lambda}$B.
The ground state doublet spacing of $^{11}_{\,\,\Lambda}$B was
determined to be 261 keV~\cite{Ma:2007zz}, 
while no $\gamma$-ray  transition in $^{10}_{\,\,\Lambda}$B was observed,
thus, suggesting the ground state doublet spacing of $^{10}_{\,\,\Lambda}$B 
is less than 100 keV~\cite{Tamura:2004pz}.
This contradiction is thought to be due to 
the $\Sigma N$-$\Lambda N$ coupling~\cite{Tamura:2004pz, Millener:2004ky}.

One of main goals of $\gamma$-ray spectroscopy of $\Lambda$ hypernuclei
at J-PARC is to extend prior studies on the effective $\Lambda N$ spin-dependent 
interaction to cases other than $p$-shell hypernuclei. 
In these analysis, the strength parameters are radial integrals associated with 
the wave functions of $s_\Lambda$ and the nucleons in the outermost orbit. 
Thus a comparison of these parameters with those for  hypernuclei of different sizes 
provides  information on the radial dependence of each interaction term.
This is also a unique test of our understanding of 
the $\Lambda N$ interaction and hypernuclear structure.
The $\Sigma N$-$\Lambda N$ coupling and the $\Lambda NN$ three-body
force are also important.
The  former gives rise to a mixture of 
$\Lambda$ hypernuclear and $\Sigma$ hypernuclear states and significantly
affects the structure of hypernuclei. 
This effect is particularly important in terms of understanding the $YN$ interaction 
in high-density matter. 
The two-body  $\Sigma N$-$\Lambda N$  interaction can be incorporated into
a two-body effective $\Lambda N$ interaction, while the three-body
$\Lambda NN$ force due to  $\Sigma N$-$\Lambda N$ coupling can not.
Therefore it may not be possible to understand hypernuclear level energies 
based solely on the two-body effective interaction in Eq.~\ref{eqn:LNint},
as suggested by the contradictory results on $^{11}_{\,\,\Lambda}$B
and  $^{10}_{\,\,\Lambda}$B.
Thus investigations concerning the $\Lambda NN$ three-body force are important
not only to examine the three-body force itself but also to determine the 
strengths of  the reliable two-body $\Lambda N$ spin-dependent interactions.

Charge symmetry breaking (CSB) in the $\Lambda N$ interaction
is another important research topics.
Nuclear force based on the strong interaction and nuclear system
are invariant under the exchange of protons and neutrons (charge symmetry)
or more generally under any rotation in the isospin space (charge independence).
This approximate but basic symmetry  holds almost
exactly in $NN$ interaction and ordinary nuclei, 
and the CSB effects are very small.
For example, the binding energy difference between $^3$H and $^3$He is
only 70 keV after the correction of a large Coulomb effect.
In contrast, there has been a long-standing puzzle of CSB in 
hypernuclei. 
The binding energies of $\Lambda$ on mirror $\Lambda$ hypernuclei $^4_\Lambda$H
and $^4_\Lambda$He obtained by emulsion experiments~\cite{Juric:1973zq}
have a significant difference of 0.35 $\pm$0.05 MeV.
The $1^+$-$0^+$ level spacing of these mirror $\Lambda$ hypernuclei measured 
by the transition $\gamma$-rays gives additional information on CSB effect.
The $^4_\Lambda$H $\gamma$-ray was measured three times~\cite{Bedjidian:1976zh, Bedjidian:1979qh, KawachiPhD1997} and its weighted average is 
1.09 $\pm$ 0.02 MeV. 
On the other hand, the $^4_\Lambda$He $\gamma$-ray was reported 
only once to be 1.15 $\pm$ 0.04 MeV.
However, the $^4_\Lambda$He $\gamma$-ray spectrum is statistically
insufficient and it is claimed that the identification of the $^4_\Lambda$He 
events seems to be ambiguous.
In order to confirm  the reported data and to resolve the puzzle,
new measurements of the $1^+$-$0^+$ level spacing of 
$A$=4 $\Lambda$ hypernuclei, especially for  $^4_\Lambda$He, are both
necessary and important.


The J-PARC E13 experiment was proposed to measure hypernuclear $\gamma$-rays from
$^4_\Lambda$He, $^7_\Lambda$Li,  $^{10}_{\,\,\Lambda}$B, 
$^{11}_{\,\,\Lambda}$B, and $^{19}_{\,\,\Lambda}$F.
Data on the $s$-shell hypernucleus, $^4_\Lambda$He~\cite{Yamamoto:2015avw}, and 
the $sd$-shell hypernucleus, $^{19}_{\,\,\Lambda}$F~\cite{Yang:2017mvm}, 
were obtained at the K1.8 beamline  in 2015.

The $^{19}_{\,\,\Lambda}$F hypernucleus was produced by the
$^{19}$F$(K^-, \pi^-)$ reaction at 1.8 GeV/{\it c} with 
a 20 g/cm$^2$-thick liquid CF$_4$ target.
The hypernuclear states were identified by the missing mass for
the reaction by analyzing the incident $K^-$ and outgoing $\pi^-$
using the K1.8 beam spectrometer and the SKS, respectively.
The missing-mass resolution was 8.7 MeV (FWHM), which 
was sufficient to reject the contributions from highly excited states of $^{19}_{\,\,\Lambda}$F
and background $^{12}_{\,\,\Lambda}$C.
The $\gamma$-rays  were measured in coincidence 
using   Hyperball-J detectors~\cite{Koike:2014jna}, 
comprising 27 coaxial-type Ge detectors each
having a crystal size of 70mm ($\phi$) $\times$ 70mm (length) and 
PWO counters for background suppression from Compton scattering
in the Ge crystals and $\pi^0 \rightarrow 2\gamma$.
After the in-beam energy calibration of the Ge detectors using
a $^{232}$Th source and known $\gamma$-rays from the target or
surrounding materials, an accuracy of 0.5 keV was achieved over the
range from 0.1 to 2.5 MeV, while the energy resolution was 
measured to be 4.5 keV (FWHM) for 1 MeV $\gamma$-rays.

\begin{figure}[htbp]
\begin{minipage}{0.65\textwidth}
\includegraphics[width=0.98\textwidth]{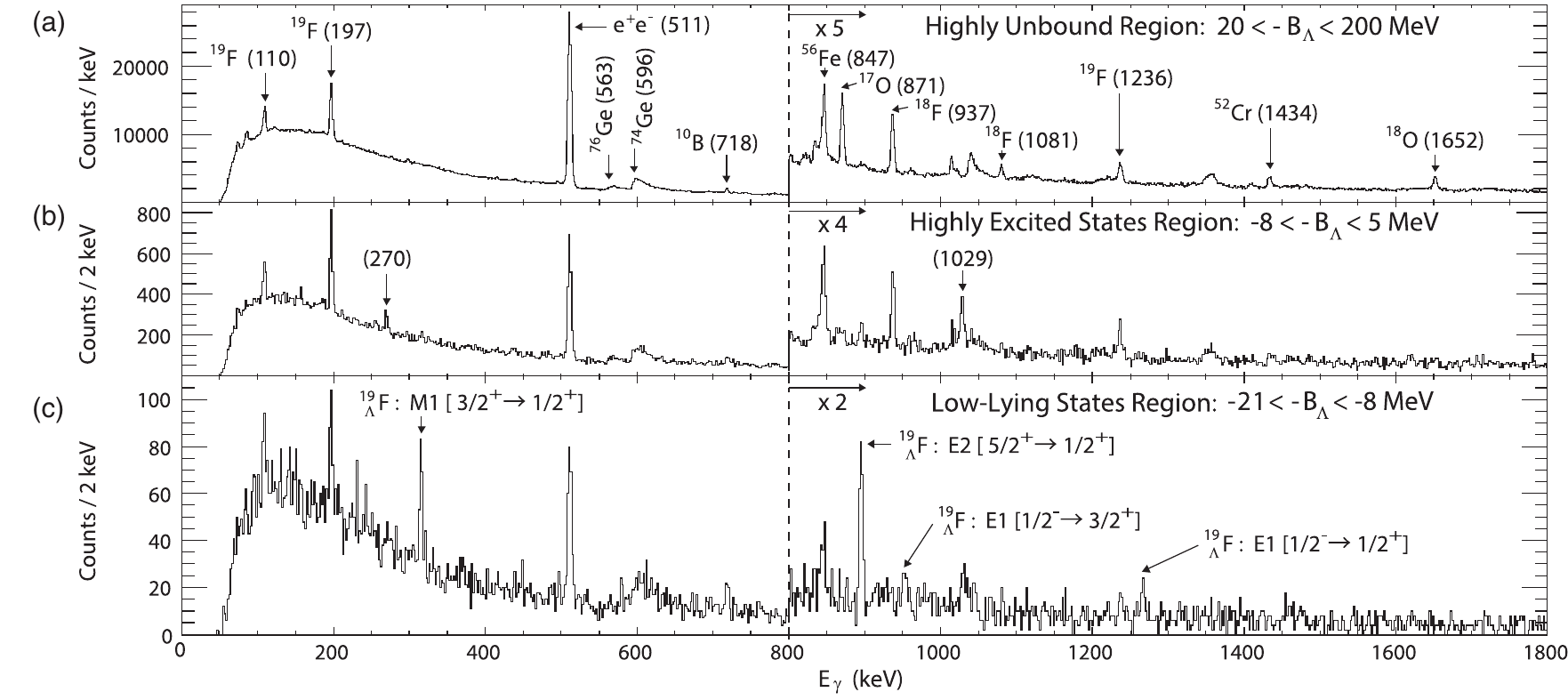}
\end{minipage}
\begin{minipage}{0.35\textwidth}
\includegraphics[width=0.98\textwidth]{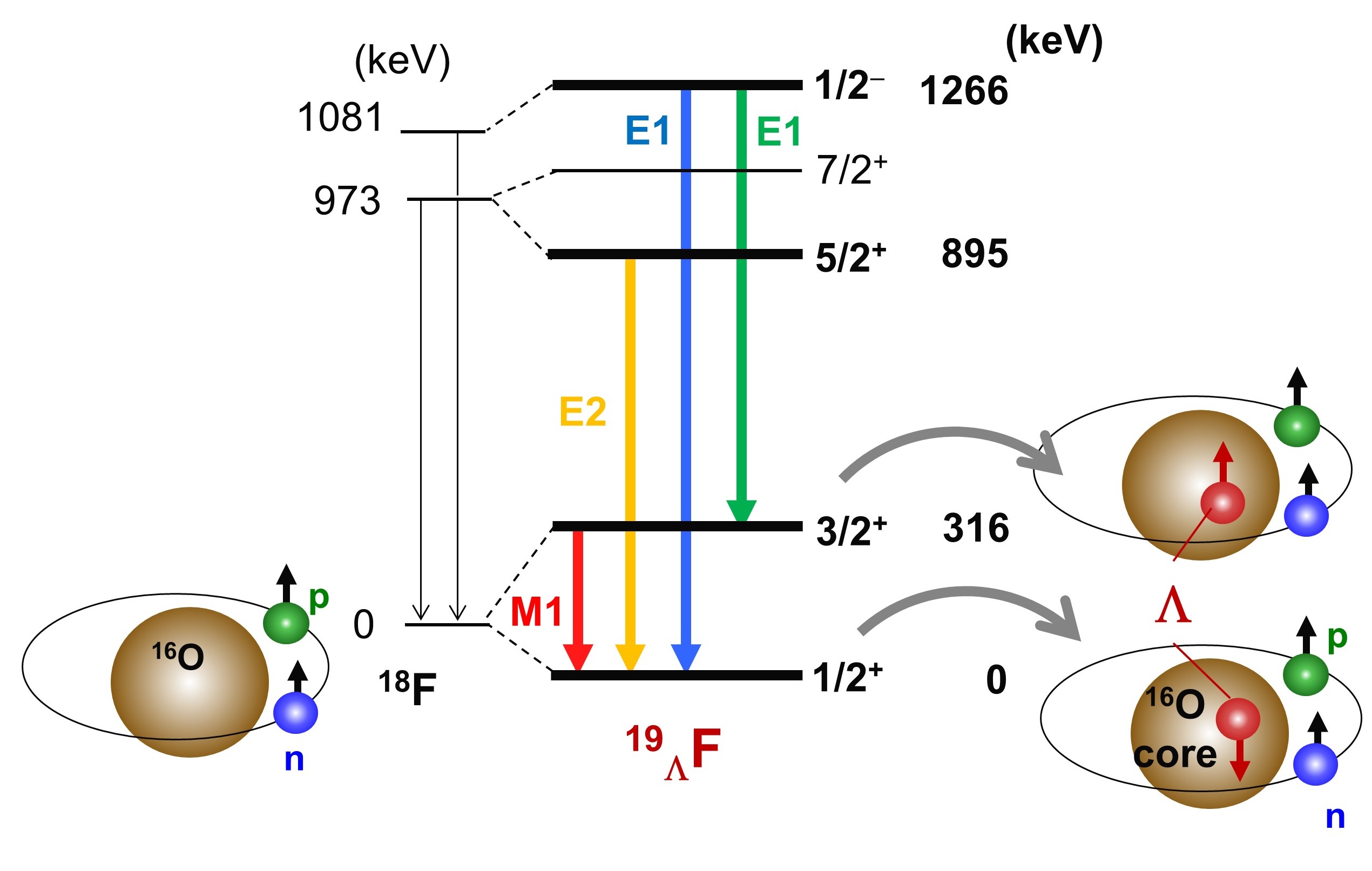}
\end{minipage}
\caption{
$\gamma$-ray spectra gated by $\Lambda$ binding energy ranges in
the missing mass of the $^{19}$F$(K^-,\pi^-)$ reaction, showing
(a) the highly unbound region ($20 < -B_\Lambda < 200$ MeV),
(b) the region associated with highly excited states ($-8 < -B_\Lambda< 5$ MeV),
(c) the region associated with low-lying states($-21 < -B_\Lambda < -8$ MeV).
The figure is taken form Ref.~\cite{Yang:2017mvm} (left).
A reconstructed level scheme of  $^{19}_{\,\,\Lambda}$F with 
the assigned $\gamma$-ray transitions (right).
}
\label{E13-19LH}
\end{figure}

Figure~\ref{E13-19LH} (left) presents the $\gamma$-ray spectrum
after selecting the region associated with low-lying states of  $^{19}_{\,\,\Lambda}$F.
In addition to the  $\gamma$-rays known to be emitted from ordinary nuclei,
four additional $\gamma$-rays can be identified with energies of 
315.5$\pm$0.4 (stat.) $^{+0.6}_{-0.5}$ (syst.),
895.2$\pm$0.3 (stat.) $\pm$0.5 (syst.), 
952.8$\pm$1.2 (stat.) $^{+0.5}_{-0.6}$ (syst.), and
1265.6$\pm$1.2 (stat.) $^{+0.7}_{-0.5}$ (syst.) keV .
Assuming a weak-coupling between a $\Lambda$ and a core nucleus and
taking into account the expected cross sections of the excited states,
a level structure of $^{19}_{\,\,\Lambda}$F is reconstructed and $\gamma$-ray 
transitions are assigned as shown in Fig.~\ref{E13-19LH} (right).

The energy spacing of the ground-state doublet is primarily determined
by the $\Lambda N$ spin-spin interaction and is of great interest
as discussed.
The obtained spacing value of 316 keV is in good agreement with
two independent shell-model calculations.
D.~J.~Millner predicted  value of 305 keV based on the phenomenological
spin-dependent $\Lambda N$ interaction strengths obtained
from $p$-shell $\Lambda$ hypernuclear data~\cite{Millener:2013mwa}.
On the other hand, the shell model calculation by A.~Umeya and T.~Motoba using
the effective $\Lambda N$ interaction by G-matrix method
based on Nijmegen SC97e and SC97f interactions~\cite{Rijken:1998yy}
provides 419 and 245 keV, respectively~\cite{Umeya:2016bbt}.
A spacing of 346 keV is expected
when the spin-spin interaction  is adjusted by mixing the SC97e and SC97f
interactions to reproduce the energy spacing
of 692 keV for the $p$-shell hypernucleus, $^7_\Lambda$Li.
The agreement between these values indicates that the theoretical framework
and $\Lambda N$ interaction as an input are good
even for heavier hypernuclei.

\begin{figure}[htbp]
\begin{minipage}{0.5\textwidth}
\includegraphics[width=0.98\textwidth]{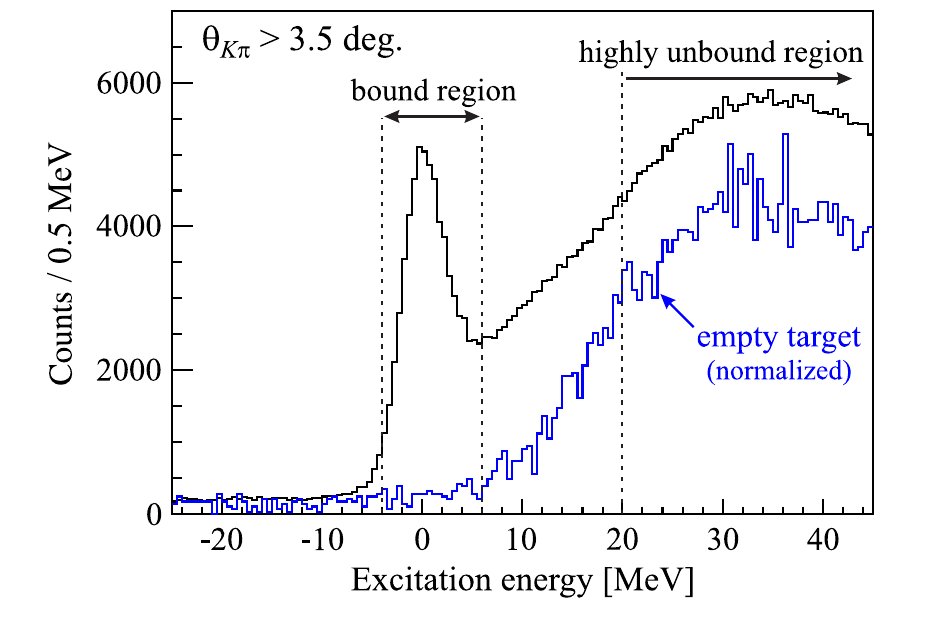}
\end{minipage}
\begin{minipage}{0.5\textwidth}
\includegraphics[width=0.98\textwidth]{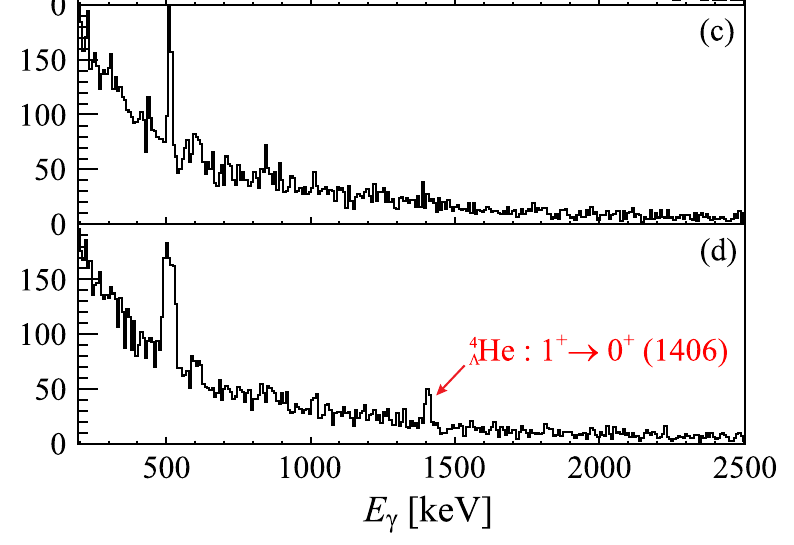}
\end{minipage}
\caption{
Missing-mass spectrum for the $^4$He$(K^-, \pi^-)$ reaction (left).
A peak shows the production of bound states for $^4_\Lambda$He.
(black), while no peak exists in the empty target run (blue).
The $\gamma$-ray spectra obtained by selecting events
in the bound region in the left spectrum without (right top) and
with (right bottom) a Doppler shift correction, respectively.
The figures are taken from Ref.~\cite{Yamamoto:2015avw}.
}
\label{E13-4LHSpectra}
\end{figure}

The $\gamma$-ray from $^4_\Lambda$He was measured in a similar way
as  the $\gamma$-rays from $^{19}_{\,\,\Lambda}$F.
The $^4_\Lambda$He was produced and identified by the $(K^-, \pi^-)$
reaction on a 2.8 g/cm$^2$-thick liquid He target.
An incident momentum of 1.5 GeV/{\it c} was chosen to~
populate the spin-flip $1^+$ state as much as possible
considering the spin-flip amplitude of the elementary
$K^+ + n \rightarrow \pi^- + \Lambda$ process and the 
beam intensity available at the K1.8 beamline.
The $^4$He$(K^-,\pi^-)$ missing-mass spectrum is shown
in Fig.~\ref{E13-4LHSpectra} (left), and exhibits
a peak corresponding to the $^4_\Lambda$He bound states ($0^+$ and $1^+$).
By selecting events in the $\Lambda$ bound region, $-4$ MeV $<$ 
excitation energy $< +6$ MeV, 
the $\gamma$-ray spectra were obtained  as shown 
in Fig.~\ref{E13-4LHSpectra}  (right top and bottom), 
for which an event-by-event Doppler-shift correction has and has not
been applied, respectively.
A 1406-keV peak is clearly visible after the Doppler-shift correction,
which is consistent with M1 transition.
Thus the excitation energy of the $^4_\Lambda$He $1^+$ state was
precisely determined to be 1.406$\pm$0.002 (stat.)$\pm$0.002 (syst.) MeV,
by taking  the nuclear recoil effect of 0.2 keV into account.

\begin{figure}[htbp]
\centerline{\includegraphics[width=0.65\textwidth]{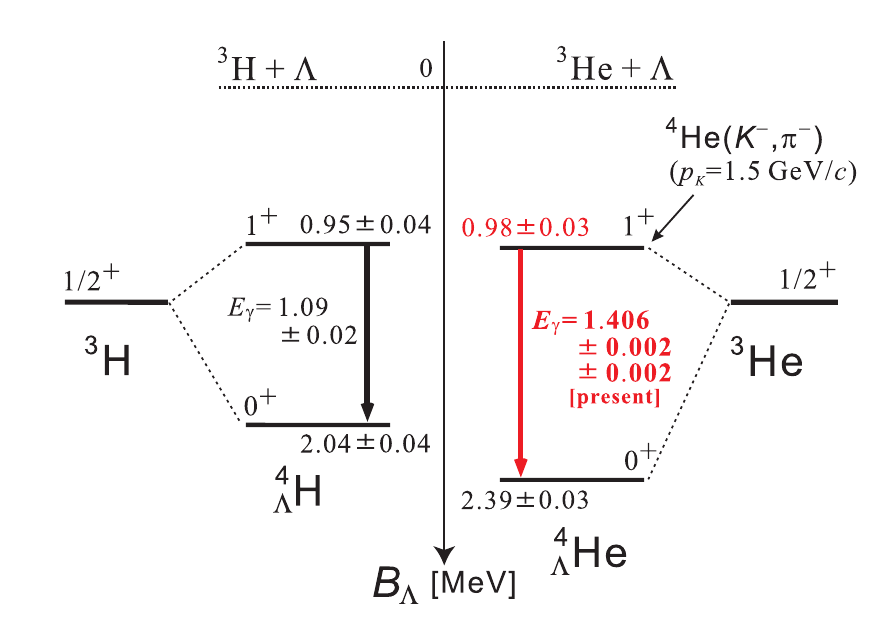}}
\caption{Level scheme for the mirror $\Lambda$ hypernuclei, 
$^4_\Lambda$H and $^4_\Lambda$He.
$\Lambda$ binding energies ($B_\Lambda$) 
for the ground states of $^4_\Lambda$H and $^4_\Lambda$He 
are taken from the emulsion data~\cite{Juric:1973zq}.
The $B_\Lambda$ for the 1$^+$ state of $^4_\Lambda$H was obtained
using  $B_\Lambda$ for the ground state of $^4_\Lambda$H  
and the weighted average of the $\gamma$-ray energies measured
in past experiments~\cite{Bedjidian:1976zh, Bedjidian:1979qh, KawachiPhD1997}.
The figure is taken from Ref.~\cite{Yamamoto:2015avw}.
}
\label{E13-4LHLevelScheme}
\end{figure}

This value for $^4_\Lambda$He  is obviously larger than that for
$^4_\Lambda$H (1.09$\pm$0.02 MeV).
This result clearly indicates the existence of a large CSB in
the $\Lambda N$ interaction.
Combining this result with the $\Lambda$ binding energies ($B_\Lambda$s) of
the ground states from the past emulsion data~\cite{Juric:1973zq},
the $B_\Lambda$ of the $1^+$ state of $^4_\Lambda$He is 0.98 $\pm$ 0.03 MeV
as shown in Fig.~\ref{E13-4LHLevelScheme}.
Comparing the present updated value for $B_\Lambda$ and those reported 
for the $1^+$ and  $0^+$ states of the mirror $\Lambda$ hypernuclei,
the $B_\Lambda$  difference for the $1^+$ states is  0.03$\pm$0.05 MeV,
which is much smaller than that for the $0^+$ state (0.35$\pm$0.05 MeV).
Therefore, the CSB effect is found to be strongly spin dependent.
This fact would provide a key to resolve the puzzle for a large CSB in $\Lambda$ 
hypernuclei and to elucidate the underlying $\Lambda N$ interaction.

It should be noted that the efforts to confirm and improve 
the experimental data
have been also performed at other facilities
in order to clarify the CSB effect in $\Lambda$  hypernuclei
and to investigate its origin.
The $\pi^-$ momentum in the
$^4_\Lambda$H $\rightarrow$ $^4$He + $\pi^-$ weak decay 
was precisely measured at MAMI-C~\cite{Esser:2015trs}, and
$B_\Lambda$ of the $0^+$ state of $^4_\Lambda$H 
was obtained to be 2.12$\pm$0.01(stat.)$\pm$0.09(syst.) MeV,
which is consistent with the emulsion data.
On the other hand, at Jefferson Lab, the $B_\Lambda$ for the 
$^7_\Lambda$He ground state 
was measured by the $^7$Li$(e, e'K^+)$ reaction and 
$B_\Lambda$ values for the  iso-triplet states of $A=7$ $\Lambda$ hypernuclei
( $^7_\Lambda$He,  $^7_\Lambda$Li$^*$, and $^7_\Lambda$Be)
were compared~\cite{Nakamura:2012hw}.
Ref.~\cite{Botta:2016kqd} summarizes $B_\Lambda$ values for 
$A\leq 16$ $\Lambda$ hypernuclei iso-multiplets including the results from the FINUDA experiment.
These results suggest very small CSB effects in the $p$-shell
$\Lambda$ hypernuclei.

At J-PARC, a measurement of $\gamma$-ray resulting from
$1^+ \rightarrow 0^+$ transition on $^4_\Lambda$H is planned
as a new experiment, E63, 
which will be performed using the Hyperball-J and the SKS 
at the K1.1 beamline~\cite{E63-proposal}.
$^4_{\Lambda}$H hypernuclei are produced by the $(K^-,\pi^-)$ reaction
on $^7$Li (or natural metal Li) target at 0.9 GeV/{\it c}
as a fragment from the highly excited states of $^7_\Lambda$Li$^*$.
In the previous experiment, a 1.1 MeV $\gamma$-ray peak
was  observed by selecting the highly unbound region
(excitation energy $E_x$=22 -- 33 MeV)
of $^7_\Lambda$Li produced by the $(K^-,\pi^-)$ reaction at 0.83 GeV/{\it c}~\cite{May:1984vm}.
The fragment production process, $^7_\Lambda$Li$^*$ $\rightarrow$ 
$^4_{\Lambda}$H + $^3$He, 
gives a larger recoil velocity of $^4_{\Lambda}$H than that for  $^7_\Lambda$Li$^*$
produced via the $(K^-,\pi^-)$ reaction,
and the Doppler broadening can not be corrected for.
However, this effect can be reduced by selecting
$^7_\Lambda$Li$^*$ events close to the  
$^4_{\Lambda}$H$^*$ + $^3$He decay threshold in the missing-mass
spectrum.

\subsubsection{ Investigation of in-medium $\Lambda$ properties
by $\gamma$-ray spectroscopy of $\Lambda$ hypernuclei \label{subsubsec:E63}}

\begin{figure}[htbp]
\centerline{\includegraphics[width=0.75\textwidth]{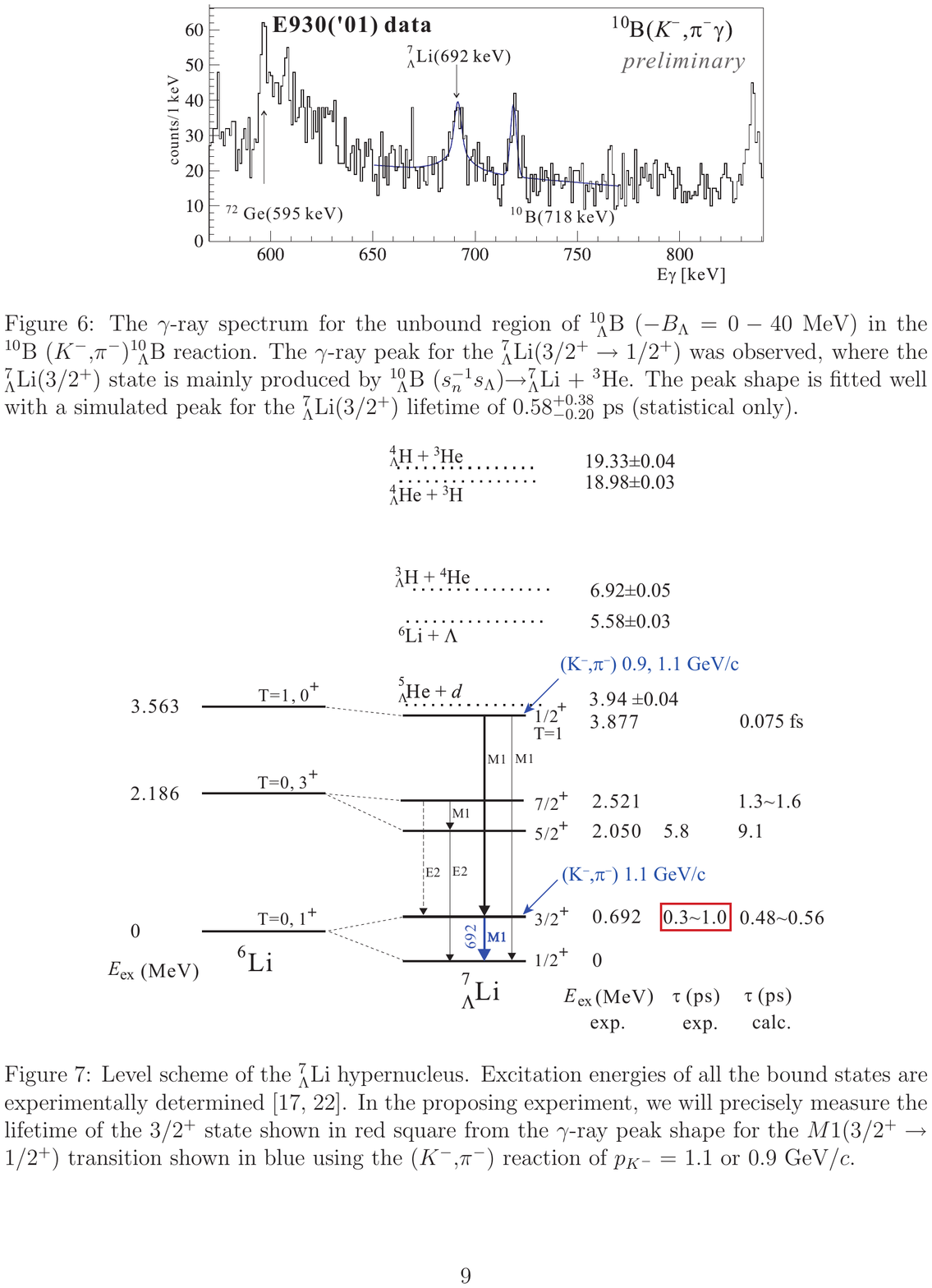}}
\caption{Level scheme of $^7_\Lambda$Li hypernucleus.
The excitation energies of  all bound states are experimentally 
determined~\cite{Tamura:2000ea, Ukai:2004zz}.
The lifetime of the $3/2^+$ state shown in the red square 
will be precisely measured from the $\gamma$-ray peak shape for
the $M1(3/2^+\rightarrow1/2^+)$ transition shown in blue using the
$(K^-,\pi^-)$ reaction at 0.9 or 1.1 GeV/{\it c}~\cite{Tamura:2010zza}.
The figure is taken from the E63 proposal~\cite{E63-proposal}.
}
\label{7LLiLevelScheme}
\end{figure}

In the E63 experiment which is a revised version of the E13 experiment,
$\gamma$-rays from the $^7_\Lambda$Li ground state doublet transition will
also be measured in order to derive the transition probability of a M1 
transition ($B(M1)$) and to examine the possible change of 
the $g$-factor of a $\Lambda$ in a nucleus.
In the picture of the constituent quark models, where the magnetic moment of
baryons is well described as a sum of a magnetic moment of a Dirac particle 
with a constituent quark mass,
the magnetic moment of the baryons may change, 
if the constituent mass is modified in a nucleus.
A $\Lambda$ hyperon in a $\Lambda$ hypernucleus is the best
probe to investigate whether such a change really occurs or not.

Instead of the direct measurement on the magnetic moment of 
a $\Lambda$ hypernucleus,  which is extremely difficult due to a short 
lifetime for spin precession,
the $g$-factor of the $\Lambda$ in the nucleus 
will be derived from the probability of a $\Lambda$ spin-flip transition.
Because the level scheme of  $^7_\Lambda$Li has been well
established experimentally as shown in Fig.~\ref{7LLiLevelScheme}~\cite{Tamura:2000ea, Ukai:2004zz}, 
the best target is the ground-state doublet of $^7_\Lambda$Li.
In the ground-state doublet of $^7_\Lambda$Li, 
spin directions of the $\Lambda$ of two members are opposite
each other to the spin direction of the core nucleus and
the $3/2^+ \rightarrow 1/2^+$ transition  corresponds to  a spin-flip of
the $\Lambda$.
In a weak coupling limit between the $\Lambda$ and the core nucleus,
the $B(M1)$ value of such a transition can be expressed as~\cite{Dalitz:1978hw},
\begin{eqnarray}
B(M1) & = & (2J_{up}+1)^{-1}  \left |\left <\phi_{low} \left\|
\textnormal{\mathversion{bold}$\mu$} \right\| \phi_{up}\right >\right |^{2} \nonumber \\ 
	& = & (2J_{up}+1)^{-1} \left | \left <\phi_{low} \left \|
 g_{c}\textnormal{\mathversion{bold}$J_c$}
+g_{\Lambda}\textnormal{\mathversion{bold}$J_\Lambda$}
\right \| \phi_{up}\right >\right |^{2} \nonumber \\
	& = & (2J_{up}+1)^{-1} \left | \left <\phi_{low} \left \|
g_{c}\textnormal{\mathversion{bold}$J$}
+(g_{\Lambda}-g_{c})\textnormal{\mathversion{bold}$J_\Lambda$}
\right \| \phi_{up}\right >\right |^{2} \nonumber \\
	& = & \frac{3}{8\pi}\frac{(2J_{low}+1)}{(2J_{c}+1)} 
\left (g_{c}-g_{\Lambda}\right )^{2}, 
\label{eq:BM1-7LLi}
\end{eqnarray} 
where $g_c$ and $g_\Lambda$ denote the effective $g$-factors of the core nucleus
and the $\Lambda$, respectively.
{\mathversion{bold}$J_c$} and  {\mathversion{bold}$J_\Lambda$}
denote their spins,
and {\mathversion{bold} $J=J_c+J_\Lambda$} is the spin of the $\Lambda$ hypernucleus. 
The spatial components of the wave functions for the lower and upper
states of the doublet, $\phi_{low}$ with spin $J_{low}$ and 
$\phi_{up}$ with spin $J_{up}$, are assumed to be identical.
The reduced transition probability $B(M1)$ can be derived 
from the lifetime $\tau$ of the upper state as,
\begin{equation}
1/\tau = \frac{16\pi}{9} E_{\gamma}^{3} B(M1),
\end{equation}
where $E_\gamma$ is the energy of  transition $\gamma$-rays.
The lifetime can be determined using Doppler shift attenuation method
(DSAM) which was successfully applied to obtain $B(E2)$ for the
$5/2^+ \rightarrow 1/2^+$ transition of  $^7_\Lambda$Li
in the previous experiment~\cite{Tanida:2000zs}.
In order to apply the DSAM, the stopping time $t_{stop}$ of the produced
excited states of the $\Lambda$ hypernucleus in the target material
should be on the same order as its lifetime $\tau$.
According to a simulation, this lifetime can be determined most
precisely when the stopping time is 2 --  3 times longer than the lifetime.
Barring any anomalous effects, the  lifetime of the 3/2$^-$ state is 
estimated to be very short, on the order of $\sim$ 0.5 ps.
Therefore a high-density Li$_2$O  target of 2.01 g/cm$^{3}$ will be used.

The $3/2^+$ state of $^7_\Lambda$Li is produced by the
$(K^-,\pi^-)$ reaction at 1.1 or 0.9 GeV/{\it c}.
The 1.1 GeV/{\it c} is the best suited to directly 
populate the spin-flip $3/2^+$ state, 
due to the large spin-flip amplitude of the elementary process at finite angles.
The highest $K^-$ beam intensity can be also obtained at 1.1 GeV/{\it c}.
Another option is the 0.9 GeV/{\it c}, at which the elementary process
has the largest non-spin-flip amplitude to produce the $1/2^+ (T=1)$ state 
and feeding process of $1/2^+ (T=1)$ $\rightarrow 3/2^+$, 
although the $K^-$ beam intensity is lower than that at 1.1 GeV/{\it c}.
The beam momentum will be determined by taking pilot data
to investigate yields and background levels.

\subsubsection{Hyperon-Nucleon scattering \label{subsubsec:E40}}

There had been long discussions on the existence of $\Sigma$ hypernuclei and
their widths.
The BNL-AGS E905 experiment 
showed that a $\Sigma$-bound state exists in the isospin $I=1/2$
channel ( $^4_\Sigma$He ), while no bound state exists in the $I=3/2$ channel
for the $A=4$ system, by comparing the $^4$He($K^-, \pi^{\pm}$) spectra~\cite{Nagae:1998tj}.
Thus, $\Sigma$-nucleus potential turned out to have a large isospin dependence (Lane-term) 
and to be repulsive in isospin average.
The spectra for quasi-free $\Sigma$ production on medium-to-heavy nuclear targets
also support a repulsive $\Sigma$ potential~\cite{Saha:2004ha, Harada:2005hs}.
Therefore, quantitative information on $\Sigma$ potential and $\Sigma N$
interaction can not be extracted via the spectroscopic methods.
Thus direct scattering experiments of $\Sigma$ hyperons yielding high
statistics data set have been desired.

\begin{figure}[htbp]
\begin{minipage}{0.5\textwidth}
\includegraphics[width=0.98\textwidth]{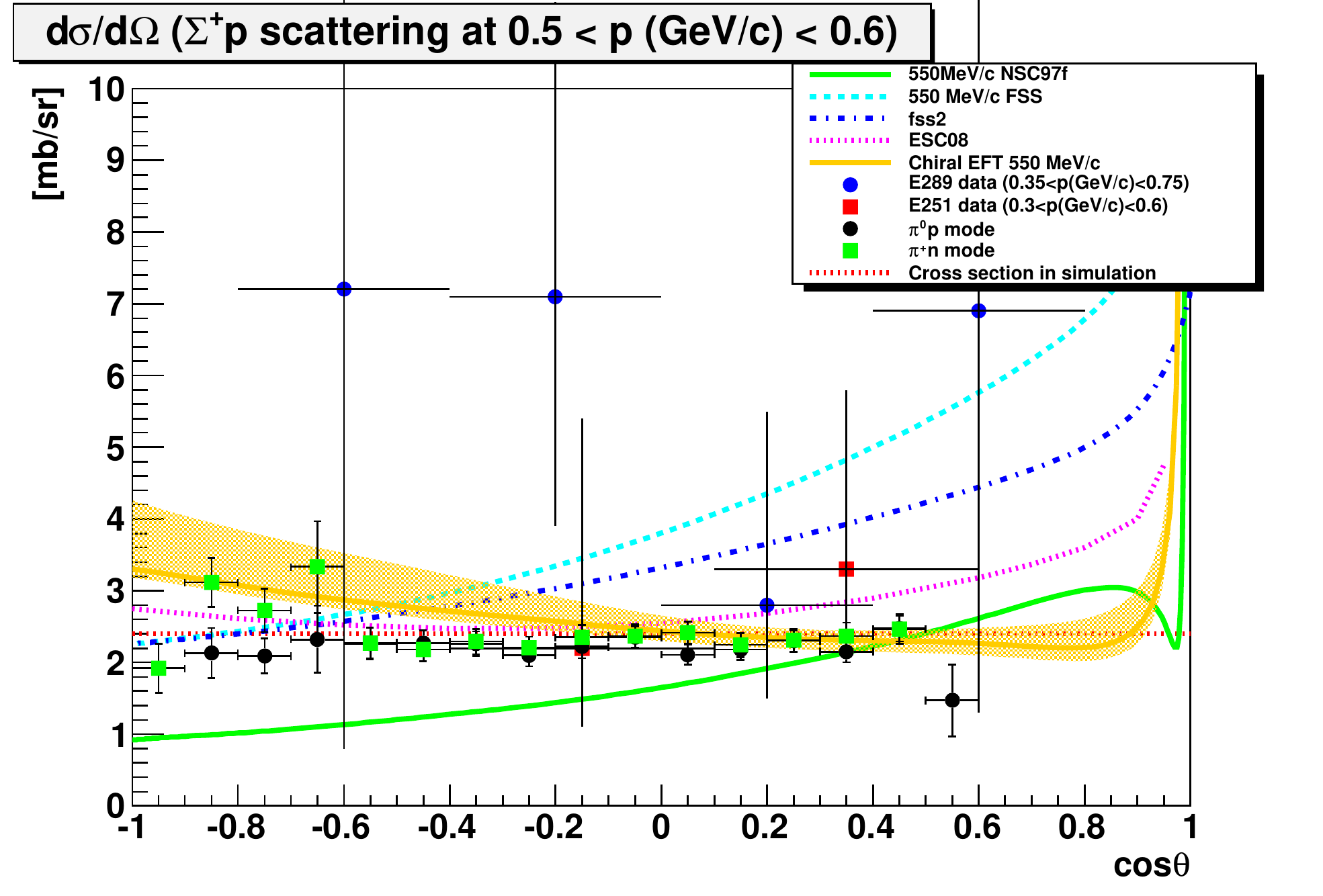}
\end{minipage}
\begin{minipage}{0.5\textwidth}
\includegraphics[width=0.96\textwidth]{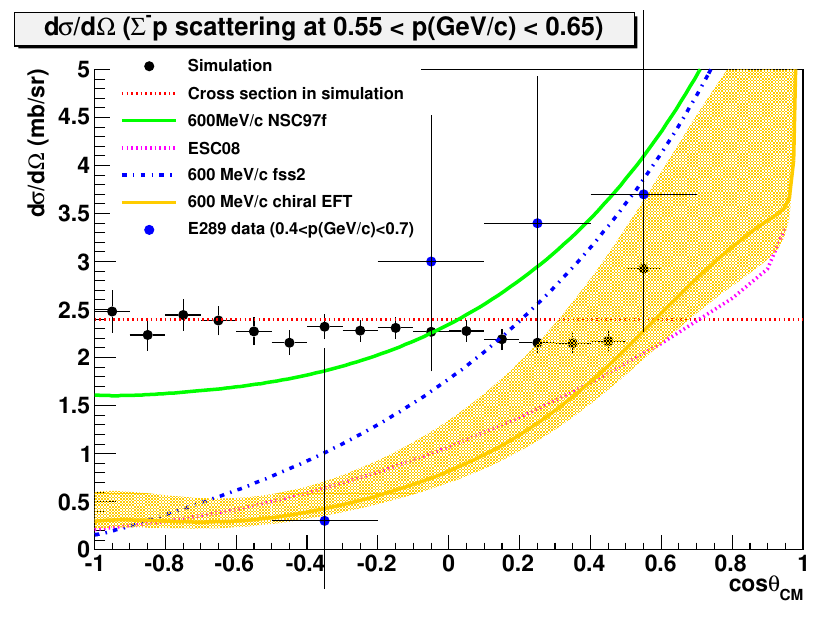}
\end{minipage}
\caption{Simulated differential cross sections showing the expected experimental errors
taken from the presentation by K.~Miwa at HYP2015~\cite{Miwa:2017cfj}.
(Left) $\Sigma^+ p$ elastic channel of the $\Sigma^+$ momentum from 0.5 to 
0.6 GeV/{\it c} for the 55$\times$10$^6$ tagged $\Sigma^+$ beam.
The black circles and green squares represent the results 
for the $p\pi^0$ and $n\pi^+$ decay modes, respectively, 
assuming the cross section with a flat angular distribution shown by a red dotted line.
The blue circles and red squares are the experimental data obtained 
in the KEK-PS E289~\cite{Ahn:2005gb} and the KEK-PS E251~\cite{Ahn:1997wa}, respectively.
The additional curves present the theoretical predictions for
meson exchange models, NSC97a (green solid)~\cite{Rijken:1998yy} 
and ESC08 (magenta dotted)~\cite{Rijken:2010zzb},
quark cluster models, FSS (light blue dotted)~\cite{Fujiwara:1996qj} 
and fss2 (blue dotted)~\cite{Fujiwara:2001jg}.
The orange band indicates the prediction by chiral effective theory~\cite{Haidenbauer:2013oca}.
(Right) $\Sigma^- p$ elastic channel of the $\Sigma^-$ momentum from 0.45 to 0.55
GeV{\it c} for the  16$\times$10$^6$ tagged $\Sigma^-$ beam.
The black circles show the results assuming the cross section 
with a flat angular distribution shown by the red dotted line.
The blue circles show the experimental data obtained 
in the KEK-PS E289~\cite{Kondo:2000hn}.
The additional curves and the band present the same information as those in the left figure.
}
\label{E40SimdSdW}
\end{figure}

The $\Sigma^+ p$ channel can be simply described by two multiplets of 
${\bf 27}$ and ${\bf 10}$ in Eq.~\ref{su3f-decomp}.
In the case of the S-wave interaction that is dominant in the low-energy region,
the spin singlet state ($^1S_0$) can be described by ${\bf 27}$ 
which is the same multiplet as appeared in the $NN(I=1)$ channel,
and is well known at least phenomenologically based on a substantial quantity of  $NN$ data.
The spin triplet state ($^3S_1$) can be described by   
${\bf 10}$ which is almost Pauli forbidden state.
Considering the spin-weight factors, the $\Sigma^+ p$ channel is expected
to be quite repulsive because of  the almost Pauli forbidden state 
as previously discussed.
As shown in Fig.~\ref{E40SimdSdW} (left), theoretical models based on
meson exchange (Nijmegen models NSC97f~\cite{Rijken:1998yy}  and 
NSC08~\cite{Rijken:2010zzb}) and quark cluster models which include
an explicit quark degree of freedom in the short range part
 (FSS~\cite{Fujiwara:1996qj} and fss2~\cite{Fujiwara:2001jg})
predict very different cross sections.
The $\Sigma^+ p$ data can be used to determine which model is superior and
whether or not our current understanding of quark-based picture in the short
range part is correct.
In addition, a large core radius can also be confirmed from the energy dependence of
the differential cross sections at 90$^\circ$, as R. Jastrow applied to the
$pp$ scattering data and estimated a hard core radius of 
nuclear force~\cite{Jastrow:1951vyc}.
 
Assuming  isospin symmetry, the $\Sigma^+ p$ and
$\Sigma^- n$ interactions should be the equivalent except for the contribution
of the electromagnetic interaction. 
Therefore  the experimental measurements on the strength on the 
repulsive force between $\Sigma^+$ and proton is essential information on
constructing EOS for  high density neutron star matter.
In such a high-density neutron star matter, negatively charged particles
can reduce the  large Fermi energies of electrons (and neutrons) via the
so-called hyperionization process.
The $\Sigma^-$ potential, the primary component of which is the $\Sigma^- n$ 
interaction, determines  whether $\Sigma^-$ hyperons emerge in
the core of a neutron star and, if so, at what density.

\begin{figure}[htbp]
\centerline{\includegraphics[width=0.7\textwidth]{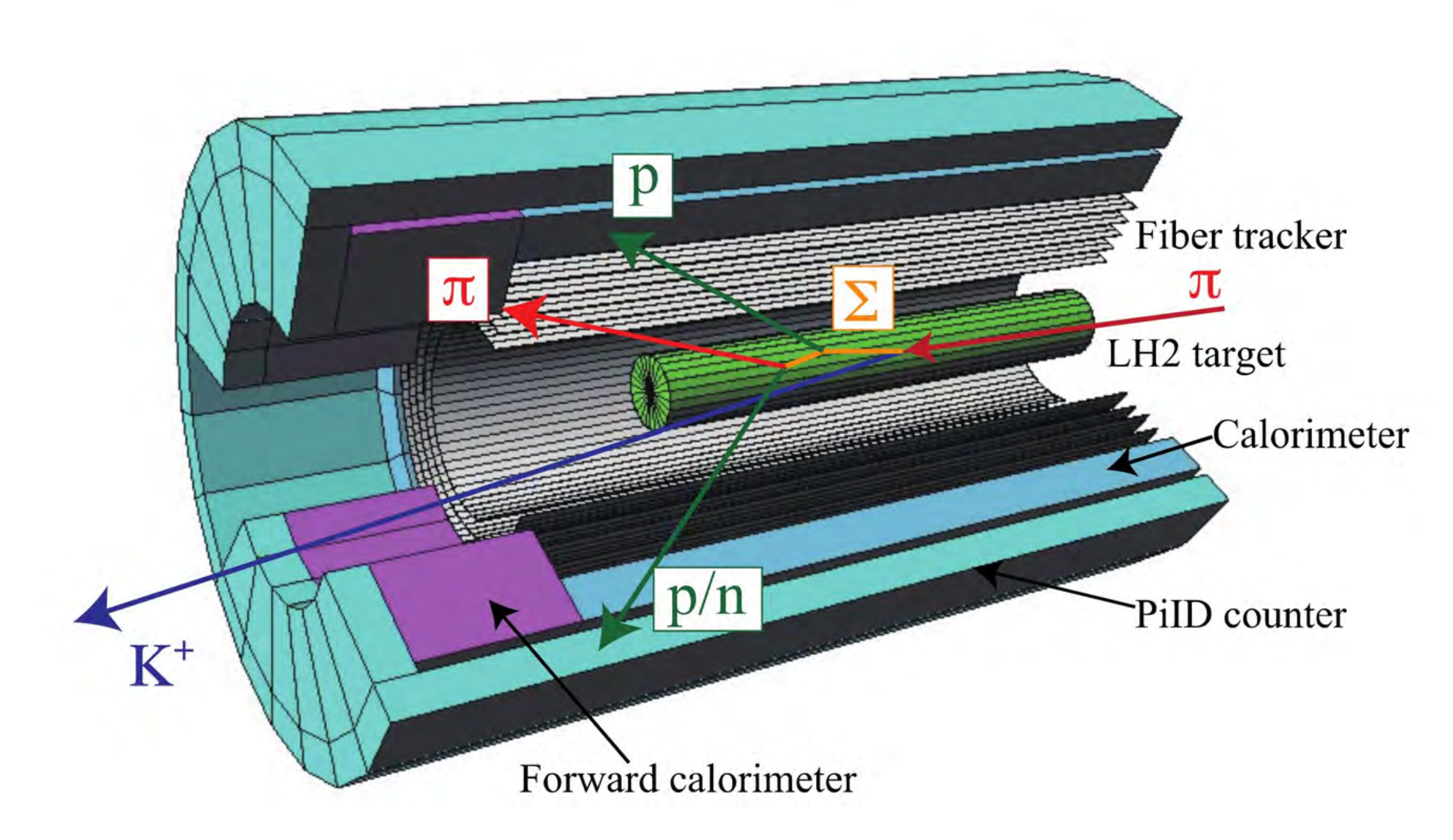}}
\caption{Schematic drawing of CATCH detector for 
the $\Sigma^{\pm} p$ scattering experiment. }
\label{CATCHfig}
\end{figure}

The J-PARC E40 experiment aims to measure  $\Sigma^{\pm}$-$p$ 
elastic scattering and $\Sigma^- p$ $\rightarrow$
$\Lambda n$ conversion with high statistics 
using  the modern  techniques.
Figure~\ref{E40SimdSdW} shows the expected statistical errors in
the differential cross sections for $\Sigma^+ p$ (left) and $\Sigma^- p$
(right) elastic scattering estimated by simulations assuming the cross section
with a flat angular distribution shown by the red dotted lines.
In this experiment,  $\Sigma^{\pm}$ hyperons are produced using high intensity
$\pi^{\pm}$ beams
via the $p (\pi^{\pm}, K^+)\Sigma^{\pm}$ reactions on a liquid hydrogen target
at the incident momenta of 1.4 GeV/{\it c} for $\pi^+$ and 1.32 GeV/{\it c}
for $\pi^-$ and identified by the K1.8 beam and KURAMA spectrometers.
In these kinematics, $\Sigma$ momentum ranges from 0.45 to 0.85 GeV/{\it c}.
$\Sigma$ hyperons are scattered by protons in the same
liquid hydrogen target. 
Charged particles are produced by the reactions, then their decays
are measured by a detector system surrounding the target which is 
called a cylindrical active tracker and calorimeter system for
hyperon-proton scattering (CATCH)~\cite{catch-detector}.
The CATCH detector consists of a cylindrical fiber tracker with 6 straight and
6 spiral layers of scintillation fibers for trajectory measurements
and BGO calorimeters for energy measurements, as shown in Fig.~\ref{CATCHfig}.
Secondary reactions such as elastic scattering and conversion process
are kinematically identified from $\Sigma$ momentum vectors 
using magnetic spectrometers, the energy and direction of each 
charged particles measured using  CATCH.

Commissioning of the beamline and the spectrometer system
for the E40 experiment was carried out  in February 2017.
In total $\sim$ 17$\times$10$^6$ tagged $\Sigma^-$ data for the $\Sigma^- p$ mode were obtained using 1.96$\times$10$^7$ $\pi^-$ 
per spill beam in June 2018, February and March 2019. 
About a half of the $\Sigma^+ p$ data ($\sim$ 40$\times$10$^6$ tagged $\Sigma^+$)
were also taken using 1.9$\times$10$^7$ $\pi^+$ per spill beam in April 2019, while the remaining $\Sigma^+ p$ data will be taken in March 2020. 
The analysis of the scattering events is in progress.
Thus high-statistic cross section data 
including their angular distributions
will be reported in the near future.

\subsubsection{Double-$\Lambda$ hypernuclei \label{subsubsec:E07}}

To date, there has been only limited 
information available regarding the $B_{8}B_{8}$ interaction and the hadron system with $S=-2$.
The $\Lambda\Lambda$ and $\Xi N$ interactions
as well as $\Xi N$ $\rightarrow$ $\Lambda\Lambda$ conversion 
interaction must be determined experimentally 
including the spin and/or isospin dependence.
Five experiments aiming to investigate the $S=-2$ system 
and to obtain these $B_{8}B_{8}$ interactions in the  $S=-2$ channel
have been approved at J-PARC;
a photographic emulsion experiment (E07), 
missing-mass spectroscopy of the $^{12}_{\,\Xi}$Be hypernucleus  (E05 and E70), 
X-ray spectroscopy of $\Xi^-$-Fe atom (E03), and
a search for the $H$-dibaryon (E42) (as already discussed in Sec.~\ref{subsubsec:E42}).
Each of these experiments requires a high-intensity 1.8 GeV/{\it c} $K^-$ beam 
at the K1.8 beamline to produce the $S=-2$ system via the $(K^-, K^+)$ reaction.
Some of these have already been carried out.

Since a $\Lambda$ hyperon can not be used as a fixed target due to its very 
short lifetime, a direct $\Lambda\Lambda$ scattering experiment can not be carried out.
Therefore, spectroscopic studies on double-$\Lambda$ hypernuclei, in which two $\Lambda$
hyperons are embedded in nuclei, are possible ways to investigate the $\Lambda\Lambda$
interaction except for using final state interaction (FSI) 
or the measurements of the $\Lambda\Lambda$ momentum-correlation in heavy ion
collision which was successfully done recently~\cite{Adamczyk:2014vca}.
The $\Lambda\Lambda$ interaction strength 
in the double-$\Lambda$ hypernucleus  $^{\ A}_{\Lambda\Lambda}Z$ 
is often expressed in terms of
$\Delta B_{\Lambda\Lambda}$($^{\ A}_{\Lambda\Lambda}Z$).
This can be deduced from the masses of the double-$\Lambda$ and single-$\Lambda$
hypernuclei, and is defined as
\begin{equation}
\Delta B_{\Lambda\Lambda} (^{\ A}_{\Lambda\Lambda}Z)
= B_{\Lambda\Lambda}(^{\ A}_{\Lambda\Lambda}Z) 
- 2B_{\Lambda}(^{A-1}_{\ \ \ \Lambda}Z),
\end{equation}
where $B_{\Lambda\Lambda}$ and $B_{\Lambda}$ are the binding energies of
two $\Lambda$ hyperons in the double-$\Lambda$ hypernucleus and
a $\Lambda$ hyperon in the single-$\Lambda$ hypernucleus, respectively.

The double-$\Lambda$ hypernucleus
was first reported by M.~Danysz {\it et al.} 
as an event with a topology of sequential  weak decays 
in the nuclear emulsion~\cite{Danysz:1963zza}.
This event was interpreted as a $^{\,\,10}_{\Lambda\Lambda}$Be or 
$^{\,\,11}_{\Lambda\Lambda}$Be, indicating a relatively strong
attractive force between $\Lambda$ hyperons of 
$\Delta B_{\Lambda\Lambda}$ $=4.5\pm$0.4 or 3.2$\pm$0.6 MeV,
respectively.
Among the several double-$\Lambda$ hypernuclear events so far 
reported in the emulsion experiments~\cite{Danysz:1963zza, Prowse:1966nz, 
Aoki:1991ip, Takahashi:2001nm, Ahn:2013poa},
the most impressive one is the {\it NAGARA} event discovered in the
KEK-PS E373 experiment using a counter-emulsion hybrid method.
This event was  uniquely identified as a $^{\ \,6}_{\Lambda\Lambda}$He 
without any ambiguity such as a possible excited state of 
daughter (hyper)nucleus~\cite{Takahashi:2001nm, Ahn:2013poa}.
From this result, the $\Lambda\Lambda$ interaction was determined
to be weakly attractive, {\it i.e.} $\Delta B_{\Lambda\Lambda}$ = 0.67
$\pm$ 0.17 MeV  by assuming  $\Xi^-$ capture from an atomic
3D state of $^{12}$C with a binding energy of 0.13 MeV.
Based on this result, the event reported by M.~Danysz {\it et al.} can be 
consistently interpreted as a $^{\,\,10}_{\Lambda\Lambda}$Be, 
the decayed daughter of which ($^9_\Lambda$Be) was in an excited state.

There have been  discussions that $\Delta B_{\Lambda\Lambda}$ is not
a good measure of the $\Lambda\Lambda$ interaction,
since the mass or $B_\Lambda$ of a single-$\Lambda$ hypernucleus
is greatly affected by the structure of the core 
nucleus~\cite{Hiyama:2010zzd, Kanada-Enyo:2018pxt}.
As an example, if the core nucleus is not spin-less, $B_{\Lambda}$
has an effect based in the $\Lambda N$ spin-dependent interactions,
which are canceled in the case of  double-$\Lambda$ hypernucleus
in the ground state.
Change of the nuclear size by adding a $\Lambda$ 
($^7_\Lambda$Li~\cite{Tanida:2000zs}) is  another example.
Because the core nucleus ($^4$He) is spin-less and stiff in the case of
$^{\ \,6}_{\Lambda\Lambda}$He,
the obtained $\Delta B_{\Lambda\Lambda}$ is thought to present 
a pure $^1S_0$ interaction between $\Lambda$ hyperons.
In order to confirm this and to obtain a bare $\Lambda\Lambda$ interaction
taking into account nuclear structure effects,
systematic studies on double-$\Lambda$ hypernuclei, especially $p$-shell ones,
are necessary.
The P-wave $\Lambda\Lambda$ interaction is also of interest in the context to
the hyperon puzzle~\cite{Togashi:2016fky}.
The P-wave interaction can be extracted by observing ``$\Lambda$-excited''
double-$\Lambda$ hypernuclear states where one $\Lambda$ is in 
the $s$-state and the other is in the $p$-state, although such states may not be
bound.

The $\Xi N$$\rightarrow$$\Lambda\Lambda$ conversion interaction 
can be obtained from the strength of the imaginary part of 
the $\Xi$-nucleus potential or the widths of $\Xi$ hypernuclear states 
as described in the following sections.
However, in case that the level spacing is similar to the width or
the width is much smaller than the experimental resolution,  
it may be difficult to obtain this interaction by those methods.
Another possible approach is to measure masses of the $A$ = 4 and 5
double-$\Lambda$ hypernuclei, $^{\ \,\,4}_{\Lambda\Lambda}$H, 
$^{\ \,\,5}_{\Lambda\Lambda}$H, and $^{\ \,\,5}_{\Lambda\Lambda}$He.
A large mixing effect  due to the
$\Xi N$-$\Lambda\Lambda$ conversion interaction is 
expected in the $S=-2$ system because of the small mass
difference of 28 MeV. 
However, this mixing effect is suppressed in double-$\Lambda$ hypernuclei
having a saturated core nucleus such as $^{\ \,\,6}_{\Lambda\Lambda}$He,
because a nucleon associated with the $\Lambda\Lambda \rightarrow$ $\Xi^- p$ or 
$\Xi^0 n$ process should be in a higher energy shell.
However, in the case of the $A$ = 4 and 5 double-$\Lambda$ hypernuclei,
$s$-shell nucleons are not fully occupied and this mixing is expect
to significantly affect the mass.


The J-PARC E07 experiment is a third generation counter-emulsion hybrid experiment 
to study double-strangeness nuclei such as double-$\Lambda$ and 
$\Xi$ hypernuclei.
In the second generation counter-emulsion hybrid experiment, E373,
$\Xi^-$ hyperons produced via the quasi-free 
$K^- + $``$p$''$ \rightarrow$ $K^+ + \Xi^-$
reaction on a diamond target were injected into the emulsion module, 
and the reaction was identified by a magnetic spectrometer system.
Based on  guidance of position and direction of the  $\Xi^-$ determined by 
scintillating-fiber detectors upstream of the emulsion module, 
the $\Xi^-$ hyperons were followed through the emulsion until they stopped, decayed, or
passed through.
Among the $\sim$ 650 $\Xi^-$ stop events~\cite{Theint:2019wkg}, 
7 {\it double events} (that exhibit  sequential decays of double-$\Lambda$ hypernuclei)
and 2 {\it twin events} (showing that two single-$\Lambda$ hypernuclei are emitted
from the $\Xi^-$ capture point) were found~\cite{Ahn:2013poa}.
The goal of the E07 experiment is to obtain ten times the double-strangeness nuclear events 
compared to the E373 experiment (that is, $\sim$ 100 double-strangeness 
nuclei among $\sim 10^4$ $\Xi^-$ stop events).
The physics runs were  carried out 
separately in 2016 and 2017.
Prior to the E07 runs, the SKS located in the K1.8 area 
was moved to the K1.1 beamline area and 
the KURAMA spectrometer used to measure the outgoing $K^+$
was installed in the K1.8 area instead of the SKS.
In order to accumulate ten times the data,  a number of improvements
and upgrades were essential. These are itemized below.
\begin{itemize}
\item A high-intensity,  high-purity $K^-$ beam. \\
Since the total number of the irradiated beam particles is limited
so as to maintain suitable efficiency in the scanning and analysis of the emulsions,
the purity of $K^-$ in the beam is essential to enhance the statistics.
During the beam exposure,  a $K^-$ beam with an intensity of 2.8$\times 10^5$
particles per spill of 2.0 s duration in every 5.52 s with the purity of 82\%
was used.
The purity and intensity were increased by factors of $\sim$ 2 and 3.3, respectively, relative to the previous E373 experiment.

\item Total quantity of the emulsion. \\
A total of 2.1 tons of emulsion gel, about 3 times of the previous one, was
prepared,
and half a year was required to produce the emulsion sheets.
A total quantity of 118 emulsion modules from $\sim$ 1500 sheets 
was used to record the $\Xi^-$-induced events.
Approximately one year was required for photographic developments of these
sheets following the beam exposures.

\item An automated $\Xi^-$ following system~\cite{MyintKyawSoe:2017gpq}.\\
To complete the emulsion scanning within a reasonable time frame
while meeting the requirement to analyze ten times as many events,
an automated track following system had been developed.
This system  automatically traces each track to the end point, such as 
stopping, decay, or passing-through, using track information
from the previous sheet of the emulsion module. 
When an interesting end point such a $\Xi^-$ stop or decay is detected
the system records photographic images around the point for further confirmation
by human-eyes.
The working speed using this system is 15 times faster than that of the 
semiautomatic system with human support used in the E373 experiment.

It is also important to reduce the scanning area and track candidates
in the first emulsion sheet for the efficient scanning.
Therefore
silicon strip detectors (SSDs) having  a four layer configuration and
much better spatial resolution than
the scintillating-fiber detectors were used as the $\Xi^-$ tracker. 
The resulting resolutions were estimated to be 15 $\mu$m and 20 mrad
(RMS) for the  position and angle, respectively.

\item Enlargement of the solid angle acceptance of the KURAMA spectrometer.\\
The gap of the KURAMA magnet was extended to 800 mm
and  the downstream detectors were enlarged.
The solid angle acceptance was 280 msr.
\end{itemize}

The scanning of these emulsions is currently in progress.
By September 2019,  82 \% emulsion modules had been  scanned
at least once and 31 double-strangeness events had been observed,
several of which have already been reported.

\begin{figure}[htbp]
\centerline{\includegraphics[width=0.7\textwidth]{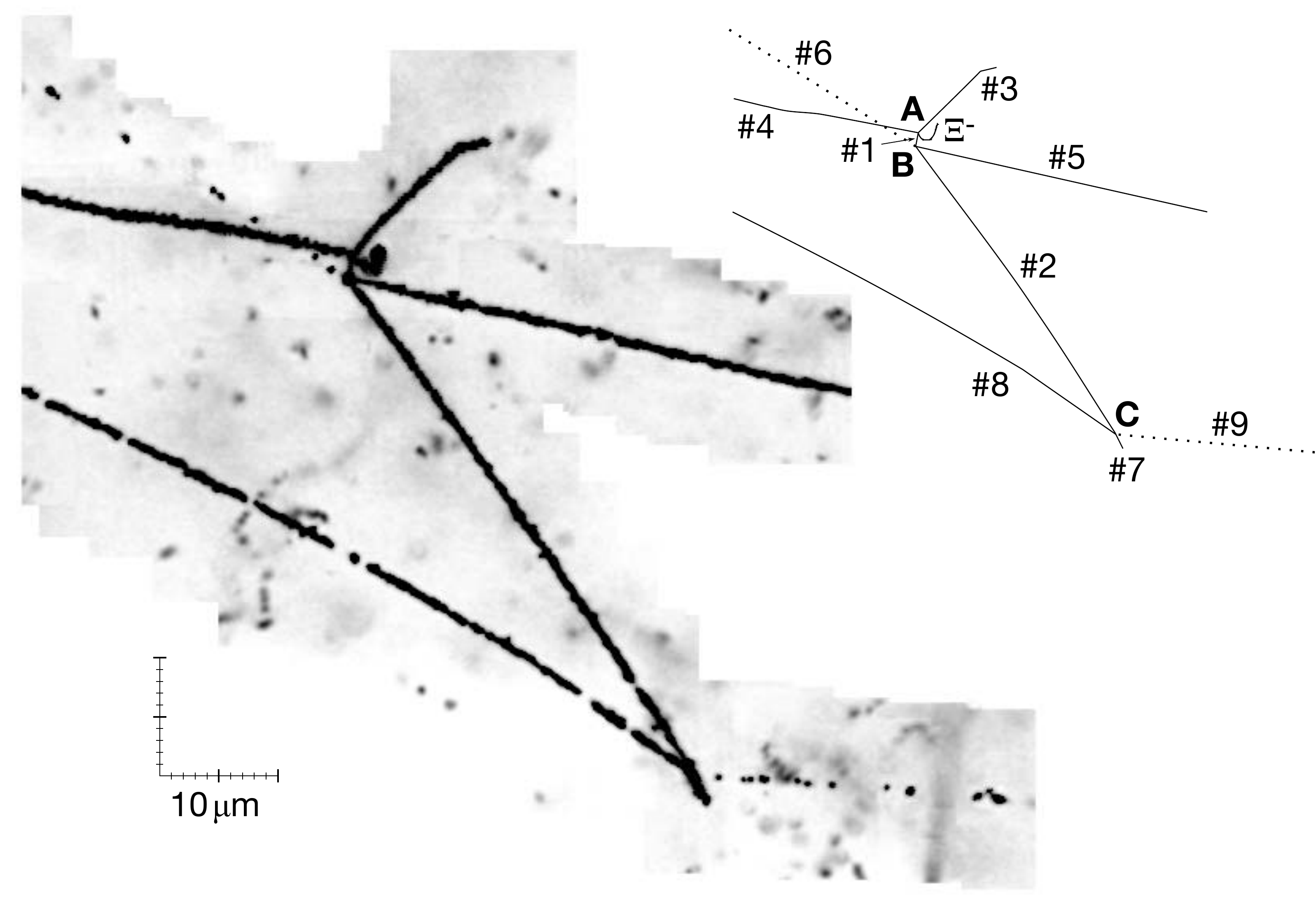}}
\caption{
Overlaid photograph and its schematic drawing of  
the {\it MINO} event observed in the J-PARC E07 experiment.
The figure is taken from Ref.~\cite{Ekawa:2018oqt}.
}
\label{MinoEventFig}
\end{figure}

One is {\it MINO} event with the topology of sequential decays 
as shown in Fig.~\ref{MinoEventFig}~\cite{Ekawa:2018oqt}.
From the detailed kinematical analysis and
assuming  that the $\Xi^-$ was captured by $^{12}$C, $^{14}$N, or $^{16}$O,
this event is interpreted as the following three processes involving
production and decay modes:

\begin{eqnarray*}
\textrm{$\Xi^{-} + ^{16}$O} 	&\rightarrow	&   
\textrm{[ $^{\,10}_{\Lambda\Lambda}$Be, $^{\,11}_{\Lambda\Lambda}$Be, 
$^{\,12}_{\Lambda\Lambda}$B ] + $^{4}$He + [ $t$,  $d$, $p$ ]  (at vertex A)}, \\
\textrm{[ $^{\,10}_{\Lambda\Lambda}$Be, $^{\,11}_{\Lambda\Lambda}$Be,
$^{\,12}_{\Lambda\Lambda}$Be ] }	& \rightarrow  & 
\textrm{$^{5}_{\Lambda}$He + [ $p$, $d$, $t$ ] + $xn$ (at vertex  B)}, \\
\textrm{$^5_\Lambda$He} &  \rightarrow &
\textrm{$^{4}$He + $p$ +$\pi^-$ (at vertex C)}. 
\end{eqnarray*}

\begin{table}[htbp]
\caption{Masses of hyperons and binding energies of hypernuclei 
used in the analysis of the {\it MINO} event.}
\label{tbl-massvalue}
\begin{center}
\begin{tabular}{lcll}
\hline
  	& \multicolumn{1}{r}{ [MeV] }	& reference 	& comment	\\
\hline
mass of $\Lambda$	& 1115.683 $\pm$ 0.006	&\cite{Tanabashi:2018oca} 	& \\
mass of $\Xi^-$		& 1321.71 $\pm$ 0.006 	&\cite{Tanabashi:2018oca} 	& \\
$B_\Lambda$($^7_\Lambda$Be)	& 5.16 $\pm$ 0.08 &\cite{Davis:1986kg} & \\
$B_\Lambda$($^8_\Lambda$Be)	& 6.84 $\pm$ 0.05 &\cite{Davis:1986kg} & \\
$B_\Lambda$($^9_\Lambda$Be)	& 6.71 $\pm$ 0.04 &\cite{Davis:1986kg} & \\
$B_\Lambda$($^{10}_{\,\,\Lambda}$Be)	& 8.60 $\pm$ 0.07 $\pm$ 0.16 &\cite{Gogami:2015tvu} & \\
$B_\Lambda$($^{11}_{\,\,\Lambda}$Be)	&  8.2 $\pm$ 0.5	&  
& a linear extrapolation from $B_\Lambda$ above 4 $_\Lambda$Be isotopes \\
\hline
$B_{\Xi^-}$ of $^{16}$O	& 0.23	&\cite{Yamaguchi:2001ip} 
& atomic 3D state of $^{16}$O	\\
\hline
\end{tabular}
\end{center}
\end{table}

By assuming that the $\Xi^-$ was captured in the atomic 3D state of $^{16}$O
with the theoretically estimated binding energy of 0.23 MeV~\cite{Yamaguchi:2001ip},
the $B_{\Lambda\Lambda}$ values for each candidate were obtained to be
15.05$\pm$0.09 (stat.) $\pm$0.07 (syst.),
19.07$\pm$0.08 (stat.) $\pm$0.07 (syst.), and
13.68$\pm$0.08 (stat.) $\pm$0.07 (syst.) MeV
for $^{\ \,\,10}_{\Lambda\Lambda}$Be, $^{11}_{\Lambda\Lambda}$Be, and
$^{12}_{\Lambda\Lambda}$Be, respectively.
The corresponding $\Delta B_{\Lambda\Lambda}$ values
are
1.63 $\pm$ 0.09 (stat.) $\pm$ 0.11 (syst.),
1.87 $\pm$ 0.08 (stat.) $\pm$ 0.36 (syst.), and
$-2.7$ $\pm$ 0.08 (stat.) $\pm$ 1.0 (syst.) MeV.
Those values were obtained 
using the mass and $B_\Lambda$ values listed in Table~\ref{tbl-massvalue}.
The negative $\Delta B_{\Lambda\Lambda}$ for $^{\,12}_{\Lambda\Lambda}$Be
seems to be inconsistent with the attractive $\Lambda\Lambda$ interaction.
However, if  $^{\,12}_{\Lambda\Lambda}$Be is produced in an excited state,
its excitation energy is added to $B_{\Lambda\Lambda}$ and
$\Delta B_{\Lambda\Lambda}$. Thus $\Delta B_{\Lambda\Lambda}$
for the ground state of $^{\,12}_{\Lambda\Lambda}$Be may become positive.
The probabilities of these three interpretations were evaluated based on the
kinematical fitting. The results are shown in Table~\ref{mino-kinfit},
together with the  $\Delta B_{\Lambda\Lambda}$ obtained from the above
analysis.
The $\chi^2$ of the kinematical fitting demonstrates that
the interpretation of the $^{\,11}_{\Lambda\Lambda}$Be production is
the most likely.

\begin{table}
\caption{Obtained $\Delta B_{\Lambda\Lambda}$ and $\chi^2$ values from 
the kinematical fitting at vertex A
for each interpretation of the {\it MINO} event. }
\label{mino-kinfit}
\begin{center}
\begin{tabular}{llr}
\hline
Interpretations	& $\Delta B_{\Lambda\Lambda}$ [MeV]	& $\chi^2$ \\
\hline
$\Xi^- + ^{16}$O $\rightarrow ^{\,10}_{\Lambda\Lambda}$Be + $^4$He + $t$	&
1.63 $\pm$ 0.14	& 11.5	\\
$\Xi^- + ^{16}$O $\rightarrow ^{\,11}_{\Lambda\Lambda}$Be + $^4$He + $d$	&
1.87 $\pm$ 0.37	&   7.3	\\
$\Xi^- + ^{16}$O $\rightarrow ^{\,12}_{\Lambda\Lambda}$Be$^*$ + $^4$He + $p$	&
$-2.7$ $\pm$ 1.0 (+$E_X$)	& 11.3 \\
\hline
\end{tabular}
\end{center}
\end{table}

Recently, a new scanning method, {\it i.e.} overall scanning,  has been
developed~\cite{Yoshida:2017oww},
in which, characteristic topologies with more than 3 vertexes are
searched for without counter's information on  the $\Xi^-$ track.
This method can detect latent $\Xi^-$ capture events caused by the 
quasi-free $K^- +$``$n$'' $\rightarrow$ $K^0 + \Xi^-$ reaction
or by the decay of an outgoing $K^+$ prior to the KURAMA spectrometer.
Thus ten times more double-strangeness events are expected to be observed.


\subsubsection{$\Xi$ hypernuclei \label{subsubsec:E05-E70}}

At present, only a few experimental data on the $\Xi N$ interaction are available.
Some signal events were observed in the $\Xi$-bound region of 
the missing-mass spectra for the $^{12}$C$(K^-, K^+)$ reaction
measured in the KEK-PS E224~\cite{Fukuda:1998bi} and BNL-AGS E885~\cite{Khaustov:1999bz} experiments.
Although the peak structure could not be observed due to the insufficient experimental
resolution of 10 MeV (FWHM) or worse and the poor statistics,
the fitting of spectrum-shape including quasi-free $\Xi$ production near the
$\Xi$ bound threshold 
suggested a weak attractive $\Xi$-nucleus 
potential with a depth of $-14$ MeV, assuming a Wood-Saxon type potential~\cite{Khaustov:1999bz}.
A new impressing event named {\it KISO} was discovered in the emulsion
exposed at the E373 experiment
by applying the overall scanning method~\cite{Yoshida:2017oww}
which had been developed for the E07 experiment.
This event was uniquely identified as
$\Xi^- + ^{14}$N $\rightarrow$ $^{10}_{\,\,\Lambda}$Be + $^5_\Lambda$He,
although there is some uncertainty as to  whether the  $^{10}_{\,\,\Lambda}$Be
was  in the ground or excited states.
Assuming that  $^{10}_{\,\,\Lambda}$Be was in the excited state,
which gives a lower $\Xi$ binding energy ($B_\Xi$),  and incorporating the excitation 
energy predicted theoretically,
$B_\Xi$ was obtained to be 1.11$\pm$0.25 MeV.
This value is much larger than
0.17 MeV for the $\Xi$ atomic 3D orbit energy~\cite{Nakazawa:2015joa}.
This value was later updated to be 1.03 $\pm$ 0.18 MeV~\cite{KisoPaper2}
by taking into account 
new experimental data on $^{10}_{\,\,\Lambda}$Be$^*$~\cite{Gogami:2015tvu}.
Thus the {\it KISO} event confirms the existence of a $\Xi$ hypernucleus, 
and strongly suggests an attractive $\Xi$-nucleus potential and an attractive $\Xi N$
interaction on average.

\begin{figure}[htbp]
\centerline{\includegraphics[width=0.55\textwidth]{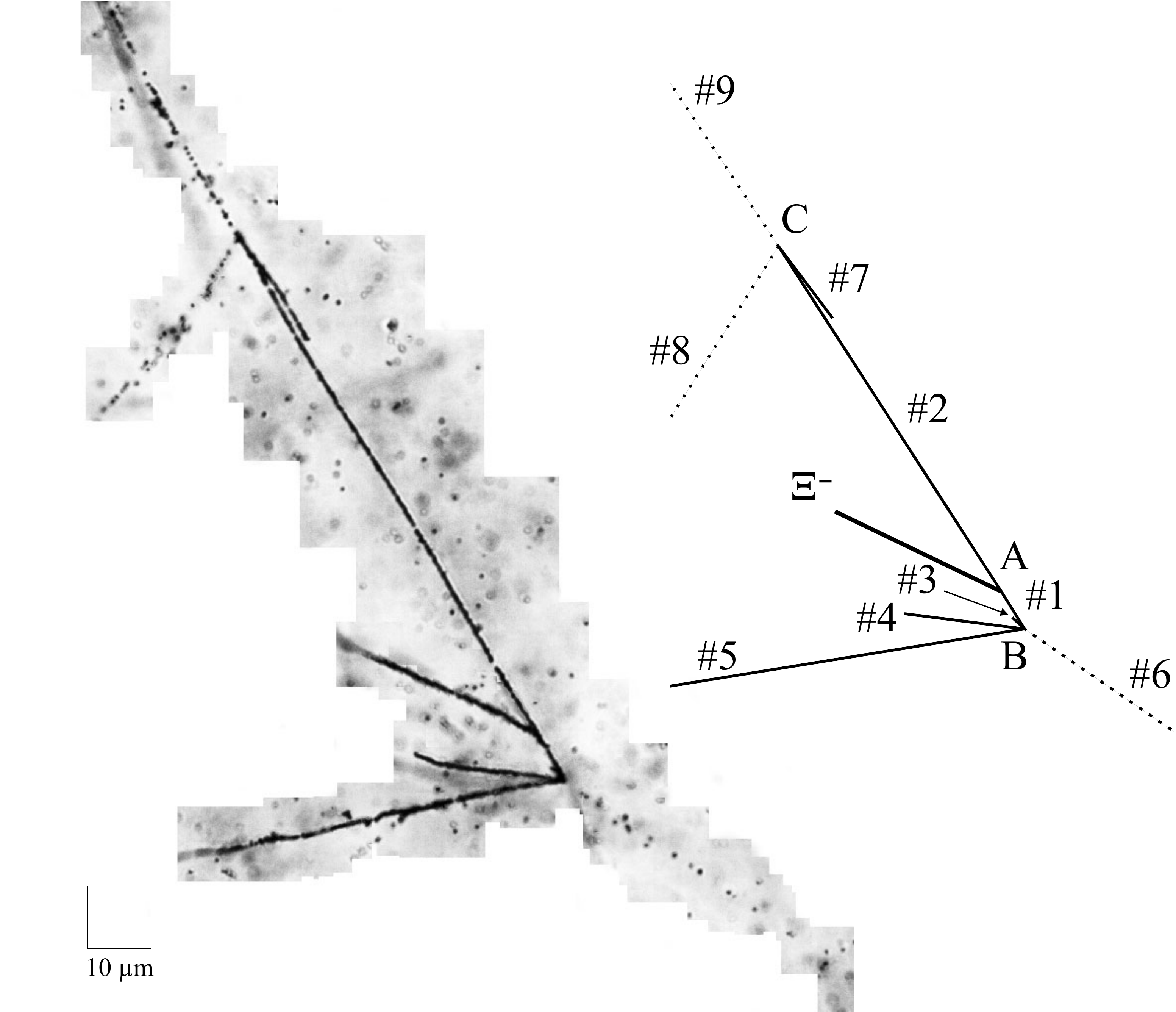}}
\caption{
Overlaid photograph and its schematic drawing  
of the {\it IBUKI} event observed in the J-PARC E07 experiment.
The figure is taken from Ref.~\cite{ibuki-reference}.
}
\label{IbukiEventFig}
\end{figure}

In the E07 experiment, a new event with a {\it twin event} topology was found
as shown in Fig.~\ref{IbukiEventFig}.
This event named {\it IBUKI} was uniquely identified as
$\Xi^- + ^{14}$N $\rightarrow ^{10}_{\,\,\Lambda}$Be (\#1)
 + $^5_{\Lambda}$He (\#2).
Only this mode is accepted at the first vertex where the $\Xi^-$ was captured.
The decays \#1 $\rightarrow$ 4-prongs  and
\#2  $\rightarrow$ 3-prongs with a $\pi^-$ track are consistent with those of 
$^{10}_{\,\,\Lambda}$Be and $^{5}_{\Lambda}$He, 
respectively~\cite{E07YoshidaQNP2018,ibuki-reference}.
This mode is the same as that for the {\it KISO} event.
However, there is no possibility of 
$^{10}_{\,\,\Lambda}$Be being in the excited state.
The $\Xi^-$ binding energy was unambiguously determined to be
1.27 $\pm$ 0.21 MeV~\cite{ibuki-reference},
which is significantly larger than 0.17 MeV for the
energy of the atomic 3D state.
Therefore this event provides additional evidence for the $^{15}_{\,\Xi}$C
hypernucleus.
 
A comparison of $B_{\Xi^-}$ obtained from different events is quite interesting.
There are two acceptable values that can be obtained from the {\it KISO} event~\cite{KisoPaper2}:
1.03 $\pm$ 0.18 MeV in the case of the excited state of $^{10}_{\,\,\Lambda}$Be
and 3.87 $\pm$ 0.21 MeV in the case of the ground state.
The present result from the {\it IBUKI} event appears to support the former case.
This value also agrees with the theoretical predictions for the 2P state, which are;
1.14 MeV based on the Ehime potential~\cite{Yamaguchi:2001ip}
and 1.22 to 1.85 MeV based on the ESC08 model~\cite{Nagels:2015dia}.
However the $\Xi$ hypernuclear state may have a large width due to
a strong $\Xi N$-$\Lambda\Lambda$ conversion in the nucleus.
Therefore the latter case can not be excluded at present.
Recently the {\it KINKA} event from the E373 experiment~\cite{Ahn:2013poa} was uniquely identified as 
$\Xi^-$ + $^{14}$N $\rightarrow$ $^9_\Lambda$Be + $^5_\Lambda$He + $n$,
although a detailed analysis to obtain $B_{\Xi^-}$ is still in progress.
By combining the results from the {\it KISO}, {\it IBUKI}, {\it KINKA}, and
other similar events that will presumably be observed in near future,
important information can be obtained regarding  not only the binding energies 
but also the widths of $^{15}_{\,\Xi}$C hypernuclear states.

It should be  emphasized that a large number of {\it twin event} samples
including both $\Xi$ hypernuclei and $\Xi^-$ absorption from atomic states 
provides an opportunity to confirm in which state 
$\Xi^-$ absorption take place (that is, the 3D state or deeper states).
This is quite important with regard to the analysis of double-$\Lambda$ hypernuclei,
because the value for $\Delta B_{\Lambda\Lambda}$ was obtained by assuming
absorption in the 3D state from the theoretical calculation.
If the $\Xi^-$ is absorbed in a deeper state such as 2P,
the corresponding $\Delta B_{\Lambda\Lambda}$ value will be a larger,
indicating a more attractive $\Lambda\Lambda$ interaction.

J-PARC E05 and E70 are experiments 
on the $(K^-, K^+)$ missing-mass spectroscopy
aiming at observing the $^{12}_{\,\Xi}$Be hypernucleus with good resolution and 
high statistics, and obtaining information on $\Xi N$ and 
$\Xi N$ $\rightarrow$ $\Lambda\Lambda$ interactions.
The original E05 plan in the proposal was to carry out the measurement
using an SksPlus spectrometer with a DD configuration by adding a dipole magnet
in front of the SKS magnet.
The solid angle acceptance of the SksPlus is 30 msr and
the missing-mass resolution is estimated to be 3.1 MeV (FWHM) with
a 5.4 g/cm$^2$ carbon target (Table~\ref{SpecList_XiHN}).
However, 
the production cross section obtained from the E885 experiment
is very small as  
89 $\pm$ 14 and 42 $\pm$ 5 nb/sr for averaged from
0$^\circ$ to 8$^\circ$ and from 0$^\circ$ to 14$^\circ$, 
respectively~\cite{Khaustov:1999bz}. 
In the early stage, the MR beam power and consequently the $K^-$ beam intensity 
were insufficient to allow the E05 experiment to be performed with the SksPlus.
Therefore the experimental plan was reconsidered to be carried out
in more realistic beam condition.

\begin{table}
\caption{
$K^+$ spectrometers and their
performance for the $(K^-, K^+)$ missing-mass spectroscopy.
The effects of target thickness are included in the missing-mass resolution
(see text). }
\label{SpecList_XiHN}
\begin{center}
\begin{tabular}{|l|c|c|c|}
\hline
Spectrometer		&	SksPlus			& SKS		& S-2S 	\\
Experiment	&   (E05 in the proposal)	& (E05)		& (E70)	\\
\hline
Magnet configuration	& D+D(SKS)	& D(SKS)	& QQD	\\
Acceptance [msr]	& 30		& 110	& 55	\\
Missing-mass resolution (FWHM) [MeV]	& 3.1	& 5.4	& $<2.0$	\\
\hline
\end{tabular}
\end{center}
\end{table}

Thus, the E05 experiment was carried out by using only the SKS with 
a large solid angle of 110 msr and a thick carbon target of 9.36 g/cm$^2$
in 2015.
In this pilot run, $K^-$ intensity of 6$\times$10$^{5}$ per spill (5.52 s beam cycle)
was available to use with an MR beam power of 39 kW.
The experimental missing-mass resolution was found to
be 5.4 MeV (FWHM) from the $\Xi^-$ spectrum of the elementary 
$p ( K^-, K^+) \Xi^-$ reaction with a 9.54 g/cm$^2$ CH$_2$ target.
This is the best resolution obtained to date for the missing-mass spectroscopy
of $\Xi$ hypernucleus.
Elementary production data
at 1.5, 1.6, 1.7, 1.8, and 1.9 GeV/{\it c} were also measured
to obtain the incident momentum dependence.
Cross section on the elementary $\Xi^-$ production was confirmed to be maximum
at 1.8 GeV/{\it c}, as pointed out by C.B.~Dover and A.~Gal~\cite{Dover:1982ng}.
This is an important result with regard to the planning of future experiments. 
The $K^-$ beam at 1.8 GeV/{\it c} is evidently the most suitable to study double-strangeness 
systems to which $\Xi^-$ production is doorway.

The analysis of carbon target data is in progress, but preliminary results
have been reported at several conferences~\cite{Nagae:2017slp, Nagae:2019uzt}.
A significant number of events ($\sim$ 50 counts) 
in the $\Xi^-$ bound region were observed, 
suggesting the existence of $\Xi$ hypernuclei.
The production yield is also consistent with the reported cross sections,
and the spectrum shape is very similar to that obtained in E885~\cite{Khaustov:1999bz},
although the experimental resolution was significantly improved
to 5.4 MeV (FWHM) from 10 MeV (FWHM).
These results suggest either multiple bound states or a broad width for the
$\Xi$ hypernuclear state, or both.

\begin{figure}[htbp]
\centerline{\includegraphics[width=0.7\textwidth]{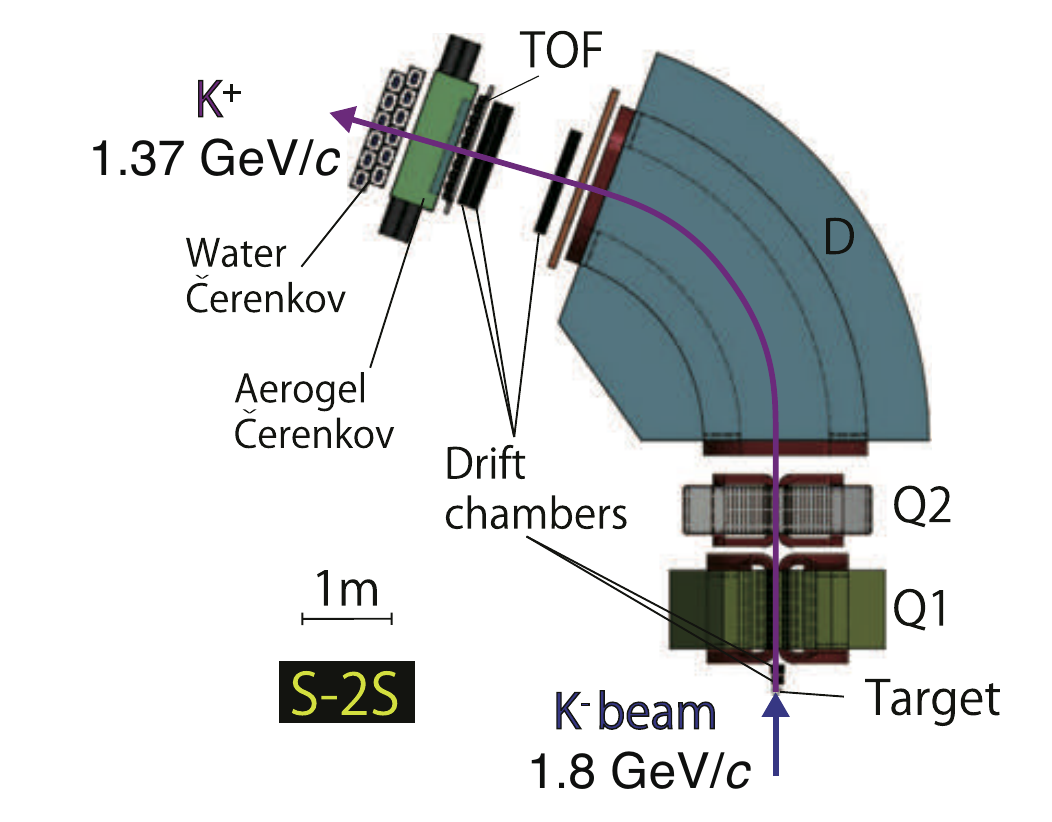}}
\caption{Schematic drawing of  the S-2S to be used in the J-PARC E70 experiment.
The figure is taken from the experimental proposal~\cite{E70-proposal}.
}
\label{S-2Sfig}
\end{figure}

In order to clarify this point and obtain as much information as possible,
the missing-mass resolution is the most important experimental issue.
Therefore in the next experiment, E70~\cite{E70-proposal}, 
a high-resolution spectrometer 
S-2S  (Strangeness $-2$ Spectrometer) will be used.
The S-2S has a QQD magnet configuration and 
good optical properties similar to those of the HKS at 
Jefferson Lab.~\cite{Fujii:2015aga, Gogami:2017wgj}.
In this case,  the experimental resolution will be primarily determined by the target
thickness due to the energy loss straggling in the target.
Even so, a reasonably thick target will be required to compensate for  the small
production cross section of $^{12}_{\,\Xi}$Be.
Therefore, an active fiber target (AFT) with a 10 cm thickness
will be employed to correct the energy-loss 
in the target  event-by-event.
The resolution in this experiment is expected to be 1.4 MeV (FWHM).
The recent  DWIA calculation by 
T.~Motoba {\it et al.}~\cite{Motoba:2010zzb} using the $\Xi N$ G-matrix interaction 
derived from the ESC08a model~\cite{Rijken:2010zzb} predicts three prominent peaks
of  the $1^-_1$, $1^-_2$, and $1^-_3$ states.
These states would be clearly observed with 1.4 MeV (FWHM) resolution in the E70.
Measurements on $^7$Li$(K^-,K^+)^7_\Xi$H and  
$^{10}$B$(K^-,K^+)^{10}_{\,\Xi}$Li are currently under consideration
in order to obtain the spin- and isospin-independent term 
of the $\Xi N$ interaction~\cite{Hiyama:2008fq}.

Depending on the coupling strength of $\Xi N$ and $\Lambda\Lambda$ due to
the $\Xi N$ $\rightarrow$ $\Lambda\Lambda$ conversion process,
a part of the production strength of $\Xi$ hypernucleus states will be used to
produce the excited states of double-$\Lambda$ hypernucleus~\cite{Harada:2013jwa}.
This is a one-step process via the $\Xi$-doorway to produce a double-$\Lambda$ hypernucleus,
similar to the one-step process via the $\Sigma$-doorway 
in the production of the neutron-rich $\Lambda$ hypernucleus.
Although the production cross section of such states is expected to be very small,
the $(K^-,K^+)$ missing-mass method is likely to be a useful means of 
measuring the excited states of double-$\Lambda$ hypernuclei..
This method can apply to even the $\Lambda$ unbound states
that can not be identified in the emulsion analysis.

The recent lattice QCD simulation at almost physical point predicts
$S = -2$ S-wave baryon-baryon interactions~\cite{Sasaki:2019qnh};
the $\Lambda\Lambda$ interaction is weakly attractive but not enough
strong to produce bound or resonance dihyperon around the $\Lambda\Lambda$
threshold. While the $\Xi N$ interaction has relatively strong attraction in
isospin-singlet, spin-singlet channel, which may lead
to light $\Xi$ hypernuclei~\cite{Hiyama:2019kpw}. 
Furthermore small $\Xi N$ $\rightarrow$
$\Lambda\Lambda$ conversion is suggested.
These features of the $S = -2$ baryon-baryon interactions would be
confirmed or checked by the future spectroscopy experiments.

Installation of the S-2S in  the K1.8 area 
is planned in near future after the completion of the experiments using 
the KURAMA spectrometer.


\subsubsection{X-ray spectroscopy of  $\Xi^-$-atoms \label{subsubsec:E03}}

X-ray spectroscopy can be applied to the negatively charged 
$\Xi^-$ hyperons to investigate the $\Xi^-$-nucleus potential.
Since the observed level shifts and broadening of the width are connected to
the $\Xi^-$ potential in the nuclear surface region or even far from the surface,   
the information on the nuclear potential that is obtained  
is complementary to that resulting from spectroscopic studies of $\Xi$ hypernuclei.

In the E07 experiment, X-rays were also measured 
in coincidence with $\Xi^-$ production
using a Hyperball-X detector, which consisted of 6 sets of clover-type Ge detectors 
surrounded by BGO counters,  located upstream of the diamond target
and the emulsion module~\cite{E07YoshidaQNP2018}.
This represented an ambitious attempts to observe $\Xi^-$-atomic X-rays, for the first time,
from the $\Xi^-$ atoms of  heavy elements such as Ag and Br in the emulsion and C 
in the diamond target.
In the case of Ar or Br, $\Xi^-$ capture can be well identified by the scanning of the
emulsion, thus, good S/N can be achieved although the X-ray yields would be very tiny.
 
In the dedicated experiment, J-PARC E03, 
the choice of targets is the most important to observe finite shifts and to 
obtain the potential information.
C.J.~Batty {\it et al.} suggested a set of 4 optimum targets, namely
F (Z=9), C (Z=17),  I (Z=53), and Pb (Z=82) for 
$(n, l)$ = (3, 2), (4, 3), (7, 6), and (9, 8), respectively~\cite{Batty:1998fn}, 
and predicted shifts on the order of 1 keV for these states.
In the E03 experiment,  Fe (Z=26)  was selected as the target based on 
considering the production rate of $\Xi^-$,
the stopping probability of the $\Xi^-$, and X-ray absorption by the
target material.

\begin{figure}[htbp]
\centerline{\includegraphics[width=0.9\textwidth]{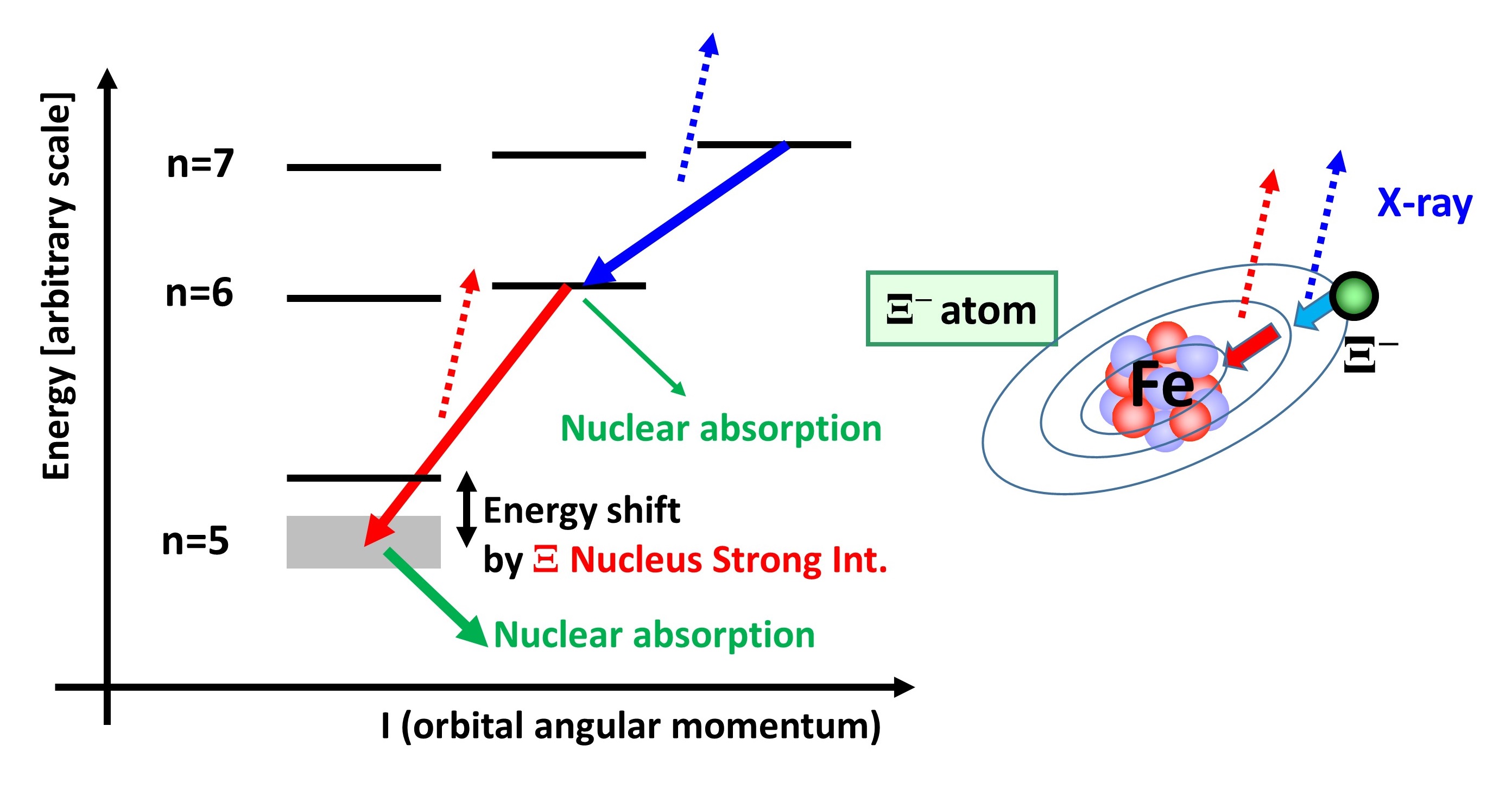}}
\caption{Schematic level scheme of $\Xi^{-}$-Fe atom.}
\label{XiFeLevel}
\end{figure}

Figure~\ref{XiFeLevel} shows a schematic level scheme for the  $\Xi^-$-Fe atom.
For the (5, 4) state, both an  energy shift and broadened width are expected 
due to the strong $\Xi^-$-nucleus interaction. 
In the case of a  Wood-Saxon potential of $-24 -3i $ MeV,
the expected shift and width ($\Gamma$) are 
$\sim$ 4 keV and $\sim$ 4 keV, respectively.
Therefore,  both 176 keV and   $\sim$ 286 keV X-rays will be measured
for the (7, 6)  $\rightarrow$ (6, 5) 
and  the  (6, 5) $\rightarrow$ (5, 4) transitions, respectively.
Neither a level shift nor width broadening are expected in the former X-ray,
while both a finite shift and width broadening are expected in the latter X-ray.
In the original proposal, more than 2500 counts
of  (6, 5) $\rightarrow$ (5, 4) transition X-ray will be measured
to obtain shift and width information
even in the case of strong absorption ($\Gamma \sim$ 4 keV).
Due to the similar reason to the E05 experiment, the original plan has been changed based on a two step strategy.
The goal of the first measurement is to obtain 10 \% of the statistics
in the original proposal.
With these data, the (7, 6) $\rightarrow$ (6, 5) transition X-rays
could be observed, although no shift would expected.
In the case of weak absorption ($\Gamma \sim$ 1 keV), 
it may be possible to observe
the  (6, 5) $\rightarrow$ (5, 4) transition X-ray and
to obtain values for the finite shift, width, and absorption strength.

The first measurement is planned in near future
using the KURAMA spectrometer for tagging the $\Xi^-$ production
and the Hyperball-X detector with a slightly different setup from that in the E07 experiment.

\section{Future Prospects \label{sec:exension}}
\begin{figure}[htbp]
\centerline{\includegraphics[width=0.95\textwidth]{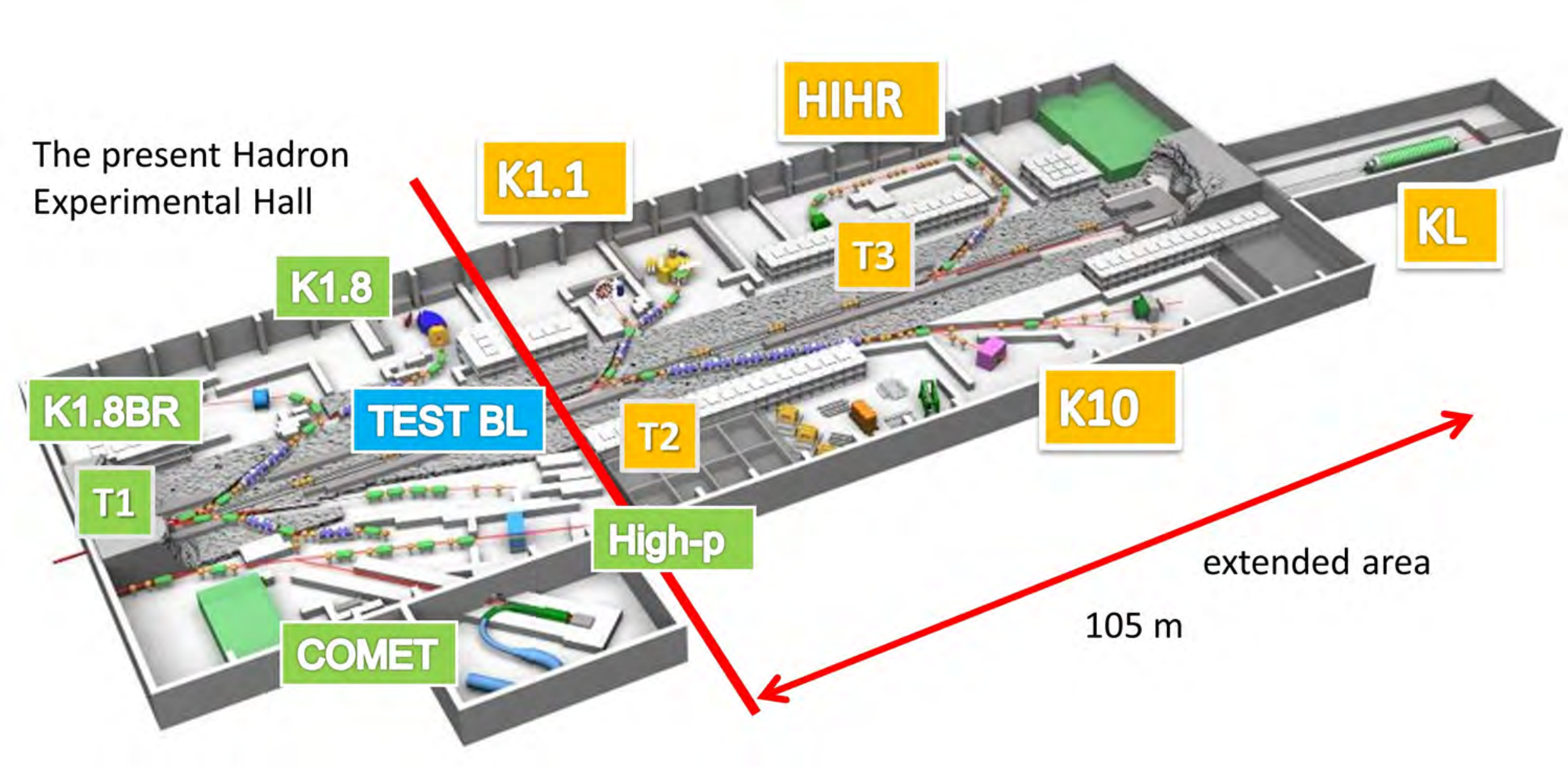}}
\caption{Layout plan of the extended HEF.}
\label{FigExtHEF}
\end{figure}

Many essential results on hadron and nuclear physics have been
obtained and will be obtained from the present HEF, as discussed in the
previous section.
However, to extend the physics programs and maintain these activities requires the capability of the HEF to be enhanced. 
Two new functions must be added, at least, 
to the HEF for future hadron and nuclear physics. 
One is a high intensity and high energy resolution beamline (HIHR) 
for high precision hadron and nuclear spectroscopy 
to utilize the high power beam from J-PARC MR. 
The other is a high intensity and high momentum particle-separated 
beamline (K10) to open a new door for heavy meson and 
baryon spectroscopy, for example.
Intensive studies and efforts have been made to realize 
these projects among the HEF users
from the beginning of HEF operation, 
and these discussions are summarized in the 
HEF Extension White Paper~\cite{HEFextWP1:2017, HEFextWP2:2019}.
Figure~\ref{FigExtHEF} shows a layout of the extended HEF together with the current HEF. 
The size of the experimental area will be three times larger and two 
new production targets will be placed.
Four beamlines plus one branched beamline, HIHR, K10, KL, K1.1, and K1.1BR, will be newly constructed. 
The specifications of the beamlines for both the present  and the extended HEF are
summarized in Table~\ref{BLspec}.

\begin{table}[htbp]
\caption{Specifications of beamlines in the present and the extended 
HEF.
Beam intensities at the present beamlines are typical and scaled to $\sim$ 80 kW.
Those at the new beamlines are the designed values with the same protons per pulse on
the production target. This table is taken from~\cite{HEFextWP1:2017} 
with modification. }
\begin{center}
\begin{tabular}{l|llll}
\hline
 & Particles	& Momentum	& Intensity	& Characteristics	\\
\hline
\multicolumn{5}{l}{
Beamlines in the present hadron experimental facility} \\
\hline
K1.8		& $K^{\pm}, \pi^{\pm}$ 		& 1.0 -- 2.0 GeV/{\it c}	
 &	$\sim$ 10$^6$ $K^-$ /spill (1.8)	& mass separated	\\
K1.8BR	& $K^{\pm}, \pi^{\pm}$		& $<$ 1.1 GeV/{\it c}
 &	$\sim$ 5$\times$10$^5$ $K^-$ /spill (1.0)	& mass separated	\\
KL		& $K_L$	& 2.1 GeV/{\it c} in ave.	& 10$^7$ $K_L$ /spill	
 & 16$^\circ$ extraction angle	\\
high-p    & $p$	&	& 10$^{10}$ $p$ /spill	& primary beam (30 GeV)	\\
		& $\pi^{\pm}$	&	$<$ 31 GeV/{\it c}	& 10$^7$ $\pi$ /spill	& secondary beam \\
COMET	& $\mu^-$	&	&	& for $\mu^- \rightarrow e^-$  experiment \\
\hline
\multicolumn{5}{l}{
Beamlines in the extended area} \\
\hline
K1.1		& $K^{\pm}, \pi^{\pm}$	& $<$ 1.2 GeV/{\it c}	
 & $\sim$ 10$^6$ $K^-$ /spill (1.1)	& mass separated	\\
K1.1BR	& $K^{\pm}, \pi^{\pm}$	& 0.7 -- 0.8 GeV{\it c}
 & $\sim$ 10$^6$ $K^-$ /spill 	& mass separated	\\
HIHR	& $\pi^\pm$	&	$<$ 2.0 GeV/{\it c}
 & $\sim$ 3$\times$10$^8$ $\pi$ /spill	& mass separated	\\
	&	&	&	& $\times$10 better $\Delta p/p$	\\
K10	& $K^\pm, \pi^\pm, \overline{p}$		& $<$ 10 GeV/{\it c}
 & $\sim$ 10$^7$ $K^-$ /spill	& mass separated \\
KL	& $K_L$	& 5.2 GeV/{\it c} in ave.	
 & 10$^8$ $K_L$ /spill	& 5$^\circ$ extraction angle \\
	&	&	&	& optimized $n/K_L$	\\
\hline
\end{tabular}
\end{center}
\label{BLspec}
\end{table}

Physics at the new functional beamlines, HIHR and K10, and
their designs are briefly described here.
Details can be obtained from the HEF Extension White 
Paper~\cite{HEFextWP1:2017, HEFextWP2:2019}.

\subsection{Physics at HIHR}

The properties of high-density nuclear or hadronic matter have attracted  
interest in connection with neutron stars, the highest density objects in the universe.
Considering that the rather high Fermi-energy with neutrons deep inside of
a neutron star can be reduced by converting neutrons to hyperons
and that the attractive hyperon interaction in a nucleus, 
at least for $\Lambda$, is experimentally confirmed,
it is naturally expected that hyperons exist deep inside of neutron stars.
The internal structure of a neutron star is theoretically studied with an equation of state (EOS)
for high-density nuclear matter.
There are several versions of the EOS that were constructed
using realistic nuclear models with the knowledge on the 
properties of nuclei and interactions of nucleons or hadrons.
The established models including hyperon interactions predict
that hyperons appear in the inner core of neutron stars and
the maximum mass of neutron stars is approximately 1.5 solar mass. 

However, this natural conclusion from nuclear physics conflicts 
with the existence of 2 solar mass neutron 
stars~\cite{Demorest:2010bx, Antoniadis:2013pzd}.
This discrepancy between the astronomical observations and 
the theoretical predictions in nuclear physics is referred to as the ``hyperon puzzle''.
The appearance of hyperons in high-density nuclear matter means
the EOS must be softened and therefore such a softened EOS can not support
heavy neutron stars.
The hyperon puzzle thus is a significant challenge to be 
resolved in nuclear physics.

One of the promising approaches to solve the hyperon puzzle is to
introduce a three-body repulsive force (TBRF) in the baryonic 
interactions with hyperons.
The idea of TBRF has been discussed to overcome the softening 
problem of the EOS, by introducing TBRF ``universally'', {\it i.e.} equally
to the $YN$ and $YY$ parts, as well as $NN$ 
part~\cite{Nishizaki:2001in, Nishizaki:2002ih}.
The effects of three-body force in ordinary nuclei with $S=0$
has already been observed, for example, in the analyzing powers for the 
$d p$ elastic scattering,~\cite{Sekiguchi:2011ku}
and the repulsive effect was suggested in the theoretical
analysis of $^{16}$O + $^{16}$O elastic scattering~\cite{Furumoto:2009zza}.

A baryonic TBRF affects the single particle
energy of a hyperon that is bound deep inside of a hypernucleus.
The effect on the $\Lambda$ binding energy is expected to be
approximately a half MeV, which is smaller than the achievable 
experimental resolution and hence the accuracy of the present
apparatus for the missing-mass spectroscopy using meson beams.
The large momentum spread of the secondary meson
beams means that the momenta of the beam particles should be directly measured
event-by-event; therefore, the acceptable beam intensity is limited
and a thick target that degrades the experimental resolution
is necessary at present.

By adopting dispersion matching techniques~\cite{Fuiita:1997dispmatch}, 
direct measurement of the beam particle is not necessary in the HIHR.
Therefore, the HIHR enables $(\pi^\pm, K^-)$ 
missing-mass spectroscopy to be performed with a resolution of a few 100 keV (FWHM) 
through the use of high-intensity pion beams and relatively thin nuclear targets.
High-resolution and high-precision spectroscopy on light to heavy
$\Lambda$ hypernuclei will reveal the existence of TBRF in
hyperon interactions and hopefully solve the hyperon puzzle.

\subsection{Physics at K10}

The K10 beamline at the extended HEF is designed to handle a secondary beam up to 10 GeV/$c$ while maintaining the intensity as high as possible. The separated beam will be available with momentum up to 4 GeV/$c$ and 6 GeV/$c$, for $K^-$ and $\bar{p}$, respectively. With the current design, 27~m of electrostatic (ES) separator will be used for particle separation. An RF-type separator is under discussion instead of the ES separator to extend the particle separation capability~\cite{HEFextWP1:2017, HEFextWP2:2019}. 

The $\Omega$ baryon nucleon interaction is one of the hot topics in hadron physics. Recently, a lattice QCD calculation suggested an attractive interaction between an $\Omega$ baryon and a nucleon~\cite{Iritani:2018sra}, which leads to the existence of the $\Omega$-nucleon bound state. An $\Omega$ baryon could be produced via $(K^-, K^+K^{0*})$ interaction by utilizing a high momentum $K^-$ beam. The first project could be a measurement of the $\Omega$ production cross section in the production threshold region, {\it i.e.} a $K^-$ beam momentum of approximately 6 GeV/$c$. At the same time, a significant amount of excited $\Xi$ baryons are also produced, which would allow for $\Xi^*$ baryon spectroscopy to reveal the hadron structure with S=-2 baryons. High intensity and high momentum $K^-$, which will only be available at the K10 beamline, will allow direct $\Omega-p$ and $\Xi^- -p$ scattering experiments to be performed. Such information is required to achieve a complete understanding of the baryon-baryon interaction in the $SU_f(3)$  framework~\cite{HEFextWP1:2017, HEFextWP2:2019}.

Investigation on charmed meson and baryon interactions is also an important subject. The formation of a charmed meson and nucleus bound state is expected~\cite{Hayashigaki:2000es,Tsushima:1998ru,Sibirtsev:1999js} due to the attractive interaction between a charmed meson and a nucleon. To produce a charmed meson efficiently requires a high intensity anti-proton beam, which is available at the K10 beamline. However, there is no knowledge regarding $D,\bar{D}$ production with antiprotons around the production threshold; therefore, measurement of the production cross section of a $D\bar{D}$ pair at the production threshold with a $\bar{p}$ beam will be the first physics program at the K10 beamline. Once the $D\bar{D}$ production cross section is determined, the feasibility of a search for the D meson bound nucleus will be clarified. Moreover, for example, 7~GeV/$c$ $\bar{p}$ on proton is above the production threshold of X(3872), which is known to be an exotic hadron candidate. Therefore, the study of X(3872) in a nucleus will also be a subject of interest.  

Details will be reviewed in Refs.~\cite{HEFextWP1:2017} and \cite{HEFextWP2:2019} 

\section{Summary \label{sec:summary}}
One of the biggest goals in hadron physics is to reveal
fruitful phenomena predicted by QCD, especially in the low-energy
non-perturbative regime.
The J-PARC HEF was constructed with an aim to explore such low-energy QCD
dynamics through the use of various high-intensity secondary beams as well as 30 GeV
primary proton beams.
Since the first beam was delivered to the J-PARC HEF in February 2009,
various experiments have been performed and proposed.
In this review, we have summarized the experimental results and plans
that have focused on the study of the fundamental components of matter and
their interactions.
These experiments have been made via a wide variety of approaches, and thus
have provided extremely broad and rich results.

To access information on the QCD vacuum structure, missing
resonances in the non-strangeness sector will be investigated at the E45 experiment
with partial wave analysis of the $\pi N \to \pi \pi N$ and $\pi N \to KY$
reactions.
In the strangeness sector, the E72 experiment aims to determine the spin and
parity of a $\Lambda^*$ resonance at 1669 MeV produced via
the $p(K^-,\Lambda)\eta$ reaction.
The exotic five-quark state with strangeness $S=+1$, pentaquark baryon
$\Theta^+$, was searched for during the E19 experiment; however, 
no corresponding structure was observed in the missing-mass spectrum 
of $p(\pi^-,K^-)X$. 
The $H$ dibaryon, six-quark state with $S=-2$ will be
searched for in the E42 experiment
using the Hyperon Spectrometer  
with statistics an order of 100 times higher than with 
the previous experiment performed at KEK.

The introduction of heavy-quark symmetry is one of the best ways to study
the effective degrees of freedom to describe a hadron itself.
The experiment to show the di-quark correlation in charmed baryons 
is planned as the E50 experiment with the $(\pi^-,D^{*-})$ reaction at
up to 20 GeV/$c$.
To investigate the origin of the QCD mass generated from the spontaneous
breaking of chiral symmetry, the E16 collaboration will conduct
a precise measurement of the vector-meson spectral function in medium via
a dielectron channel with the 30 GeV primary proton beam.
The E26 and
E29 experiments are also planned to search for the $\omega$- and
$\phi$-nucleus bound states with the $(\pi^-,n)$ and $(\bar p, \phi)$
reactions, respectively.

The kaon-nucleon interaction ($\bar K N$) close to the mass thresholds
provides crucial information on the interplay between spontaneous and
explicit chiral symmetry braking.
To determine the individual isoscaler and isovector scattering lengths
of the $\bar K N$ interaction, the E57 experiment will measure the
shift and width of the kaonic-deuterium $1s$ state using many arrays of
silicon drift detectors (SDDs) covering a large solid angle.
For the purpose of obtaining the isospin dependence of the $\bar K N$
interaction, high-resolution measurement of the kaonic-$^4$He
and $^3$He atoms was performed by the E62 experiment with a
superconducting transition-edge-sensor (TES) microcalorimeter having 
resolution one order of magnitude better than conventional semi-conductor
detectors.

The $\Lambda(1405)$ state, which is widely recognized as a $\bar K N$
quasi-bound state in the $I=0$ channel, provides essential information on the
$\bar K N$ interaction below the mass threshold.
The E31 experiment was conducted to exclusively demonstrate the
$\Lambda(1405)$ line shape via the $\bar K N \to \pi \Sigma$ channels
using the $(K^-,n)$ reaction on a deuterium target at 1.0 GeV/$c$.
The result clearly shows the interference between the $I=0$ and $I=1$
amplitudes in the $\pi^{\pm}\Sigma^{\mp}$ spectra, as theoretically
expected.
The $\bar K NN$ quasi-bound state, which is one of the kaon-nucleus quasi-bound
systems from the strongly attractive $\bar K N$ interaction in the
$I=0$ channel, was searched in two experiments with different reactions.
The E27 collaboration measured the exclusive reaction of $\pi^+ d \to
K^+X$ followed by $X \to \Sigma^0 p$ and $\Lambda p$ decays at 1.69
GeV/$c$.
A broad enhancement in the $d(\pi^+,K^+)\Sigma^0 p$ missing-mass
spectrum was observed around 2.27 GeV/$c^2$, which
corresponds to a $\bar KNN$ binding energy of $\sim$ 100 MeV.
With the $(K^-,n)$ reaction on a $^3$He target at 1.0
GeV/$c$, the E15 collaboration  observed a distinct peak with a binding
energy of $\sim$ 50 MeV
in the $\Lambda p$ invariant mass of the $\Lambda pn$ final state.
Further investigation of the kaonic nuclei is planned using a new
large 4$\pi$ spectrometer with neutral-particle detectors to
systematically explore the light kaonic nuclei, from the most
fundamental ``$\bar K N$'' state ({\it i.e.}, the $\Lambda(1405)$ state) to the
``$\bar K NNNN$'' via the $(K^-,n)$ reactions.

Extension of the nuclear physics to the strangeness sector is essential to
obtain further understanding of the interactions between octet
baryons under flavor SU(3) symmetry.
In the $S=-1$ nuclear system, a neutron-rich hypernucleus $^6_{\Lambda}$H
was investigated in the E10 experiment via the $^6$Li$(\pi^-,K^+)$ reaction
to reveal the effect of the $\Sigma N - \Lambda N$ coupling and to obtain
information on the $\Lambda$ potential in a neutron-rich environment.
However, no peak was observed in the bound region of $\Lambda$.
In the E13 experiment, hypernuclear $\gamma$-rays from $^4_{\Lambda}$He
and $^{19}_{\Lambda}$F were measured by Ge detectors in coincidence with
hypernuclear production via the ($K^-,\pi^-$) reaction identified by the
superconducting kaon spectrometer (SKS).
The excitation energy of the $^4_{\Lambda}$He $1^+$ state obtained 
clearly indicated a large charge symmetry breaking
between $^4_{\Lambda}$H and $^4_{\Lambda}$He.
The energy spacing of the ground-state doublet of $^{19}_{\Lambda}$F 
is in good agreement with theoretical calculations, which shows that the
theoretical framework of the $\Lambda N$ interaction is valid, even in
heavier hypernuclei.
A modern $\Sigma^{\pm} p$ scattering experiment is being performed
as the E40 experiment with very high intensity $\pi^\pm$ beams and 
liquid H$_2$ for the $\Sigma^{\pm}$ production and scattering targets.
The experiment aims to investigate the quark Pauli effect
in the $\Sigma N (I=3/2)$ interactions by measurement of the
differential cross sections and phase shifts.

In the J-PARC HEF, efficient production of $S=-2$ systems is
possible using the world's highest-intensity $K^-$ beam.
To clarify the $\Xi N$ interaction including $\Xi N$-$\Lambda \Lambda$
coupling, the spectroscopic study of $\Xi$ hypernuclei via the
$(K^-,K^+)$ reaction (E05 and E70) and the study of $\Xi$-atomic
X-rays (E03 and E07) were proposed.
Both the E05 and E70 experiments aims to measure the $^{12}_{\Xi}$Be
hypernucleus; the E05 experiment was conducted as a pilot run using 
the existing SKS. A new high-resolution S-2S spectrometer will be
constructed for the E70 experiment.
The E07 experiment is a third-generation counter-emulsion hybrid
experiment to study double strangeness nuclei such as double-$\Lambda$
and $\Xi$ hypernuclei.
The newly observed {\it MINO} event shows the binding (bonding) energy
of two $\Lambda$ hyperons to be 19.07$\pm$0.11 (1.87$\pm$0.37) MeV
for the $^{11}_{\Lambda\Lambda}$Be hypernucleus, which is the most
likely explanation for the observed event.
The {\it IBUKI} event was also uniquely identified as a $\Xi^- + ^{14}$N bound
nuclear system that gives $B_{\Xi} = 1.27 \pm 0.21$ MeV.
Many double strangeness nuclear events are expected to be found in
future analysis.
The E03 experiment will also measure the X-ray transitions 
from the $\Xi$-Fe atom to obtain the
$\Xi$-nucleus potential around the nuclear surface region.

At the current HEF, many essential results on hadron and nuclear
physics have been achieved and will be obtained from a wide variety of
physics programs.
However, enhancement of the HEF capabilities is mandatory to expand the physics programs to a region that has never been explored.
In the HEF extension project, construction of two new beamlines is
planned with the highest priority:
a high-intensity and high-energy-resolution beamline (HIHR) and
a high-momentum particle-separated beamline (K10).
The HIHR and K10 beamlines are aimed at the high precision spectroscopy
of $\Lambda$ hypernuclei to provide keys to resolve the ``hyperon
puzzle'', and the heavy meson and baryon spectroscopies together with
the investigation of $\Omega$-nucleon interaction as well as charmed
meson-nucleon interactions.
Intensive study and efforts
have thus been made among the HEF user community toward the realization of this project.

\section{Acknowledgments}
We are grateful to all the staff members of J-PARC/KEK/JAEA for their
extensive efforts on the successful operation of the facility, and thank
all of collaborators and colleagues from the experiments performed in
the J-PARC HEF.
We are also grateful to the fruitful discussions with Prof. Tomokazu
Fukuda and Prof. Makoto Oka.
This work is partly supported by the MEXT Grants-in-Aid for Scientific
Research on Innovative Areas ``Clustering as a window on the hierarchical
structure of quantum systems''  Group-A02 (18H05402) and Group-B01 (18H05403),
and ``Nuclear matter in neutron stars investigated by experiments and astronomical
observations'' Group-A01 (No.24105002),  
and JSPS Grants-in-Aid for Scientific Research (C) 17K05481 and (B) 18H01237,
in Japan.
Finally we would like to thank the Editor-in-Chief of PPNP
for giving us the opportunity to contribute this review paper to PPNP
in the 10th anniversary year of J-PARC.

\input{hd.bbl}

\end{document}

%% file: hd.bbl
\providecommand{\href}[2]{#2}\begingroup\raggedright\endgroup